\documentclass[a4paper,11pt]{article}
\pdfoutput=1 

\usepackage{jheppub} 

\usepackage[T1]{fontenc} 

\usepackage{lmodern}
\usepackage{enumitem}
\usepackage{slashed}
\usepackage{xcolor}
\usepackage{ textcomp }
\usepackage{amsmath} 
\usepackage{ mathrsfs }
\usepackage{ stmaryrd }
\usepackage{ tipa }
\usepackage{ upgreek }
\usepackage{ dsfont }
\usepackage{braket}
\usepackage{tikz}
\usepackage{comment}
\usepackage{bm}

\newcommand{\tn}[1]{\textnormal{#1}}

\title{The Two Scales of New Physics in Loop-Induced Higgs Couplings}

\author[a]{Florian Nortier,}
\author[b]{Gabriele Rigo}
\author[b]{and Pablo Sesma}

\affiliation[a]{Université Claude Bernard Lyon 1, CNRS/IN2P3, IP2I Lyon, UMR 5822,\\ Villeurbanne, F-69100, France}
\affiliation[b]{Université Paris-Saclay, CNRS, CEA, Institut de Physique Théorique,\\ 91191, Gif-sur-Yvette, France}

\emailAdd{f.nortier@ip2i.in2p3.fr}
\emailAdd{gabriele.rigo@ipht.fr}
\emailAdd{pablo.sesma@ipht.fr}

\abstract{Probing new physics through precise measurements of Higgs boson couplings is a central objective of the particle collider program at the high-energy frontier. An anomaly in Higgs couplings induced solely by new fermions allows one to compute an upper bound on the mass scale of new bosons. This new bosonic scale is necessary to prevent Landau poles or vacuum instability. Consequently, a single anomalous measurement can provide insight into two distinct new physics scales. In this article, we apply this approach to the loop-induced couplings of the Higgs boson to digluons ($gg$), diphotons ($\gamma \gamma$), and $Z \gamma$, and we compare our results to the projected sensitivities of the HL-LHC and future lepton colliders. This work naturally extends our previous analysis of Higgs couplings to weak dibosons ($WW$ and $ZZ$).}


\arxivnumber{2412.14237}

\begin{document}

\maketitle
\flushbottom


\section{Introduction}
 \label{introduction}
The discovery of a Higgs-like boson, $h$, by the ATLAS and CMS experiments at the CERN LHC~\cite{ATLAS:2012yve, CMS:2012qbp} represents a landmark achievement in validating the Electroweak (EW) Theory~\cite{Glashow:1961tr, Weinberg:1967tq, Salam:1968rm}, a cornerstone of the Standard Model (SM) of particle physics~\cite{Quevedo:2024kmy}. This success is grounded in the Higgs mechanism~\cite{Nambu:1960tm, Schwinger:1962tn, Anderson:1963pc, Higgs:1964ia, Englert:1964et, Higgs:1964pj, Guralnik:1964eu, Higgs:1966ev, Migdal:1966tq, Kibble:1967sv, Guralnik:2011zz}. To date, the observed properties of the $h$-boson are consistent with those predicted for the SM Higgs boson~\cite{CMS:2022dwd, ATLAS:2022vkf, ParticleDataGroup:2024cfk}, whose phenomenology has been extensively studied since the SM’s inception~\cite{Gunion:1989we, Djouadi:2005gi, Dawson:2018dcd}. Detecting any deviations in the Higgs couplings from SM predictions remains a central objective of the CERN experimental program over the coming decades, both at the LHC and in future projects such as the FCC, a linear $e^+e^-$ facility and/or a circular muon collider~\cite{Dawson:2022zbb}.

However, the lack of evidence for physics beyond the SM (BSM) at the TeV scale~\cite{ParticleDataGroup:2024cfk} has led to what some refer to as the “Physicists’ Nightmare Scenario: The Higgs and Nothing Else”~\cite{Cho:2007cb}. In this scenario, the long-standing EW hierarchy problem~\cite{Wilson:1970ag, Weinberg:1975gm, Gildener:1976ai, Susskind:1978ms, tHooft:1979rat, Veltman:1980mj, Kolda:2000wi} remains unresolved. Consequently, the naturalness principle~\cite{Giudice:2008bi, Giudice:2013yca, Giudice:2017pzm, Craig:2022uua}, which has guided model-building for four decades, faces significant challenges. In light of this situation, the focus has shifted to a model-independent approach for probing new physics at the LHC, driven by the Effective Field Theory (EFT) framework~\cite{Falkowski:2023hsg}. In Quantum Field Theory (QFT), any deviation from SM predictions can be parametrized by higher-dimensional operators added to the SM. These operators are then matched to specific ultraviolet (UV) completions. This methodology enables estimation of the interaction scale $\Lambda_\tn{NP}$, below which new physics must emerge to account for observed anomalies.
 
 In some cases, we can gain deeper insights by taking one step beyond the EFT approach. Consider purely fermionic extensions of the SM at a mass scale $\Lambda_\tn{F}$ that satisfy the following criteria~\cite{Bizot:2015zaa}: (i) they are self-consistent low-energy theories; (ii) they can modify the Higgs couplings to other SM particles; and (iii) they remain compatible with current experimental constraints. If these new fermions produce observable anomalous Higgs couplings at present or future colliders (i.e., $\Lambda_\tn{F} \lesssim \Lambda_\tn{NP}$), their presence implies a UV cutoff scale beyond which the theory faces significant issues. Specifically, the UV completion could either lose perturbative control (Landau pole) or induce an unacceptable level of metastability in the Higgs potential, conflicting with the observed stability of the Universe over its lifetime\footnote{See also Refs.~\cite{Branchina:2013jra, Branchina:2014usa, Branchina:2015nda, Branchina:2016bws, Bentivegna:2017qry, Branchina:2018xdh, Branchina:2018qlf, Branchina:2019tyy} for destabilization of the Higgs potential via Planck-suppressed higher-dimensional operators in an EFT framework.}~\cite{Gogoladze:2008ak, Chen:2012faa, Joglekar:2012vc, Kearney:2012zi, Reece:2012gi, Batell:2012ca, Fairbairn:2013xaa, Altmannshofer:2013zba, Xiao:2014kba, Ellis:2014dza, Angelescu:2016mhl, Goswami:2018jar, Gopalakrishna:2018uxn, Borah:2020nsz, Bandyopadhyay:2020djh, Hiller:2022rla, Arsenault:2022xty, Cingiloglu:2023ylm, Adhikary:2024esf, Cingiloglu:2024vdh}. The associated new scale, $\Lambda_\tn{B} \geq \Lambda_\tn{F}$, serves as an upper bound for the energy scale at which new bosonic states must emerge to resolve these instabilities in the running of the couplings\footnote{This conclusion assumes the UV completion is either a standard QFT or a perturbative string theory. For example, if $\Lambda_\tn{B}$ corresponds to the string scale, then higher-spin bosonic excitations of SM particles are expected to appear near this scale.}. These new bosons could provide definitive evidence for a symmetry protecting the weak scale, or instead indicate an underlying unnaturalness, particularly if they themselves are unstable under radiative corrections. This reasoning has been previously applied to the Higgs couplings for select fermion representations, as explored in Refs.~\cite{Arkani-Hamed:2012dcq, Blum:2015rpa}.
 
In our companion article~\cite{DAgnolo:2023rnh}, we applied this strategy to the Higgs couplings $hWW$ and $hZZ$, leading to two key findings: (i) any deviation detectable within the sensitivity of the upcoming HL-LHC would require contributions from new bosons, with $\Lambda_\tn{B} \sim \Lambda_\tn{NP}$; and (ii) deviations arising solely from new fermions could be observable at future circular colliders (e.g., the FCC). In this article, we extend the analysis to the couplings $hgg$, $h\gamma\gamma$, and $hZ\gamma$, comparing our results to the projections~\cite{deBlas:2022ofj} for the HL-LHC and future lepton colliders (FLCs). A shared characteristic of these three Higgs couplings is their absence at tree level, meaning they arise from loop processes involving virtual particles and are UV-finite at 1-loop. At hadronic colliders like the LHC, mono-Higgs production via gluon fusion ($gg \to h$) is the dominant production mechanism~\cite{Gunion:1989we, Djouadi:2005gi, Dawson:2018dcd}. Historically, the diphoton decay ($h \to \gamma\gamma$) served as the discovery channel for the $h$-boson~\cite{ATLAS:2012yve, CMS:2012qbp}, while evidence for the rare decay $h \to Z\gamma$ has only recently been reported by the ATLAS and CMS experiments~\cite{ATLAS:2023yqk}. We emphasize that the effects of new fermions on these loop-induced Higgs couplings have been extensively studied by numerous authors for various purposes~\cite{Djouadi:2007fm, Krauss:2007bz, Cacciapaglia:2009ky, Bouchart:2009vq, Gopalakrishna:2009yz, Bhattacharyya:2009nb, Casagrande:2010si, Azatov:2010pf, Alves:2011kc, Azatov:2011qy, Goertz:2011hj, Ishiwata:2011hr, Reece:2012gi, Arkani-Hamed:2012dcq, Batell:2012ca, Azatov:2012rj, Bonne:2012im, Moreau:2012da, Joglekar:2012vc, Carena:2012xa, Wang:2012gm, Ajaib:2012eb, Kearney:2012zi, Voloshin:2012tv, Frank:2012nb, Carmi:2012yp, Basso:2012nh, Kumar:2012ww, Feng:2013mea, Chen:2013dh, Frank:2013un, Englert:2013tya, Malm:2013jia, Dey:2013cqa, Hahn:2013nza, Altmannshofer:2013zba, Dermisek:2013gta, Aguilar-Saavedra:2013qpa, Carmona:2013cq, Delaunay:2013iia, Ellis:2014dza, Xiao:2014kba, Malm:2014gha, Dey:2015pba, Blum:2015rpa, Angelescu:2015kga, Bizot:2015zaa, Lalak:2015xea, Dermisek:2015hue, Angelescu:2016mhl, Arhrib:2016rlj, Hashimoto:2017jvc, Chen:2017hak, Poh:2017tfo, Aboubrahim:2018hll, DAgnolo:2023rnh, Barducci:2023zml, Adhikary:2024esf}.
 
This article is organized as follows. In Section~\ref{Setup}, we outline the main framework of our study, describe the specific class of fermionic models under consideration and discuss the key constraints on the new particles. In Section~\ref{upper_bos}, we determine the upper bound, $\Lambda_\tn{B}$, on the mass scale of new bosons for each relevant model, focusing on the $hgg$, $h\gamma\gamma$, and $hZ\gamma$ couplings. Finally, in Section~\ref{conclusion}, we summarize our findings and their implications.

\section{From New Fermions to New Bosons}
\label{Setup}

\subsection{Goal \& Strategy}
\label{Idea}
In this section, we present the objective and approach of our study. A more comprehensive discussion of the supporting arguments is provided in our companion article~\cite{DAgnolo:2023rnh}.

Consider a scenario in which an experiment uncovers deviations in certain Higgs couplings relative to SM predictions. Within the framework of an EFT analysis, such deviations are conventionally parameterized by augmenting the SM Lagrangian with higher-dimensional operators constructed from SM fields:
\begin{equation}
\sum_i \dfrac{c_i}{M_\tn{NP}^{d_i-4}} \, \mathcal{O}_i^{(d_i)} \, ,
\end{equation}
where $\mathcal{O}_i^{(d_i)}$ represents an operator of dimension $d_i \in \mathbb{N}$, $M_\tn{NP}$ denotes the mass scale of new physics, and $c_i \in \mathbb{C}$ are dimensionless Wilson coefficients determined by fitting the observed anomalies.

Now, let us focus on the case where the new degrees of freedom at the scale $M_\tn{NP}$, which are integrated out to generate the aforementioned higher-dimensional operators, consist solely of fermions, with $\Lambda_\tn{F} \equiv M_\tn{NP}$. It is well-established that such a UV completion of the SM, for sufficiently large couplings to the $h$-boson, is valid only within a finite energy range~\cite{Arkani-Hamed:2012dcq, Blum:2015rpa, DAgnolo:2023rnh}. In particular, these fermions can induce a Landau pole in the Yukawa couplings,
\begin{equation}
\dfrac{d y}{d \log \mu} \sim \dfrac{y^3}{16 \pi^2} \, ,
\end{equation}
or change the sign of the Higgs quartic coupling,
\begin{equation}
\dfrac{d \lambda}{d \log \mu} \sim - \dfrac{y^4}{16 \pi^2} \, .
\end{equation}
Therefore, within a QFT framework, new bosons are expected to emerge below a scale $\Lambda_\tn{B}$ to resolve such instabilities. The primary objective of this work is to identify, for various pure fermionic extensions of the SM, the models that allow for a hierarchy between $\Lambda_\tn{F}$ and $\Lambda_\tn{B}$, even if such a hierarchy exists only in a fine-tuned region of the parameter space of the fermionic model(s). This approach corresponds to a conservative perspective, as opposed to a naturalness-driven one.

In this analysis, we focus on the couplings $hgg$, $h \gamma \gamma$, and $h Z \gamma$, and quantify deviations as~\cite{deBlas:2022ofj}
\begin{equation}
\delta \mu_{hVV^\prime} =  \sqrt{\dfrac{\Gamma(h  \rightarrow VV^\prime)}{\Gamma^\mathrm{SM}(h \rightarrow VV^\prime)}}-1 \, ,
\label{delta_HVV}
\end{equation}
with $V, V^\prime \equiv g, \gamma, Z$. One can then compute the upper bound $\Lambda_\tn{B}$ on the mass scale of new bosons. This upper bound is determined as the minimum of the following scales:

\begin{itemize}
\item \emph{Landau pole:} $y^{(c)}(\mu = \Lambda_\tn{B}) = 4\pi$, which we define as the threshold beyond which perturbative control over the Yukawa couplings is lost\footnote{This perturbative criterion is based on dimensional analysis: if we are interested in perturbativity of the Yukawa couplings, the 1-loop diagrams studied in Section~\ref{upper_bos} scale as $yy^c/16\pi^2$. The exact threshold is somewhat arbitrary, since one expects to lose control of the Feynman expansion before reaching $4\pi$. Nevertheless, the running is fast when the Yukawa couplings approach $4\pi$, and changing at $\mathcal{O}(1)$ the upper bound on $y^{(c)}$ does not significantly impact the value of $\Lambda_\tn{B}$. This was checked numerically with the 2-loop RGEs in Ref.~\cite{Blum:2015rpa} by choosing e.g.\ $\sqrt{4\pi}$ as the upper bound instead of $4\pi$.}.
\item \emph{Vacuum instability:} $1/\lambda (\mu = \Lambda_\tn{B}) = -14.53 + 0.153 \log(\text{GeV}/\Lambda_\tn{B})$. This criterion is based on the gauge-invariant condition\footnote{In our companion paper~\cite{DAgnolo:2023rnh}, we checked numerically that using the 2-loop improved effective potential (which is gauge dependent) does not significantly modify the value of $\Lambda_\tn{B}$.} that the quartic self-interaction becomes negative at the renormalization scale $\mu = \Lambda_\tn{B}$, indicating a Universe lifetime shorter than its current age~\cite{Isidori:2001bm, Elias-Miro:2011sqh, Degrassi:2012ry, Buttazzo:2013uya, Devoto:2022qen}.
\end{itemize}

\subsection{Extension with Vectorlike Fermions}
\label{Fermions}
Pure fermionic extensions of the SM that are phenomenologically viable, free from gauge anomalies, and influence the Higgs couplings were classified in Ref.~\cite{Bizot:2015zaa}. We adopt the notation $(a,b)_Y$, where $a$ denotes the dimension of the $SU(3)_C$ representation, $b$ represents the dimension of the $SU(2)_W$ representation, and $Y$ is the hypercharge. In our companion article~\cite{DAgnolo:2023rnh}, we demonstrated that our objective can be achieved by focusing on a model with $N_{\mathrm{F}}$ copies (flavors) of two vectorlike fermions\footnote{We use the 2-component notation for spinors, as described in Ref.~\cite{Dreiner:2008tw}, e.g., for a Dirac spinor
\begin{equation}
\Psi =
\begin{pmatrix}
\psi_\alpha\\
\psi^{c\dagger\dot{\alpha}}
\end{pmatrix}.
\end{equation}
}:
\begin{equation}
L = (r, n)_Y \, , \ \ \ L^c = (\bar{r}, n)_{-Y} \, , \ \ \ E = (r, n-1)_{Y^\prime} \, , \ \ \ E^c = (\bar{r}, n-1)_{-Y^\prime} \, .
\end{equation}
The renormalizable Lagrangian for one copy has the following mass terms and perturbative Yukawa couplings:
\begin{equation}
- M_L \ L L^c - M_E \ E E^c - y \ L H E^c - y^c \ L^c H^{\dagger} E +\text{H.c.} \, ,
\label{Lag1}
\end{equation}
where $H$ is the Higgs field $(1, 2)_{1/2}$. In Appendix~\ref{su2irreps}, we describe how to systematically combine
the different representation dimensions of the VLFs in this model. Gauge invariance imposes the following relation between hypercharges:
\begin{equation}
Y^\prime=Y+\frac{1}{2} \, .
\end{equation}
For simplicity, we do not consider Yukawa couplings that mix different VLF flavors, as this approach allows us to treat all possible representation dimensions in full generality for any $N_\mathrm{F}$. In the perturbative regime, the decoupling limit is defined by $M_{L/E} \gg y^{(c)} v$, where $v$ is the vacuum expectation value (VEV) of $H$. To avoid stringent constraints from $CP$ violation and flavor physics, we assume the parameters in Eq.~\eqref{Lag1} to be real and consider no mixing with the SM fermions (or only a negligible one with the correct quantum numbers). Once again, our aim is to determine the most conservative upper bounds on the scale $\Lambda_\tn{B}$ for different fermion representations, rather than proposing a natural model. Appendix~\ref{app_Lag} provides the detailed computation of the fermion mass spectrum and their couplings to the SM bosons.

\subsection{Constraints on Vectorlike Fermions}
\label{constraints}
In this section, we provide a brief summary of the theoretical and experimental constraints on VLFs, as discussed in greater detail in our companion article~\cite{DAgnolo:2023rnh}, along with some updates.

\subsubsection{Representation Dimensions}
\begin{figure}[t]
\begin{center}
\includegraphics[width=7.5cm]{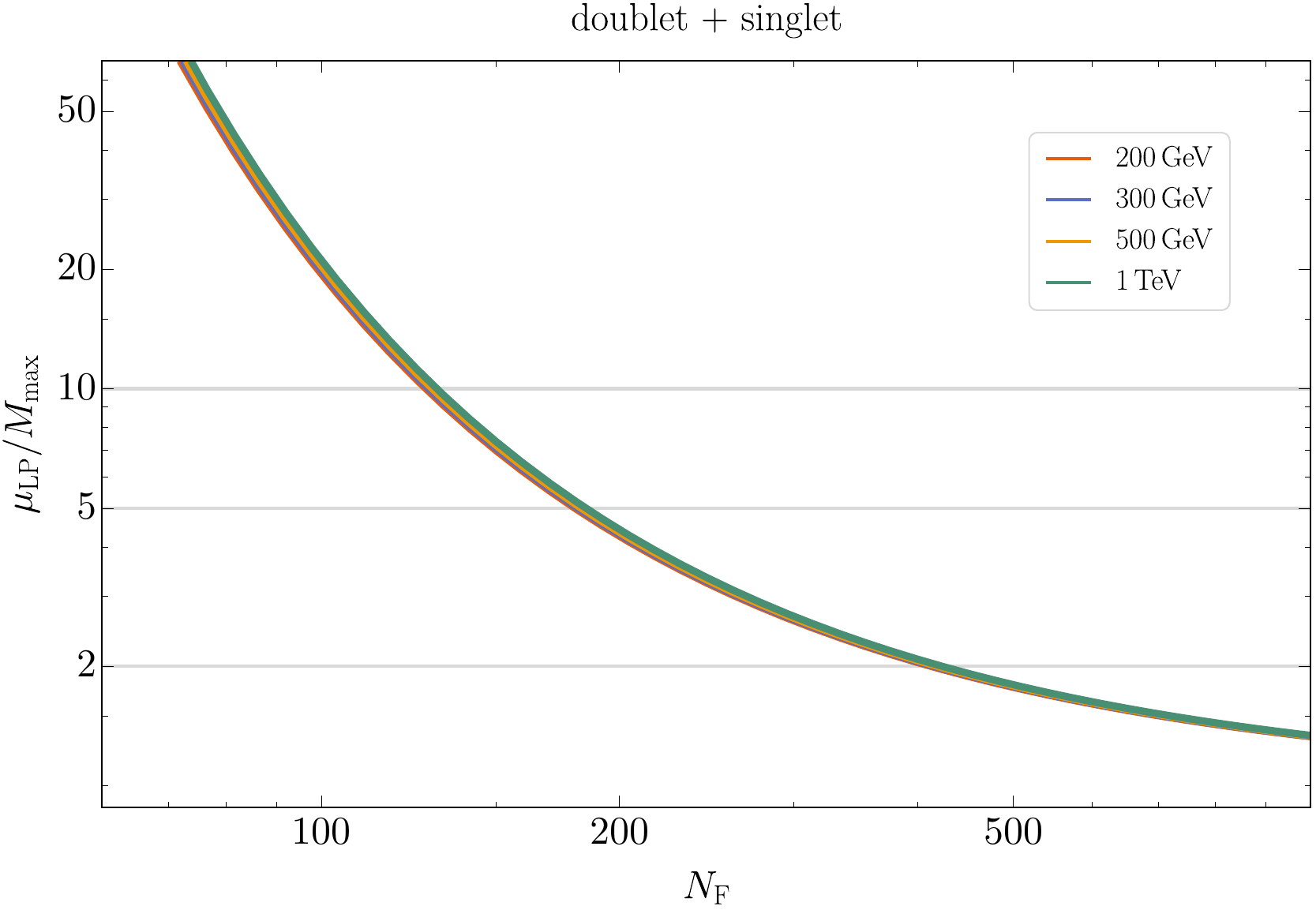}
\includegraphics[width=7.5cm]{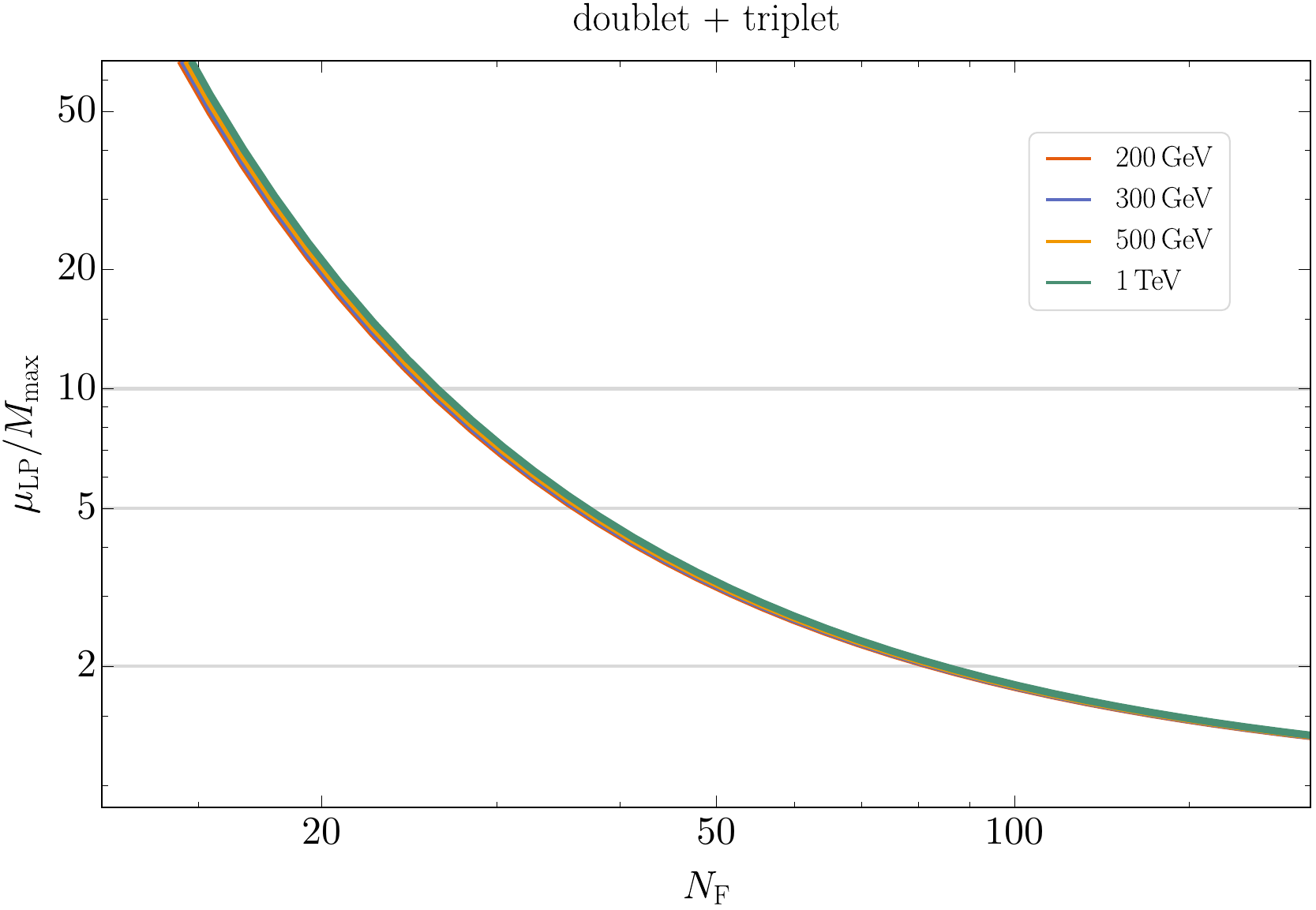}
\end{center}
\caption{Location of the $SU(2)_W$ Landau pole for $n=2$ (left panel) and $n=3$ (right panel), normalized to the heaviest mass $M_\mathrm{max}$ of the VLFs, as a function of $N_\mathrm{F}$.}
\label{Landau_pole_plots}
\end{figure}

Conservative upper bounds on the dimensions of the representations of the new fermions under the SM gauge group can be established by requiring the absence of Landau poles in the SM gauge couplings near $\Lambda_\tn{F}$. If such poles were to arise, $\Lambda_\tn{B} \lesssim \Lambda_\tn{F}$, and it would be necessary to consider models that include new bosons (in addition to new fermions) from the outset. Based on the studies in Refs.~\cite{Blum:2015rpa, DAgnolo:2023rnh}, we adopt the constraints $r \leq 8$, $n \leq 7$, and $|Y| \leq 5$. Similarly, from Fig.~\ref{Landau_pole_plots}, we infer that $N_\mathrm{F} \lesssim 130$ for $n=2$ and $N_\mathrm{F} \lesssim 30$ for $n=3$. This suggests that the number of flavors is less constrained than the dimensions of the irreducible VLF representations. Additionally, such upper bounds can also be derived using perturbative unitarity; Ref.~\cite{Milagre:2024wcg} provides an analysis based on partial-wave criteria. In this article, we rely on Landau pole bounds as conservative constraints on the theory space of VLF models.

\subsubsection{Electroweak Precision Tests}
New fermions near the weak scale introduce corrections to EW observables and are therefore subject to constraints from EW precision tests (EWPTs)~\cite{Wells:2005vk} conducted at LEP, Tevatron, and LHC. For sufficiently heavy VLFs (with masses above $100$ GeV), these constraints can be analyzed through deviations in the Peskin–Takeuchi (or oblique) $STU$ parameters\footnote{For oblique parameters involving new fermions in arbitrary representations of the EW gauge group, see Ref.~\cite{Albergaria:2023nby}.}. We define these parameters following the convention used in the PDG book~\cite{ParticleDataGroup:2024cfk}. In this scenario, corrections to the $S$ and $T$ parameters dominate over those to $U$. Thus, we compare our results to the $S$-$T$ ellipse at $95\%$ confidence level (CL) derived from the fit by the Gfitter Group~\cite{Haller:2018nnx}, assuming fixed $U = 0$. All computations are performed at the 1-loop level, with the full analytic expressions obtained using the \texttt{Wolfram Mathematica} extension \texttt{Package-X}~\cite{Patel:2015tea, Patel:2016fam}.

We emphasize that the constraints derived from EWPTs (depicted as shaded gray areas in the plots\footnote{If a shaded area is missing for certain masses in the plots, it means the corresponding region lies outside the plot range.} of Section~\ref{results_sec}) should be regarded as indicative. For readers interested in analyzing a specific model in detail, a more thorough examination of these EWPT constraints is necessary. For instance, one could relax the fine-tuned relation $y = (-1)^n y^c$ assumed in our plots.

Additionally, it is possible to mitigate the EWPT constraints through model-building, such as by introducing extended VLF models with custodial symmetry\footnote{In this case, the couplings in the new fermionic sector would need to be fine-tuned to maintain custodial symmetry, which is otherwise broken in the SM sector by Yukawa and hypercharge interactions.}~\cite{Sikivie:1980hm}. Therefore, rather than asserting that a specific model is excluded by EWPTs, we instead note that it appears to be in tension with them. As the VLF mass scale $\Lambda_\tn{F} \sim M_L$ increases relative to $y^{(c)}v$, the EWPT constraints become less stringent, consistent with the decoupling behavior of new physics~\cite{Appelquist:1974tg, Wilson:1983xri}.

\subsubsection{Collider Constraints}
To investigate collider constraints, one can safely set $M_L = M_E$ without significantly affecting the discussion~\cite{DAgnolo:2023rnh}. The objective is not to provide detailed bounds from direct searches for each model, but rather to establish conservative benchmark bounds for the plots in Section~\ref{upper_bos}. Comprehensive reviews of the current constraints on VLFs from the ATLAS and CMS experiments are available in Refs.~\cite{ATLAS:2024fdw, CMS:2024bni}.

The least constrained scenario from direct searches is the colorless doublet $+$ singlet model, which features two neutral states with masses $M_{1,2}$ and one charged state with mass $M_L$. This model is qualitatively similar to a Higgsino-Bino system, which is weakly constrained at the LHC. By exploring the parameter space $(M_L, y, y^c)$, one can identify different cases and establish an experimental lower bound on the mass scale of the new fermions, $M_{\text{exp}}$:
\begin{itemize}
\item The lightest state is charged and stable (or long-lived) $(M_L < M_{1,2})$. Since searches for such particles have very low background, the constraints are quite stringent. CMS places a bound of $M_{\text{exp}} \gtrsim 1.14$ TeV based on Drell-Yan (DY) pair production of long-lived lepton-like fermions, using $\mathrm{101\ fb^{-1}}$ of $\sqrt{s} = 13$ TeV data~\cite{CMS:2024nhn}.

\item The lightest state is charged but can decay promptly into SM particles through a tiny mixing with SM lepton doublets (small enough to avoid violating flavor constraints). ATLAS provides a limit of $M_\text{exp} \gtrsim 900$ GeV, derived from 139 fb$^{-1}$ of $\sqrt{s} = 13$ TeV data, for a new $SU(2)_W$ lepton doublet that predominantly decays into third-generation SM leptons~\cite{ATLAS:2023sbu}. Decays into first- and second-generation leptons are expected to yield stronger bounds. A small gap between this analysis and the older LEP limits is addressed by CMS~\cite{CMS:2018szt, CMS:2020bfa, CMS:2021edw, CMS:2021cox, CMS:2022sfi} and ATLAS~\cite{ATLAS:2018eui, ATLAS:2019lff, ATLAS:2020pgy, ATLAS:2019lng, ATLAS:2021moa, ATLAS:2021yqv, ATLAS:2022zwa, ATLAS:2022hbt} searches for charginos, which exclude masses around 100 GeV.
\end{itemize}
In the case where $M_L \gtrsim \mathcal{O}(yv)$, the constraints can be effectively summarized by focusing on the region of the parameter space with $M_2 > M_L > M_1$, where the lightest particle is neutral. In Fig.~9 of our companion paper~\cite{DAgnolo:2023rnh}, we provide an overview of the existing constraints, which shows that a substantial portion of the parameter space remains unexplored.

For higher representations of $SU(3)_C$, $SU(2)_W$, and/or larger hypercharges, the LHC imposes stronger constraints:

\begin{itemize}
\item Using searches for electroweakinos~\cite{CMS:2018szt, CMS:2020bfa, CMS:2021edw, CMS:2021cox, CMS:2022sfi, ATLAS:2018eui, ATLAS:2019lff, ATLAS:2020pgy, ATLAS:2019lng, ATLAS:2021moa, ATLAS:2021yqv, ATLAS:2022zwa, ATLAS:2022hbt}, the colorless doublet $+$ triplet model is excluded for $M_L \gtrsim 240$ GeV, even with mass splittings as low as 8 GeV (see Fig.~6 of Ref.~\cite{ATLAS:2021moa}). In our plots, this bound will be used as a conservative constraint on the lightest particle in models with higher $SU(2)_W$ representations.

\item Fermions with higher hypercharges cannot mix with SM particles and are therefore constrained by searches for long-lived multi-charged particles. CMS excludes $M_{\mathrm{exp}} \gtrsim 1.41$ TeV from Drell-Yan (DY) production using $\mathrm{101 \ fb^{-1}}$ of $\sqrt{s} = 13$ TeV data~\cite{CMS:2024nhn}, while ATLAS provides bounds of $M_{\mathrm{exp}} \gtrsim 1.39$, 1.52, 1.59, 1.60, and 1.57 TeV for electric charges $|Q|=3$, 4, 5, 6, and 7, respectively, based on DY plus photon-fusion production with $\mathrm{139 \ fb^{-1}}$ of $\sqrt{s} = 13$ TeV data~\cite{ATLAS:2023zxo}.

\item For colored representations, reviews of direct searches for vectorlike quarks (i.e., color triplets) by ATLAS and CMS~\cite{ATLAS:2024fdw, CMS:2024bni} indicate a lower bound of $M_\text{exp} \gtrsim 1$ TeV on the mass of the lightest state. For stable colored representations, searches for $R$-hadrons, made of (meta)-stable gluinos, by ATLAS and CMS~\cite{CMS:2016ybj, ATLAS:2018yey} yield a stricter bound of $M_\mathrm{exp} \gtrsim 1.8$ TeV.
\end{itemize}

In Section~\ref{results_sec}, we often focus on the doublet $+$ singlet model $(n=2)$, which has three mass eigenstates under the conditions $y = y^c$ and $M_L = M_E$, maximizing $\delta \mu_{hVV^\prime}$ for a given $M_L$. Two of these eigenstates (the heaviest and the lightest) mix through Yukawa couplings to the $h$-boson, while the third does not couple to the Higgs boson.

The VLFs that can affect the $h\gamma\gamma$ and $hZ\gamma$ couplings must couple to the $h$-boson and be electrically charged. In the simplest model we consider, this implies that the lightest VLF must have $Q \neq 0$, subject to the previously discussed constraints on a lightest charged state. Therefore, the optimal case previously discussed for the collider bounds (with two neutral states of mass $M_{1,2}$ and a charged one of mass $M_L$, with the hierarchy $M_2>M_L>M_1$) does not fulfill this requirement. Nevertheless, a lightest neutral VLF can still be accommodated with additional model-building. Specifically, for $Y = 1/2$, the VLF that lacks a Yukawa coupling to the $h$-boson is neutral. Adding a singlet VLF to this model, mixing with this neutral particle via a new Yukawa coupling to the $h$-boson, allows one to tune the coupling such that the lightest VLF is neutral. The trade-off is slightly stronger constraints from EWPTs and lower instabilities, in favor of relaxing the collider bounds.

In the doublet $+$ triplet model, the situation is simpler because the lightest particle can naturally be neutral with $Y = 0$, while still having VLFs that couple to both the $h$-boson and the photon. This configuration minimizes collider bounds.

This discussion highlights the vast theory space available for exploration. In this work, we restrict our analysis to the simplest benchmark models presented in Section~\ref{Fermions}, which allow for a general study of coupling deviations, treating $r$, $n$, $Y$, and $N_\mathrm{F}$ as free parameters. Therefore, in the computations for plots with $n=2$ and $r=1$, we did not introduce a second singlet, while allowing the values of $M_1$ to extend below the collider constraints for new charged fermions in the colorless case. As we will see, light $M_1$ values are still in significant tension with EWPTs if they induce sizable deviations in the Higgs couplings.

\section{Upper Bound on a New Bosonic Scale}
\label{upper_bos}

\begin{figure}[t]
\begin{center}
\includegraphics[width=5cm]{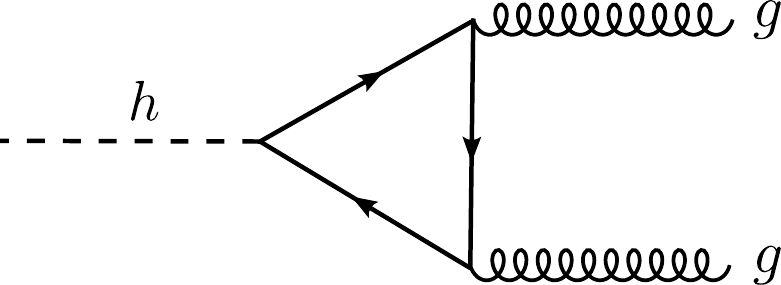}\hfill
\includegraphics[width=5cm]{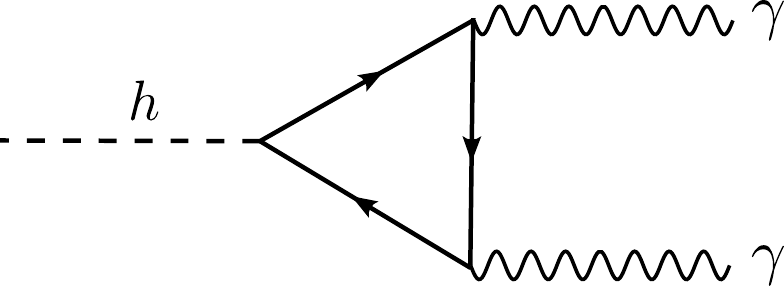}\hfill
\includegraphics[width=5cm]{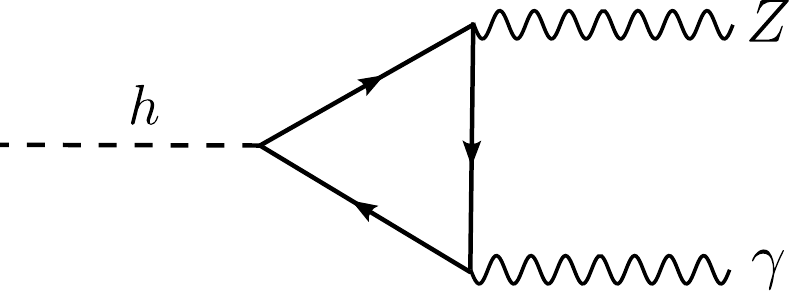}\hfill
\end{center}
\caption{Feynman diagrams of the Higgs boson decays into digluons (left), diphotons (middle) and $Z\gamma$ (right) via 1-loop of VLFs.}
\label{H_decays}
\end{figure}

\begin{figure}[t]
\begin{center}
\includegraphics[width=7.5cm]{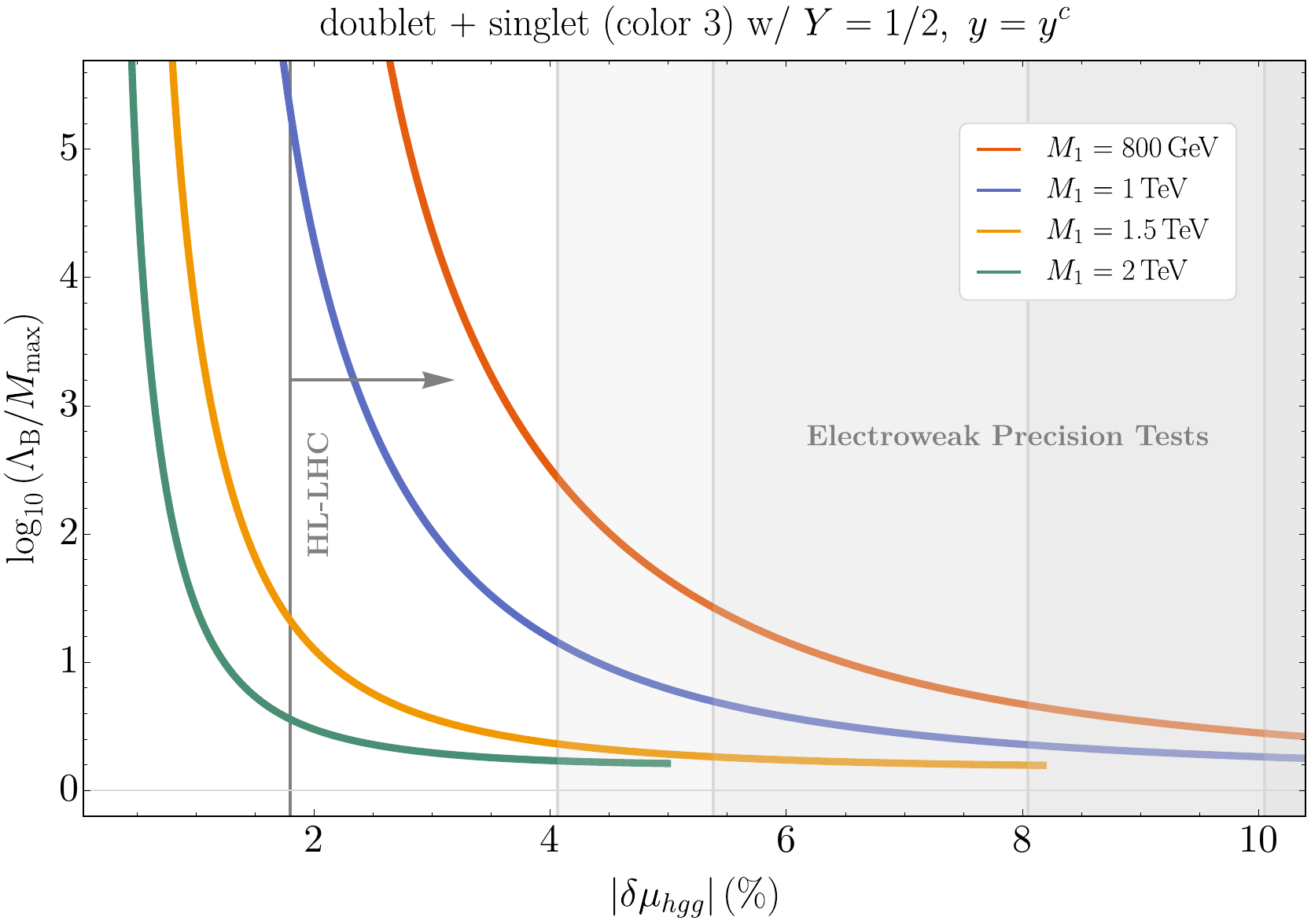}\hfill
\includegraphics[width=7.5cm]{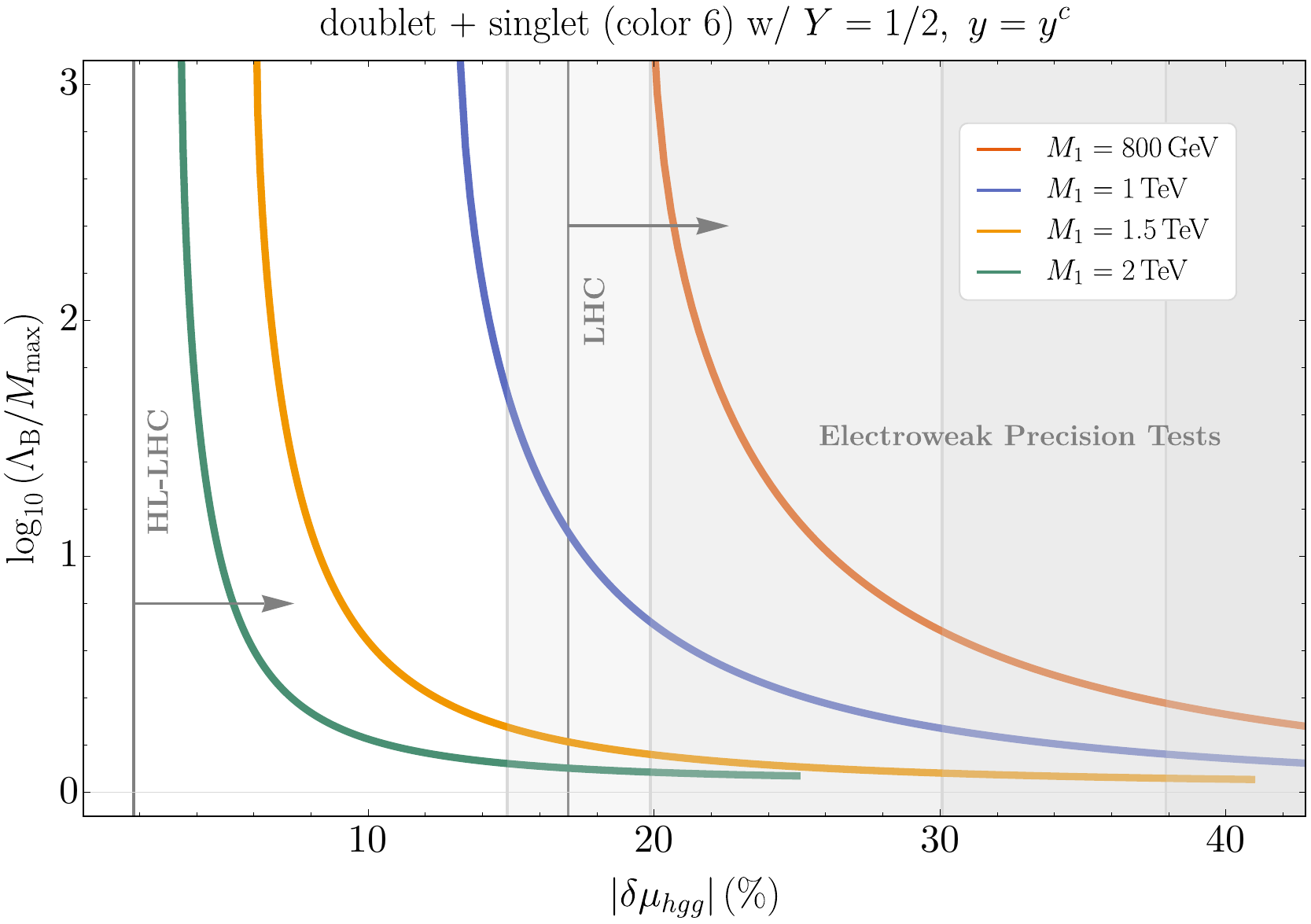}\hfill
\includegraphics[width=7.5cm]{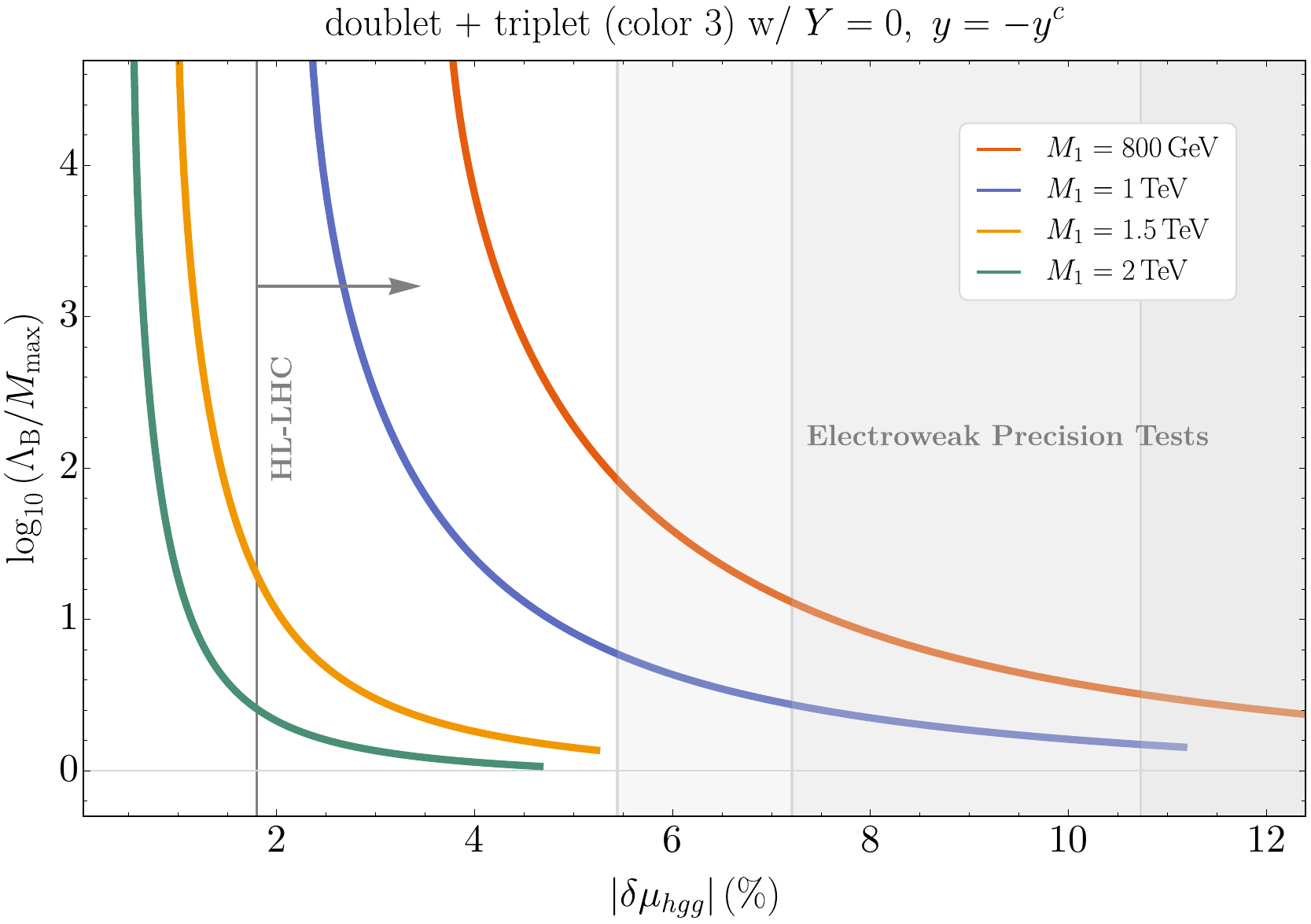}\hfill
\end{center}
\caption{Upper bound on the scale of new bosons $\Lambda_\tn{B}$ as a function of the coupling deviation $\delta \mu_{hgg}$.
The choice $y= (-1)^n y^c$ indicated in the title of each plot maximizes $\Lambda_\tn{B}$. $M_\mathrm{max}$ is the largest of the VLF masses, while $M_1$ is the smallest one. The gray shaded areas represent the constraint from EWPTs which, at lower $\delta \mu_{hgg}$, is on the line of lowest $M_1$.
Top-left: Model $(r=3, n=2, Y=1/2, N_\mathrm{F} = 1)$. Top-right: Model $(r=6, n=2, Y=1/2, N_\mathrm{F}=1)$. Bottom: Model $(r=3, n=3, Y=0, N_\mathrm{F} = 1)$.}
\label{hgg_plots}
\end{figure}

\begin{figure}[t]
\begin{center}
\includegraphics[width=7.5cm]{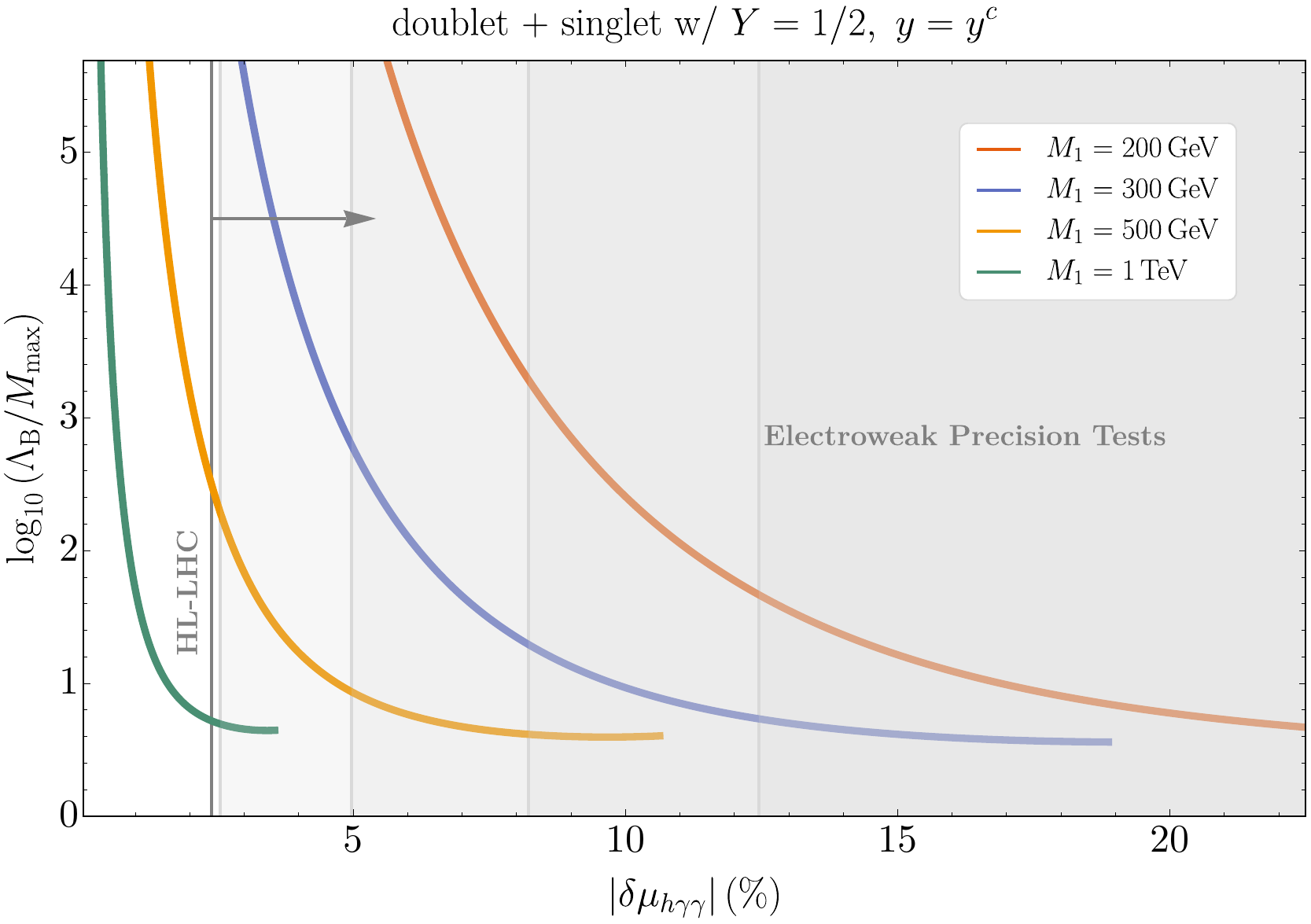}\hfill
\includegraphics[width=7.5cm]{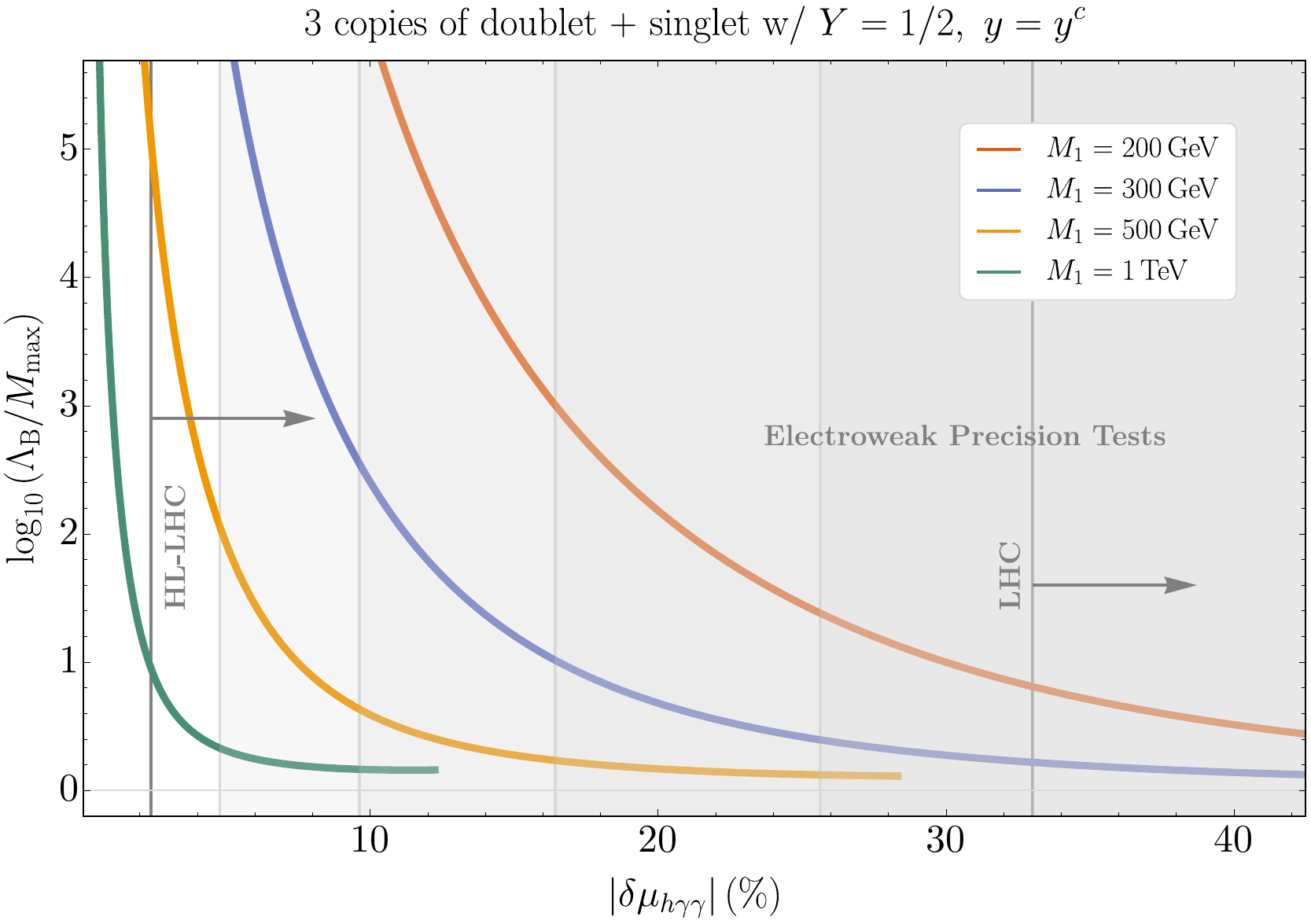}\hfill
\includegraphics[width=7.5cm]{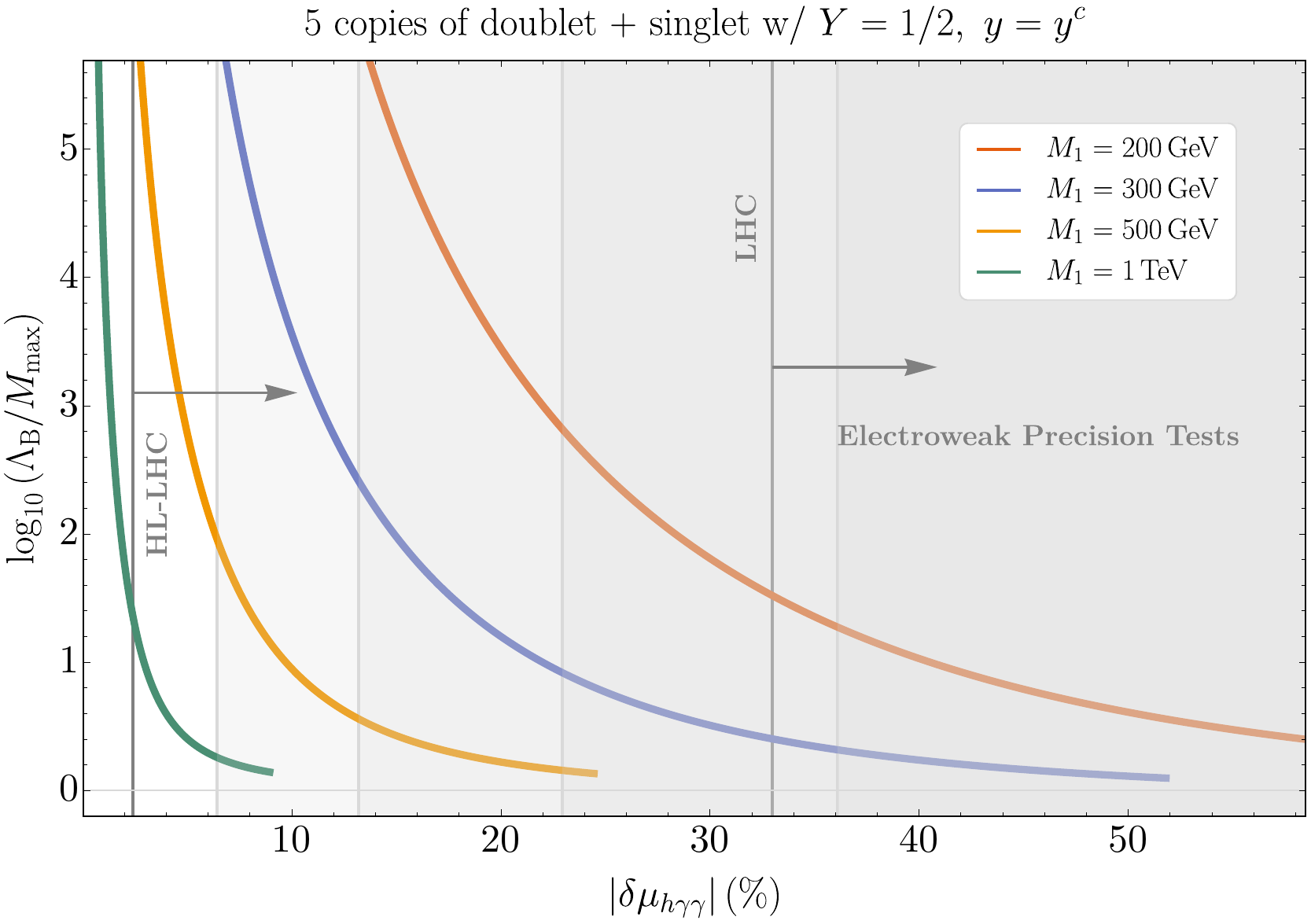}\hfill
\end{center}
\caption{Upper bound on the scale of new bosons $\Lambda_\tn{B}$ as a function of the coupling deviation $\delta \mu_{h\gamma\gamma}$ in the models $(r=1, n=2, Y=0)$ with $N_\mathrm{F} = 1, 3, 5$ (top left, top right, and bottom panels, respectively). The choice $y=(-1)^n y^c$ indicated in the title of each plot maximizes $\Lambda_\tn{B}$. $M_\mathrm{max}$ is the largest of the VLF masses, while $M_1$ is the smallest one. The gray shaded areas represent the constraint from EWPTs which, at lower $\delta \mu_{h\gamma\gamma}$, is on the line of lowest $M_1$.} 
\label{hgamgam_12_NF}
\end{figure}

\begin{figure}[t]
\begin{center}
\includegraphics[width=7.5cm]{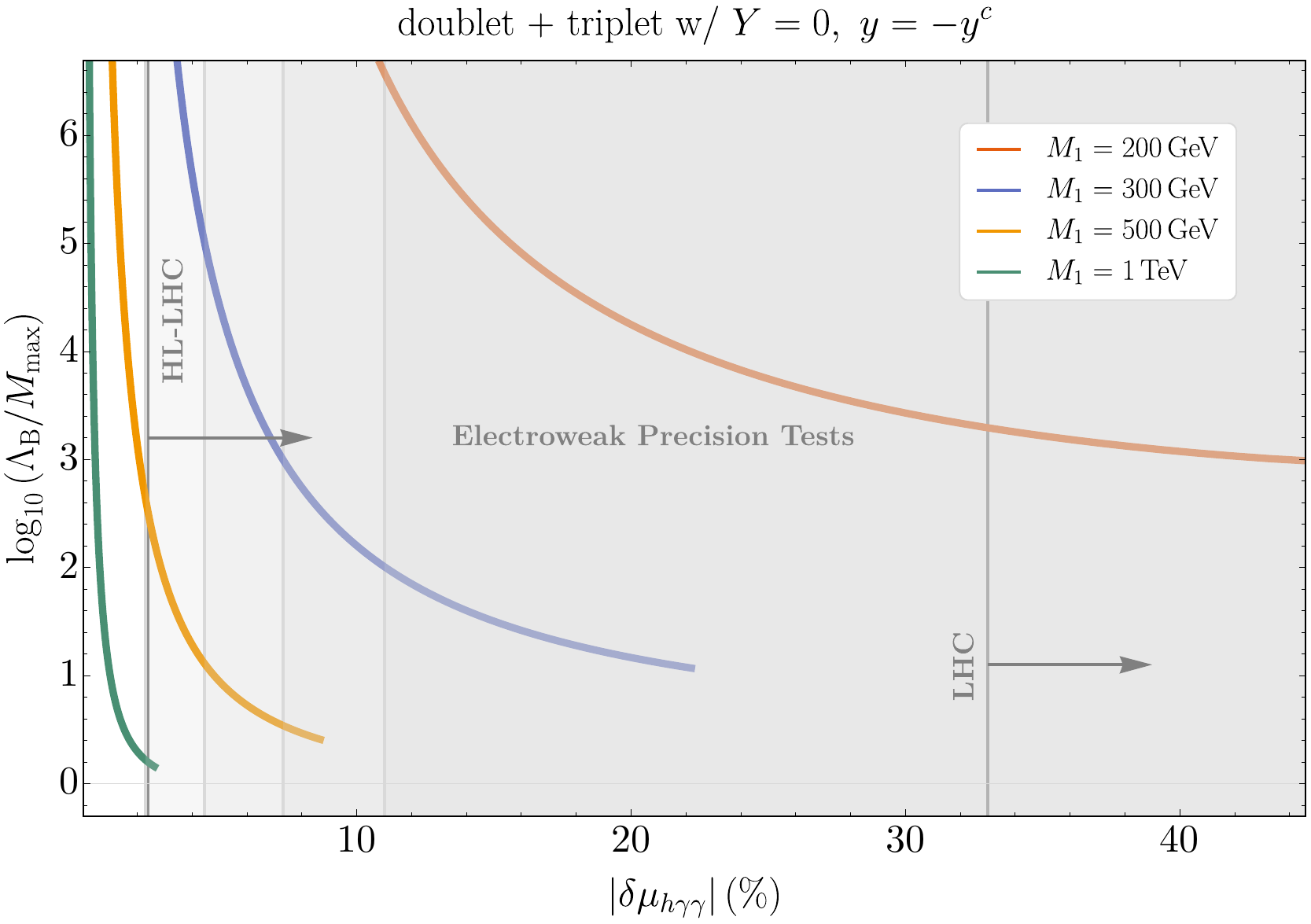}\hfill
\includegraphics[width=7.5cm]{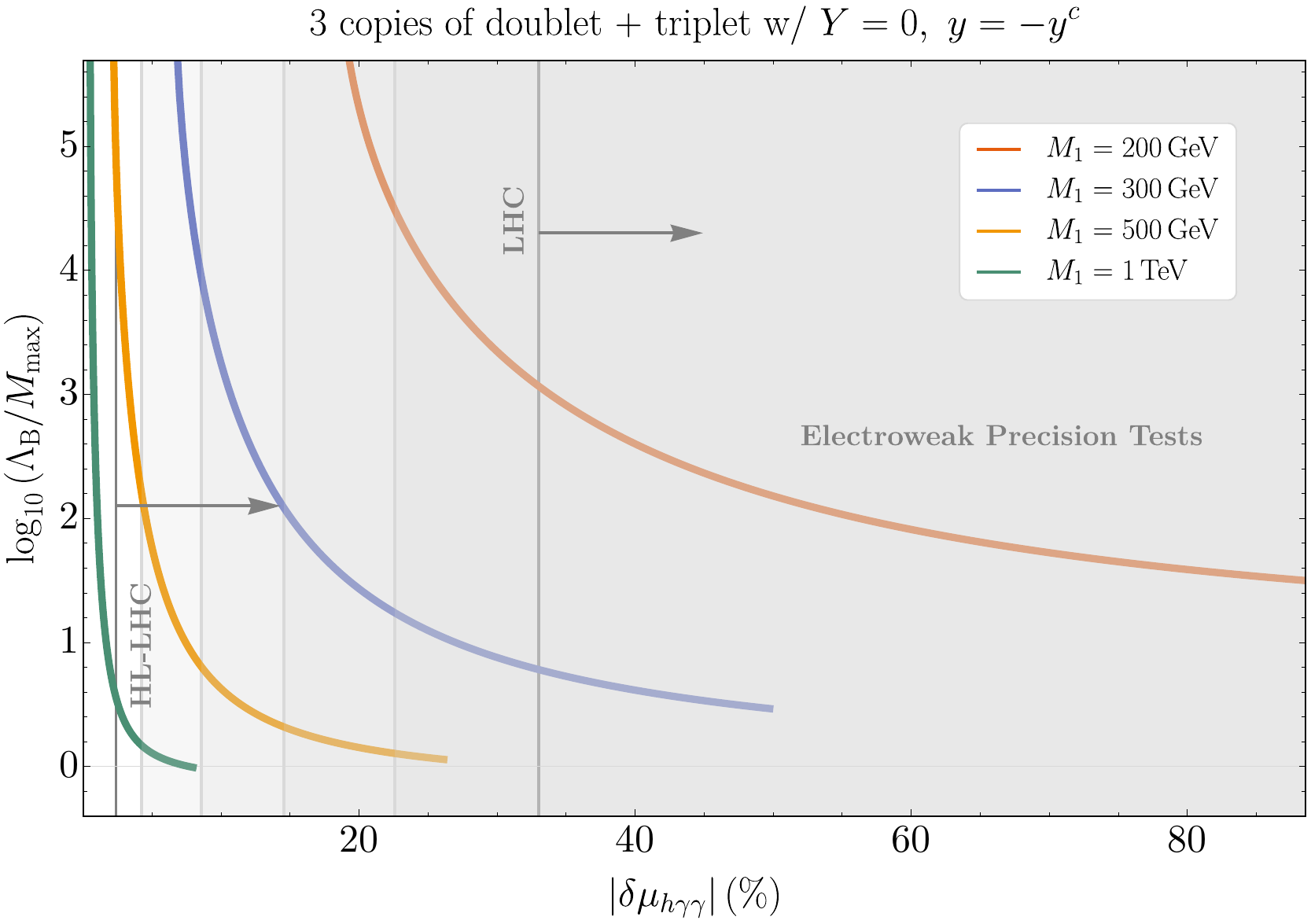}\hfill
\includegraphics[width=7.5cm]{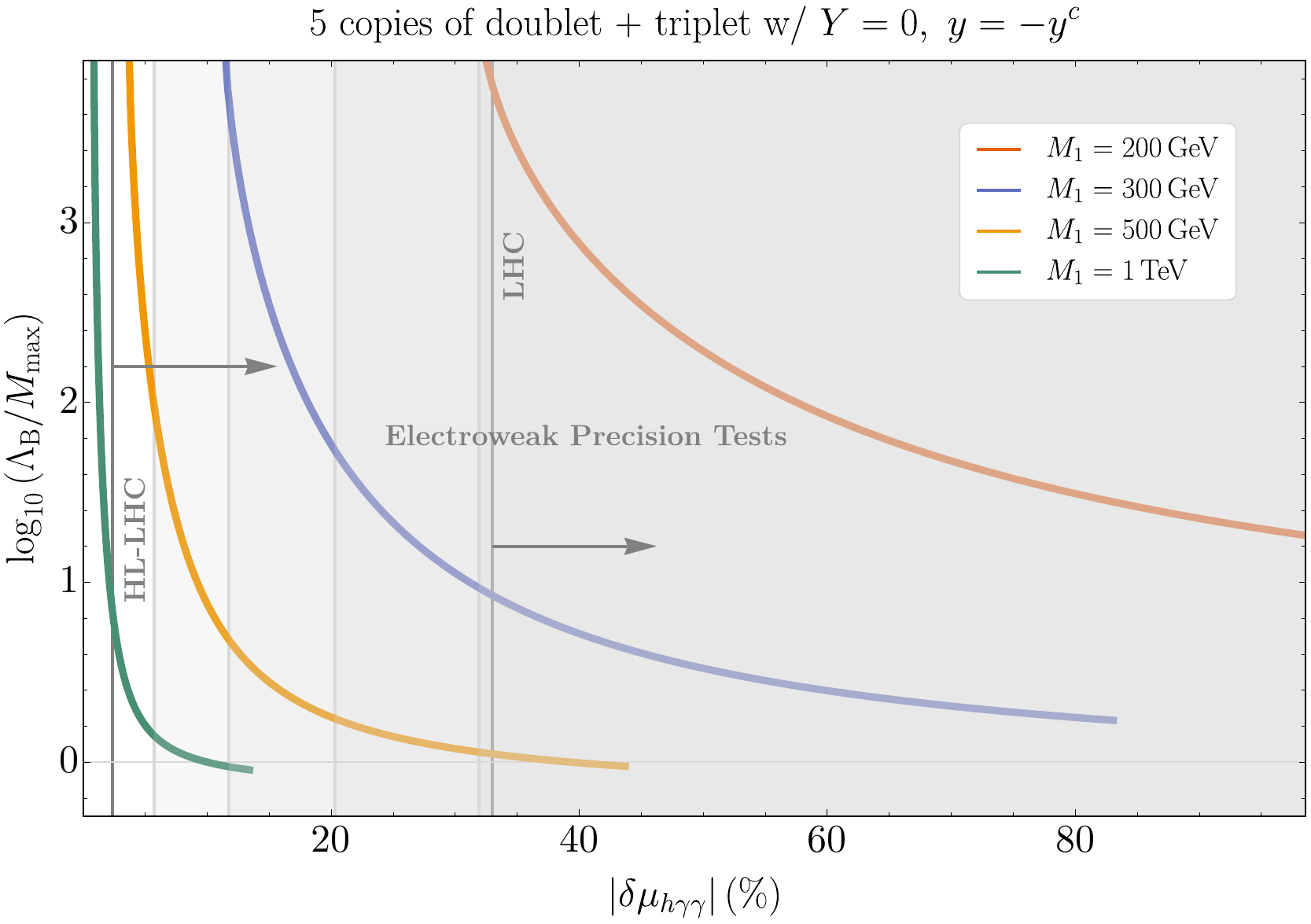}\hfill
\end{center}
\caption{Upper bound on the scale of new bosons $\Lambda_\tn{B}$ as a function of the coupling deviation $\delta \mu_{h\gamma\gamma}$ in the models $(r=1, n=3, Y=0)$ with $N_\mathrm{F} = 1, 3, 5$ flavors (top left, top right, and bottom panels, respectively). The choice $y=(-1)^n y^c$ indicated in the title of each plot maximizes $\Lambda_\tn{B}$. $M_\mathrm{max}$ is the largest of the VLF masses, while $M_1$ is the smallest one. The gray shaded areas represent the constraint from EWPTs which, at lower $\delta \mu_{h\gamma\gamma}$, is on the line of lowest $M_1$.} 
\label{hgamgam_23_NF}
\end{figure}

\begin{figure}[t]
\begin{center}
\includegraphics[width=7.5cm]{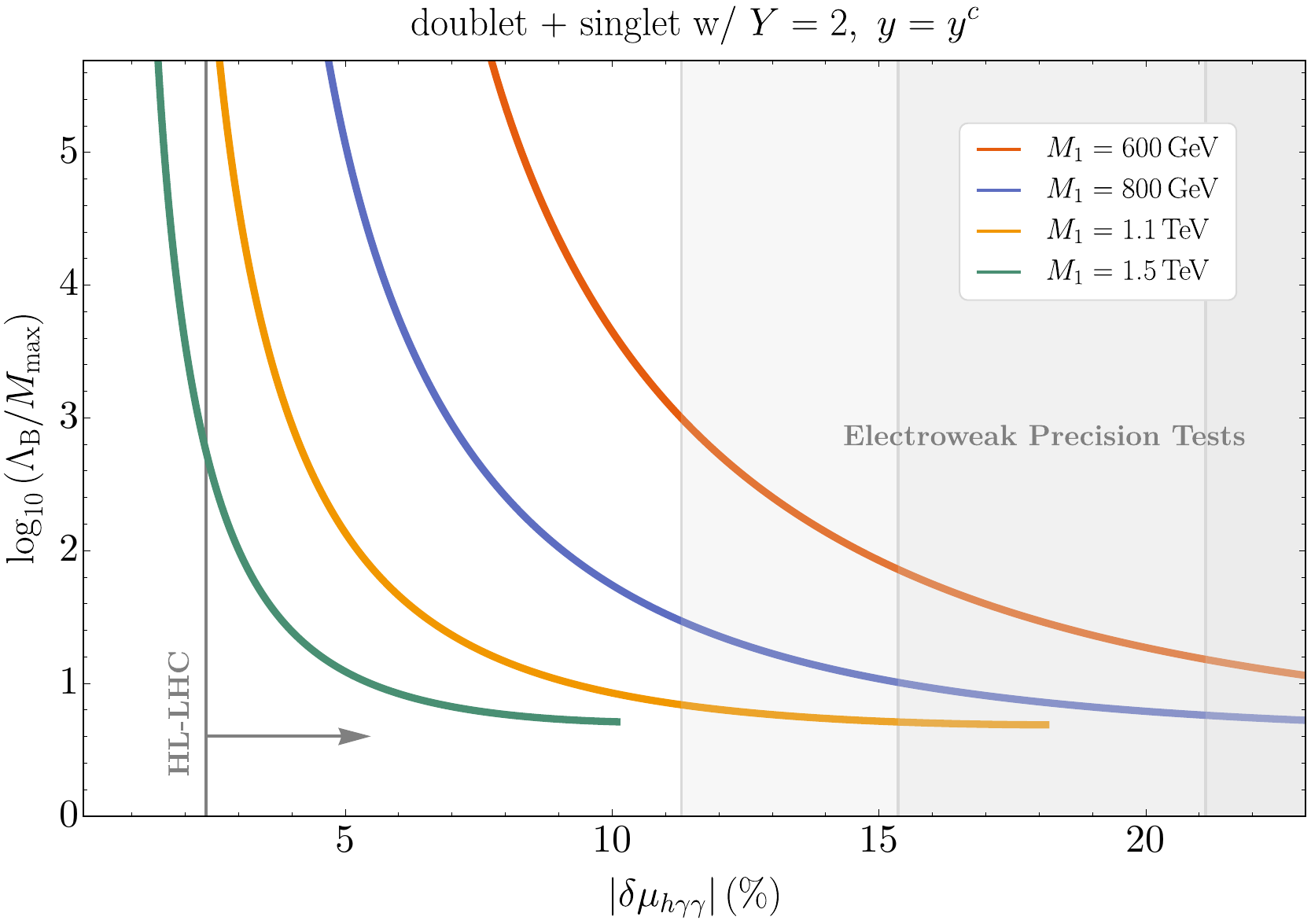}
\includegraphics[width=7.5cm]{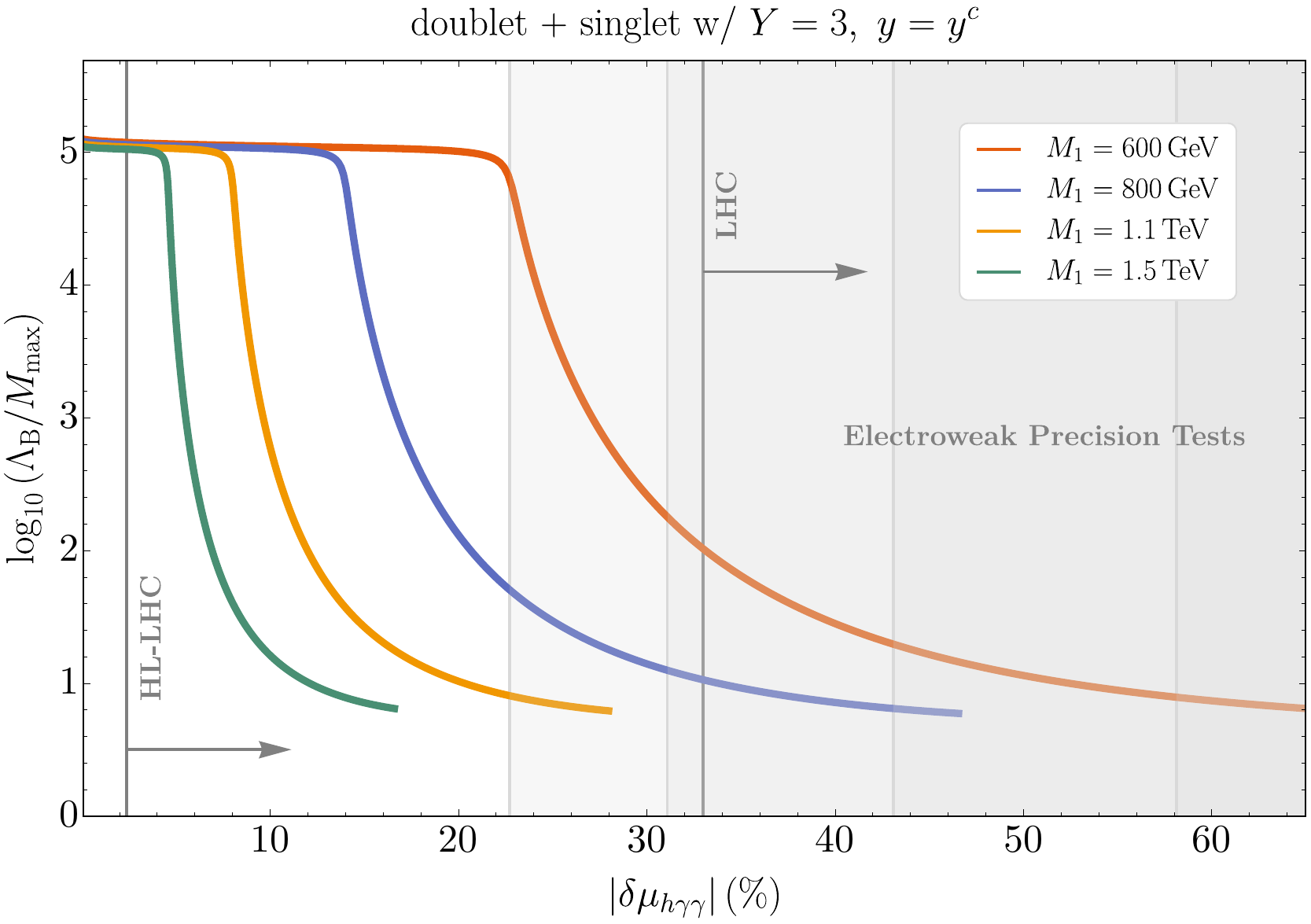}
\end{center}
\caption{Upper bound on the scale of new bosons $\Lambda_\tn{B}$ as a function of the coupling deviation $\delta \mu_{h\gamma\gamma}$ in the models $(r=1, n=2, N_\mathrm{F}=1)$ with $Y = 2, 3$ (left and right panels, respectively). The choice $y=(-1)^n y^c$ indicated in the title of each plot maximizes $\Lambda_\tn{B}$. $M_\mathrm{max}$ is the largest of the VLF masses, while $M_1$ is the smallest one. The gray shaded areas represent the constraint from EWPTs which, at lower $\delta \mu_{h\gamma\gamma}$, is on the line of lowest $M_1$.}
\label{hgamgam_Y}
\end{figure}

\begin{figure}[t]
\begin{center}
\includegraphics[width=7.5cm]{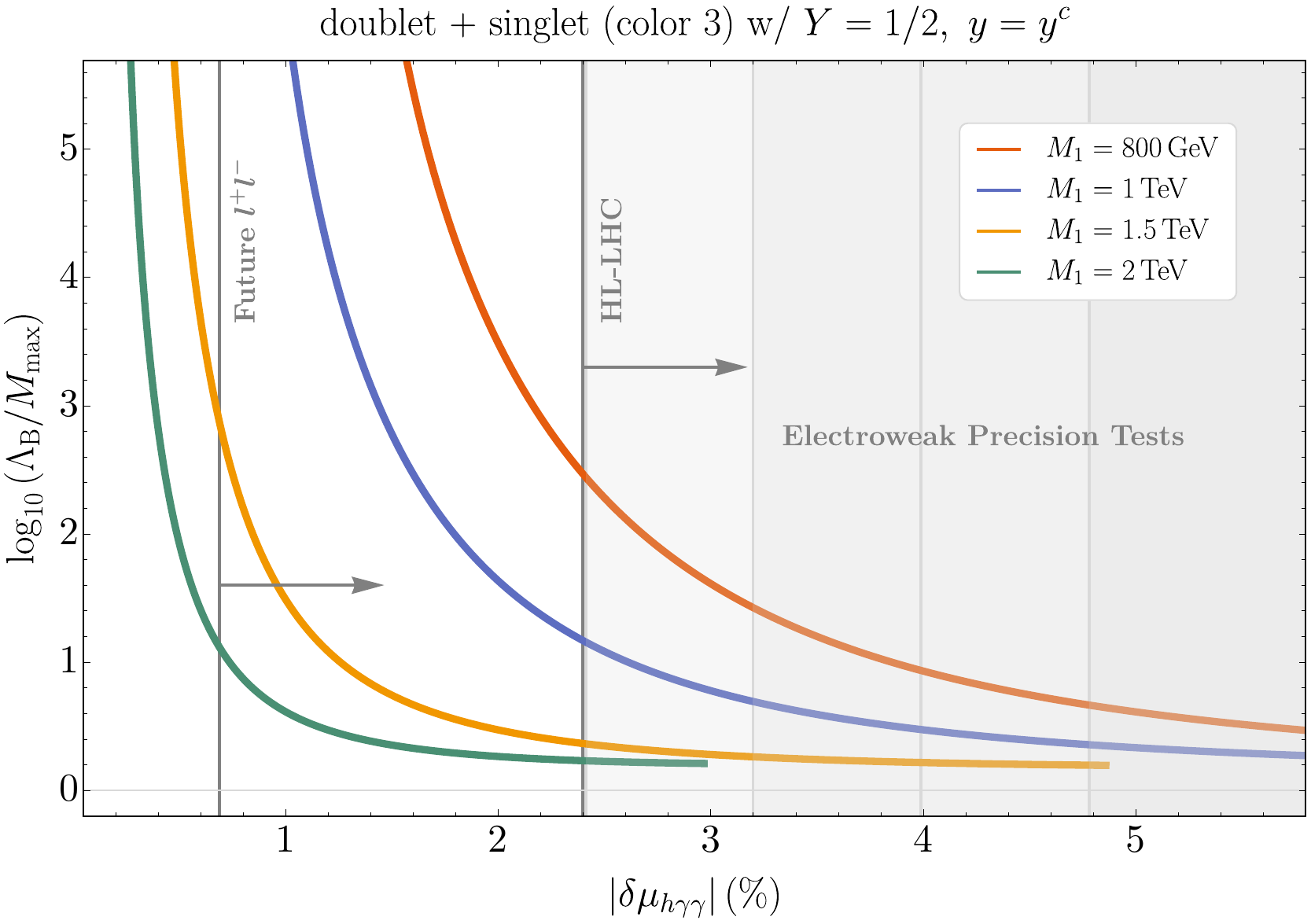}\hfill
\includegraphics[width=7.5cm]{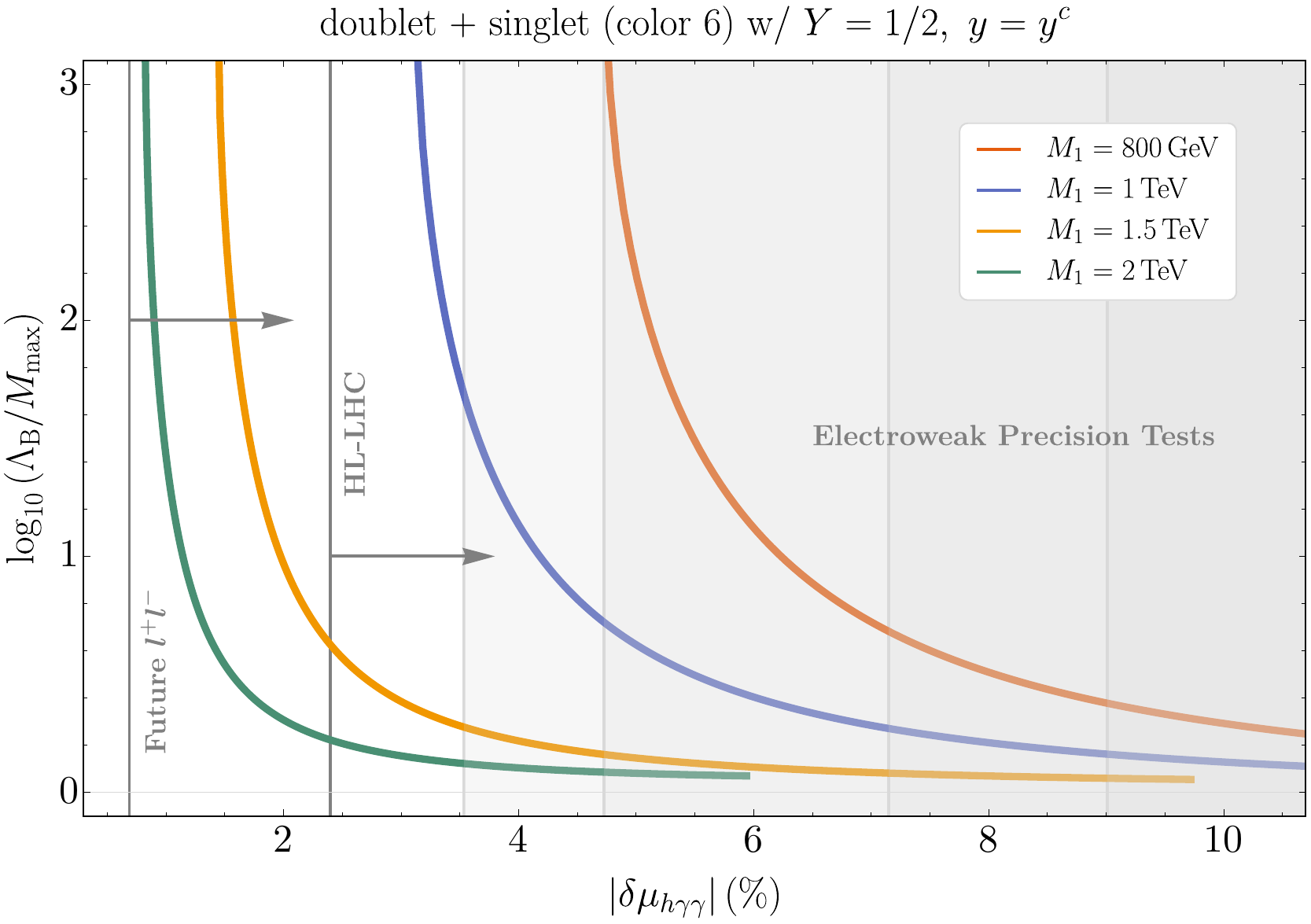}\hfill
\includegraphics[width=7.5cm]{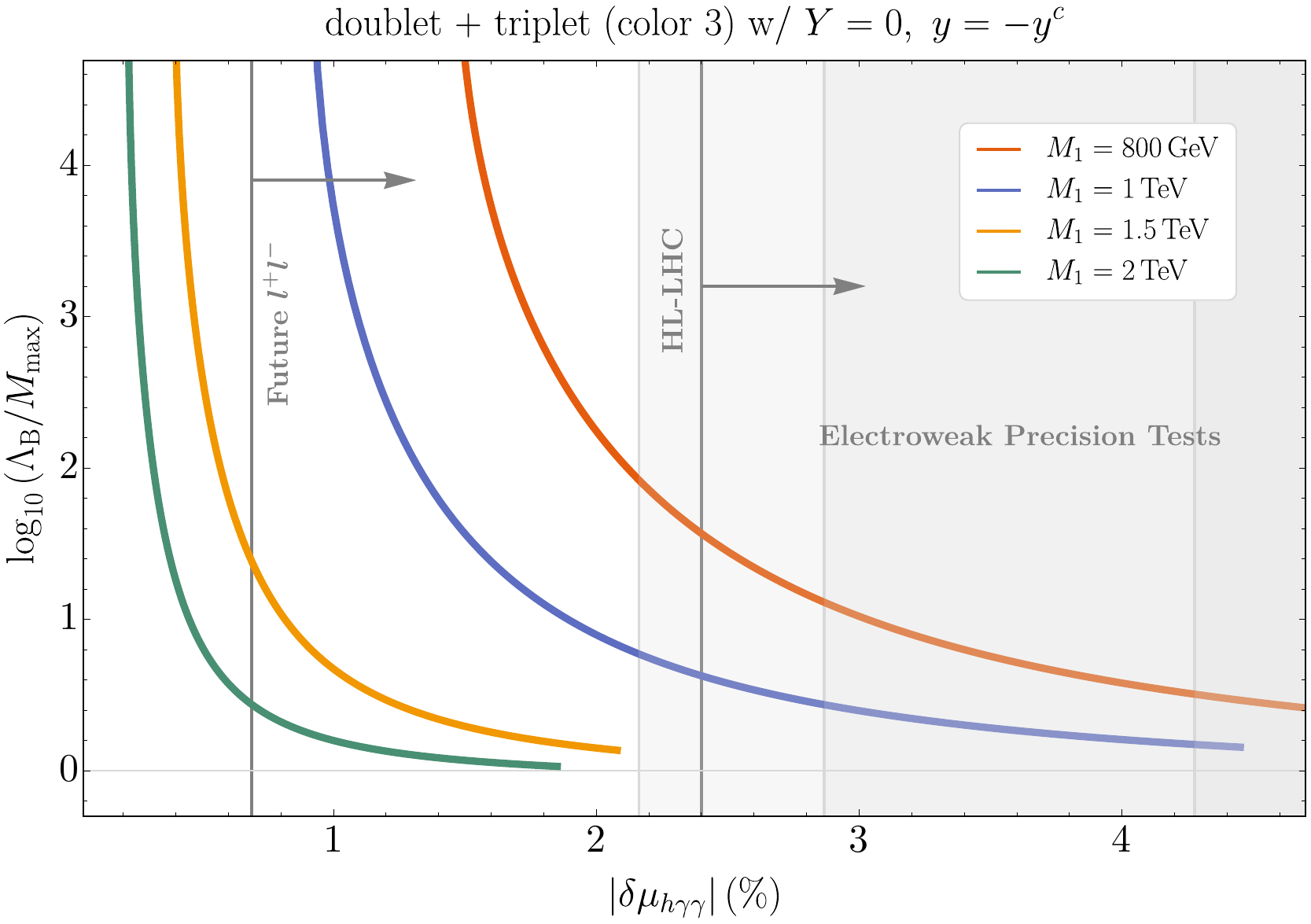}\hfill
\end{center}
\caption{Upper bound on the scale of new bosons $\Lambda_\tn{B}$ as a function of the coupling deviation $\delta \mu_{h\gamma\gamma}$.
The choice $y= (-1)^n y^c$ indicated in the title of each plot maximizes $\Lambda_\tn{B}$. $M_\mathrm{max}$ is the largest of the VLF masses, while $M_1$ is the smallest one. The gray shaded areas represent the constraint from EWPTs which, at lower $\delta \mu_{h\gamma\gamma}$, is on the line of lowest $M_1$.
Top-left: Model $(r=3, n=2, Y=1/2, N_\mathrm{F} = 1)$. Top-right: Model $(r=6, n=2, Y=1/2, N_\mathrm{F}=1)$. Bottom: Model $(r=3, n=3, Y=0, N_\mathrm{F} = 1)$.}
\label{hgamgam_Col}
\end{figure}

\begin{figure}[t]
\begin{center}
\includegraphics[width=7.5cm]{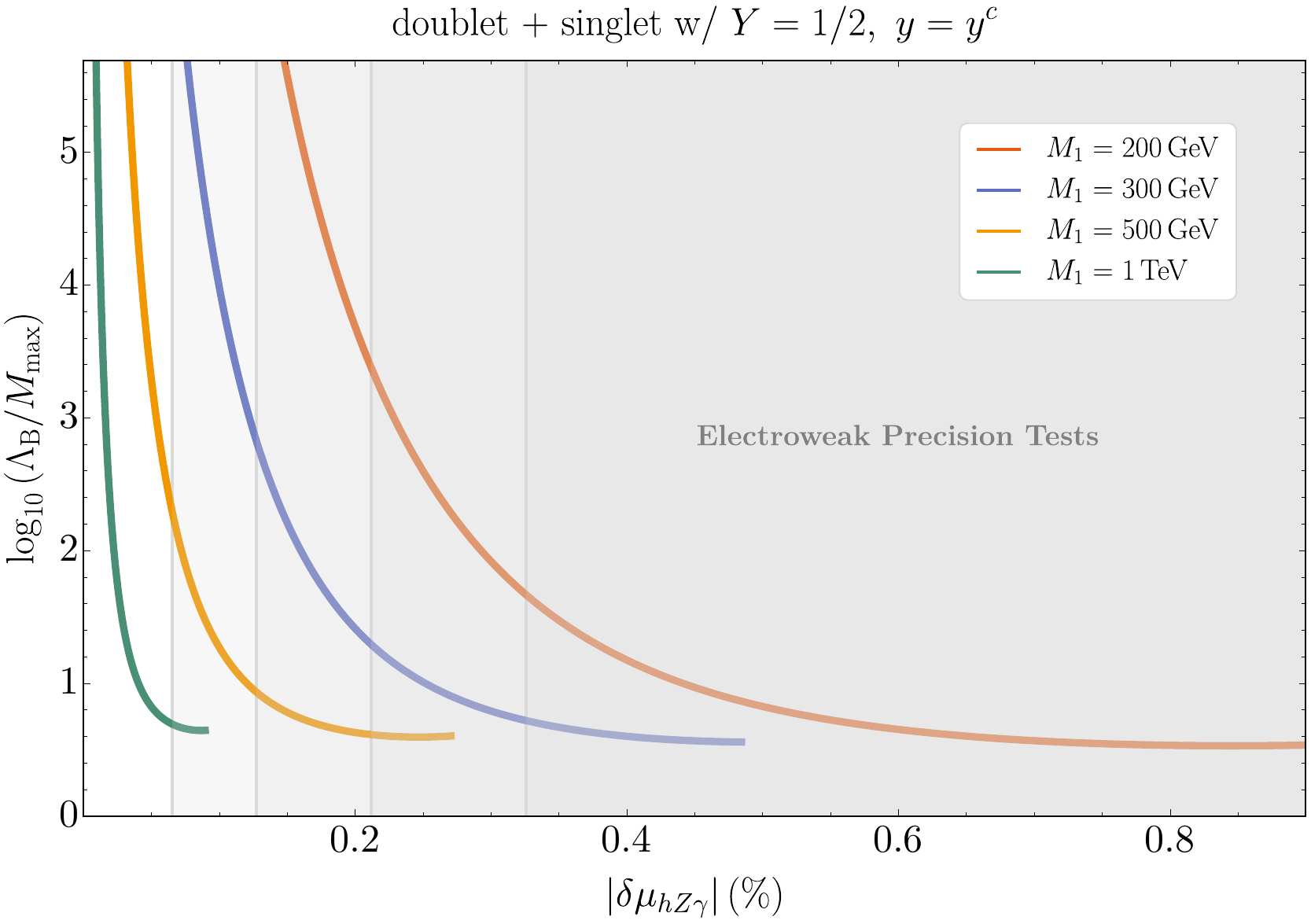}\hfill
\includegraphics[width=7.5cm]{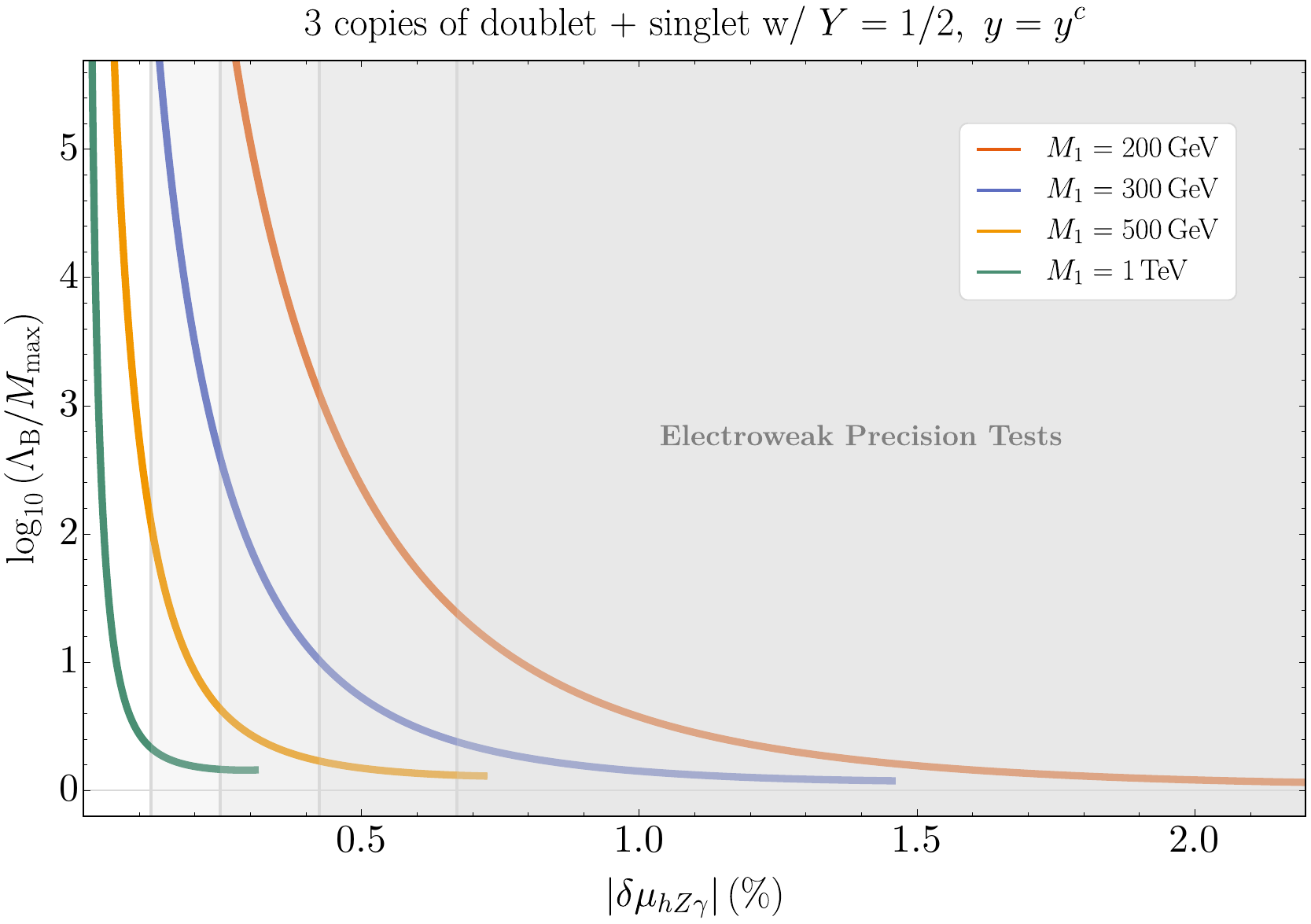}\hfill
\includegraphics[width=7.5cm]{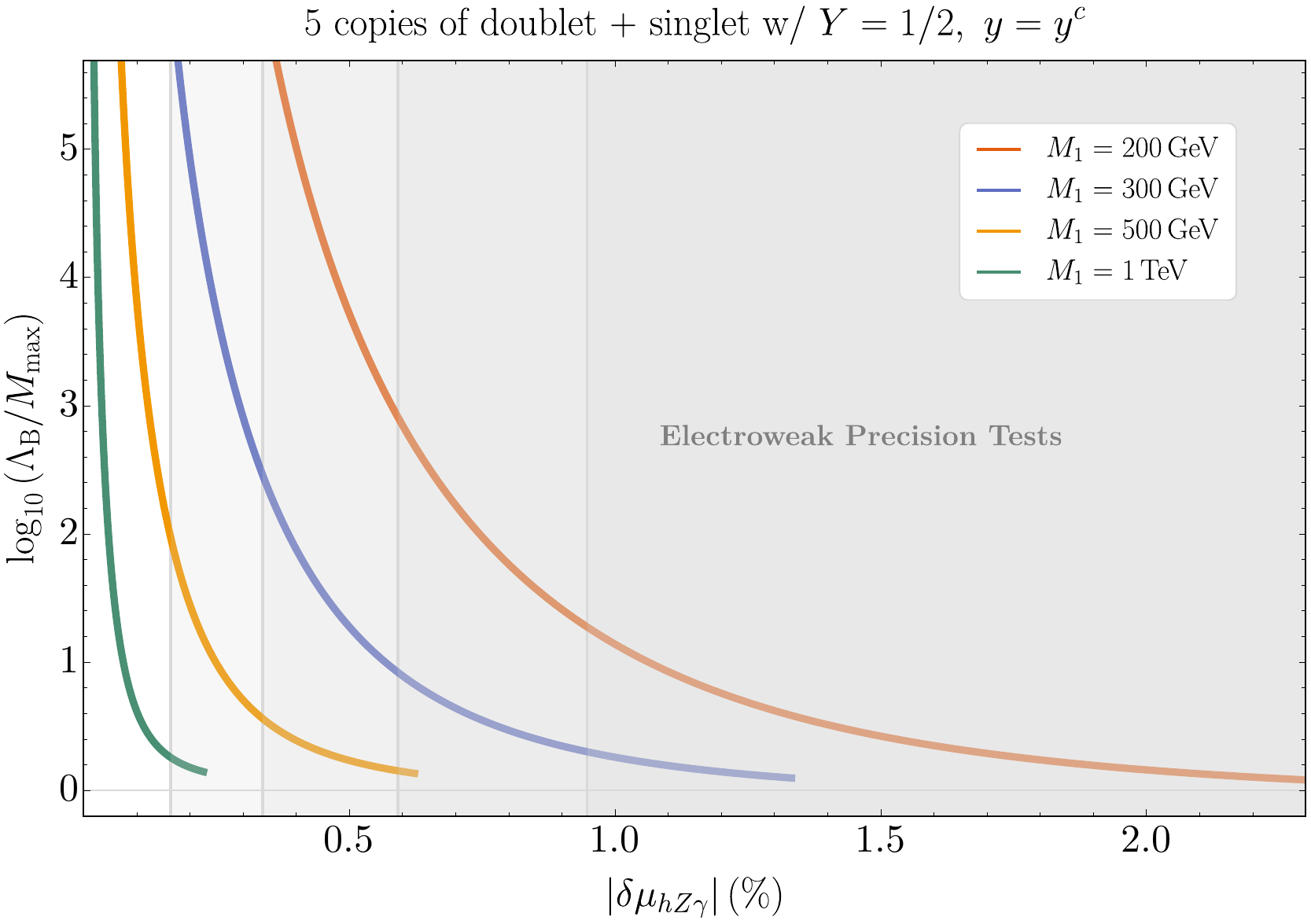}\hfill
\end{center}
\caption{Upper bound on the scale of new bosons $\Lambda_\tn{B}$ as a function of the coupling deviation $\delta \mu_{hZ\gamma}$ in the models $(r=1, n=2, Y=0)$ with $N_\mathrm{F} = 1, 3, 5$ flavors (top left, top right, and bottom panels, respectively). The choice $y=(-1)^n y^c$ indicated in the title of each plot maximizes $\Lambda_\tn{B}$. $M_\mathrm{max}$ is the largest of the VLF masses, while $M_1$ is the smallest one. The gray shaded areas represent the constraint from EWPTs which, at lower $\delta \mu_{hZ\gamma}$, is on the line with lowest $M_1$.} 
\label{hzgam_12_NF}
\end{figure}

\begin{figure}[t]
\begin{center}
\includegraphics[width=7.5cm]{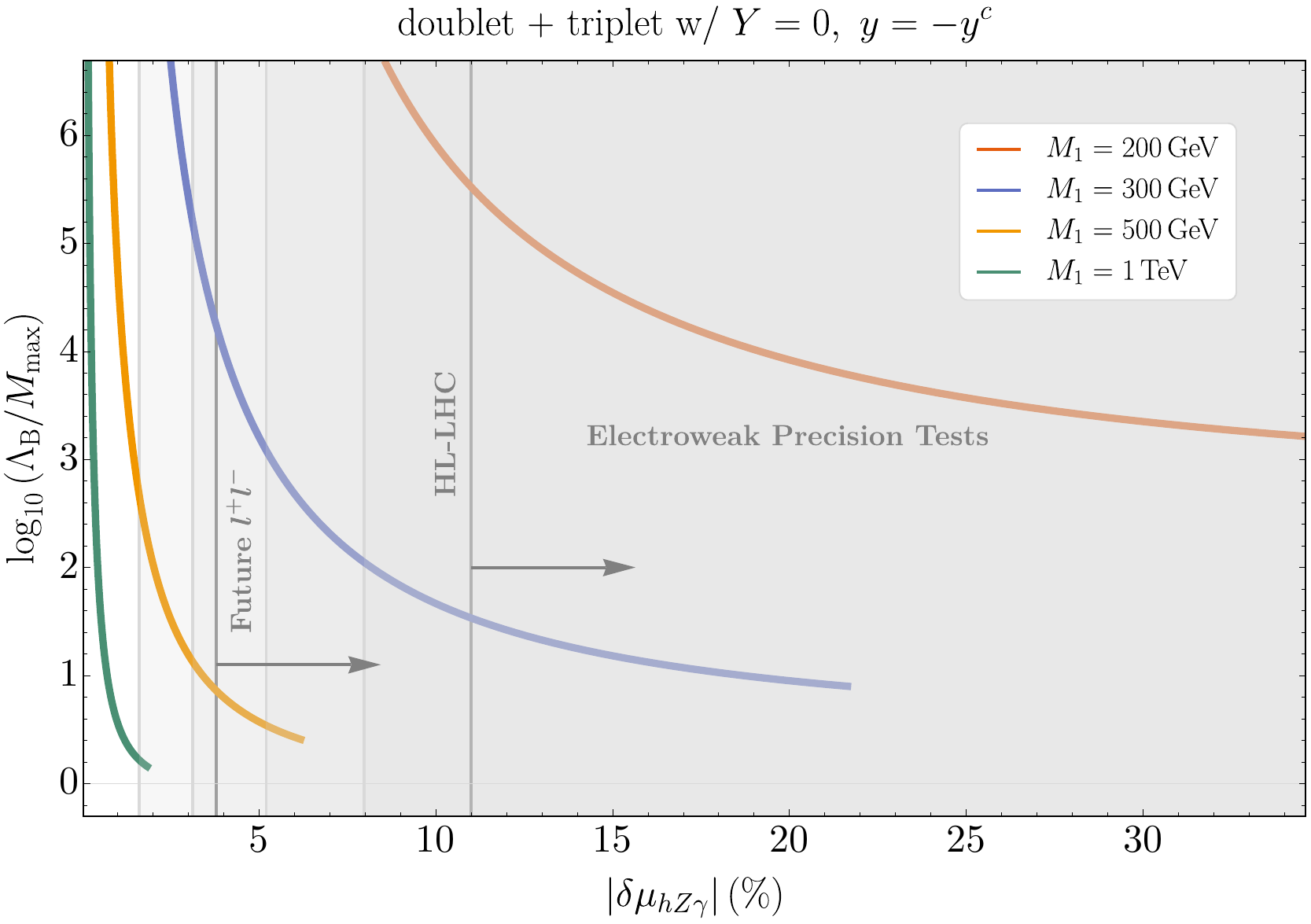}\hfill
\includegraphics[width=7.5cm]{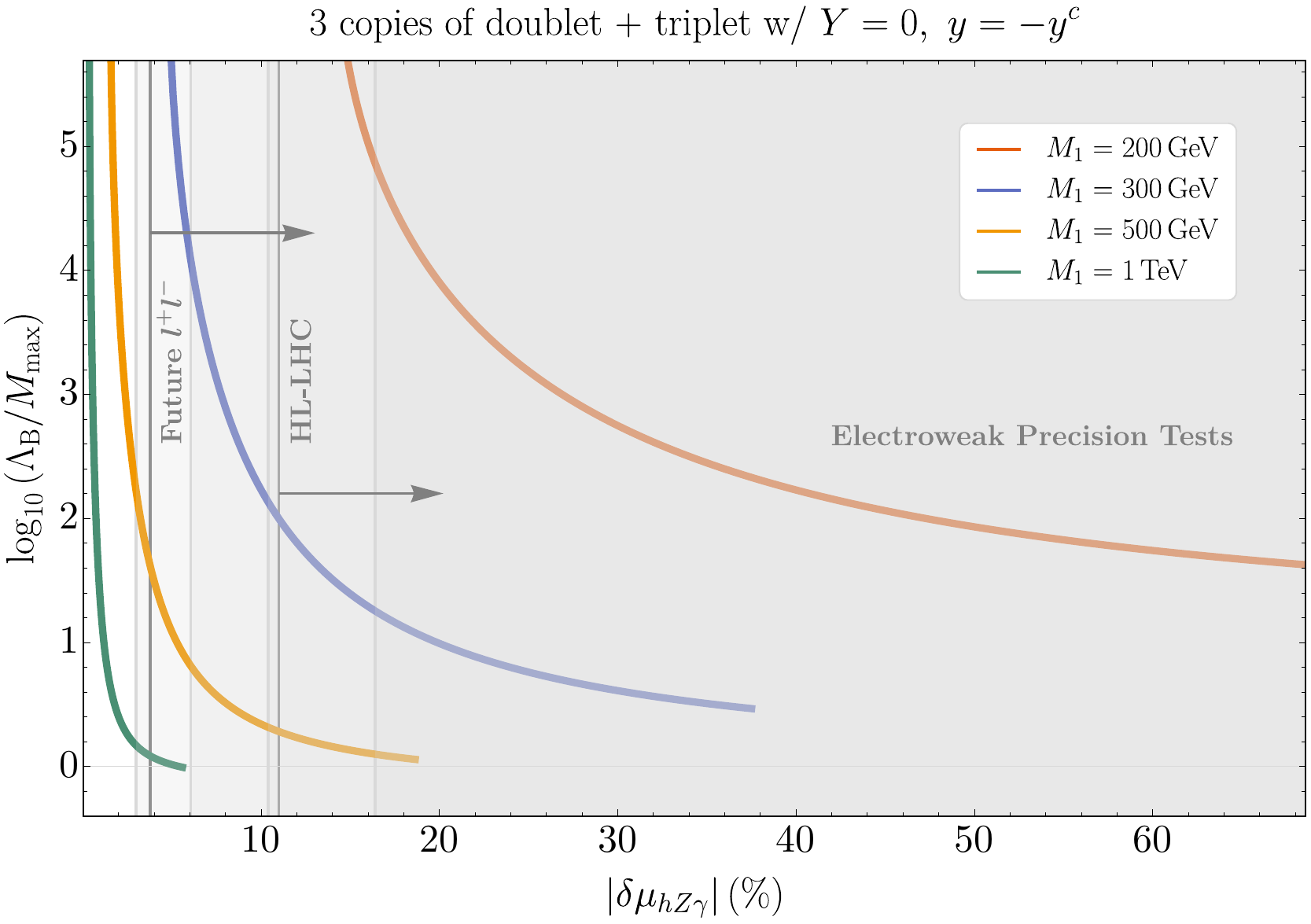}\hfill
\includegraphics[width=7.5cm]{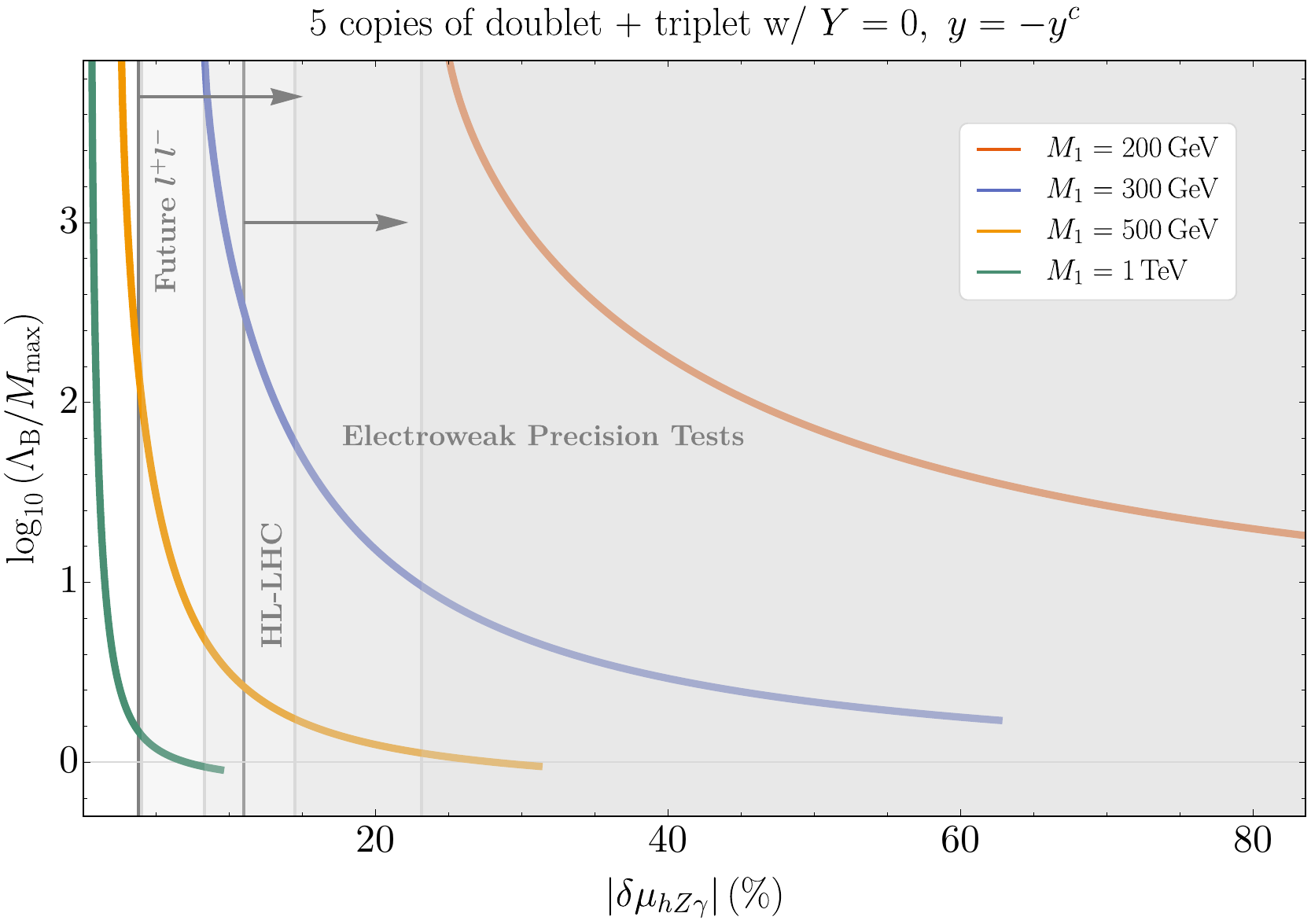}\hfill
\end{center}
\caption{Upper bound on the scale of new bosons $\Lambda_\tn{B}$ as a function of the relative coupling deviation $\delta \mu_{hZ\gamma}$ in the models $(r=1, n=3, Y=0)$ with $N_\mathrm{F} = 1, 3, 5$ flavors (top left, top right, and bottom panels, respectively). The choice $y=(-1)^n y^c$ indicated in the title of each plot maximizes $\Lambda_\tn{B}$. $M_\mathrm{max}$ is the largest of the VLF masses, while $M_1$ is the smallest one. The gray shaded areas represent the constraint from EWPTs which, at lower $\delta \mu_{hZ\gamma}$, is on the line with lowest $M_1$.} 
\label{hzgam_23_NF}
\end{figure}

\begin{figure}[t]
\begin{center}
\includegraphics[width=7.5cm]{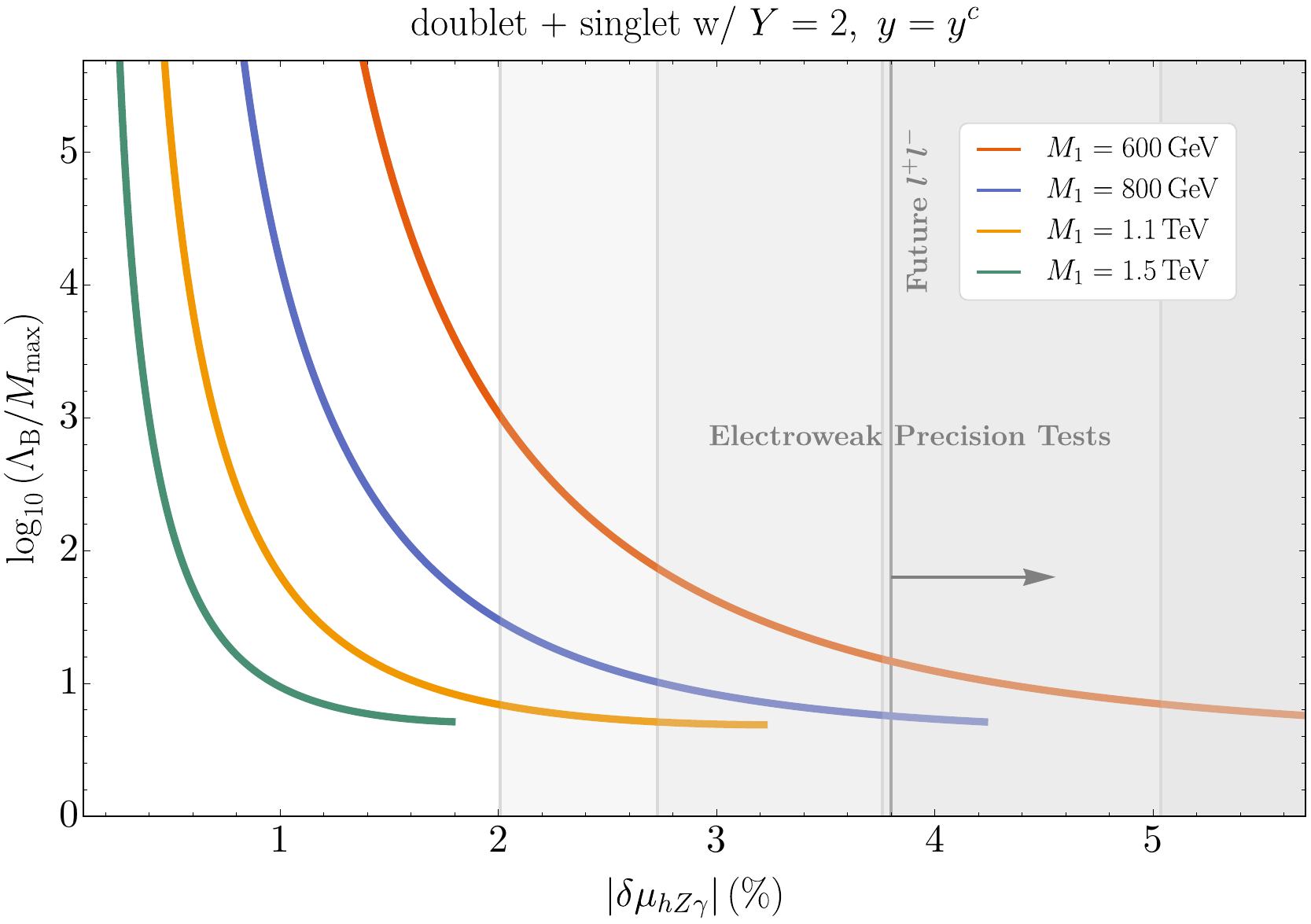}\hfill
\includegraphics[width=7.5cm]{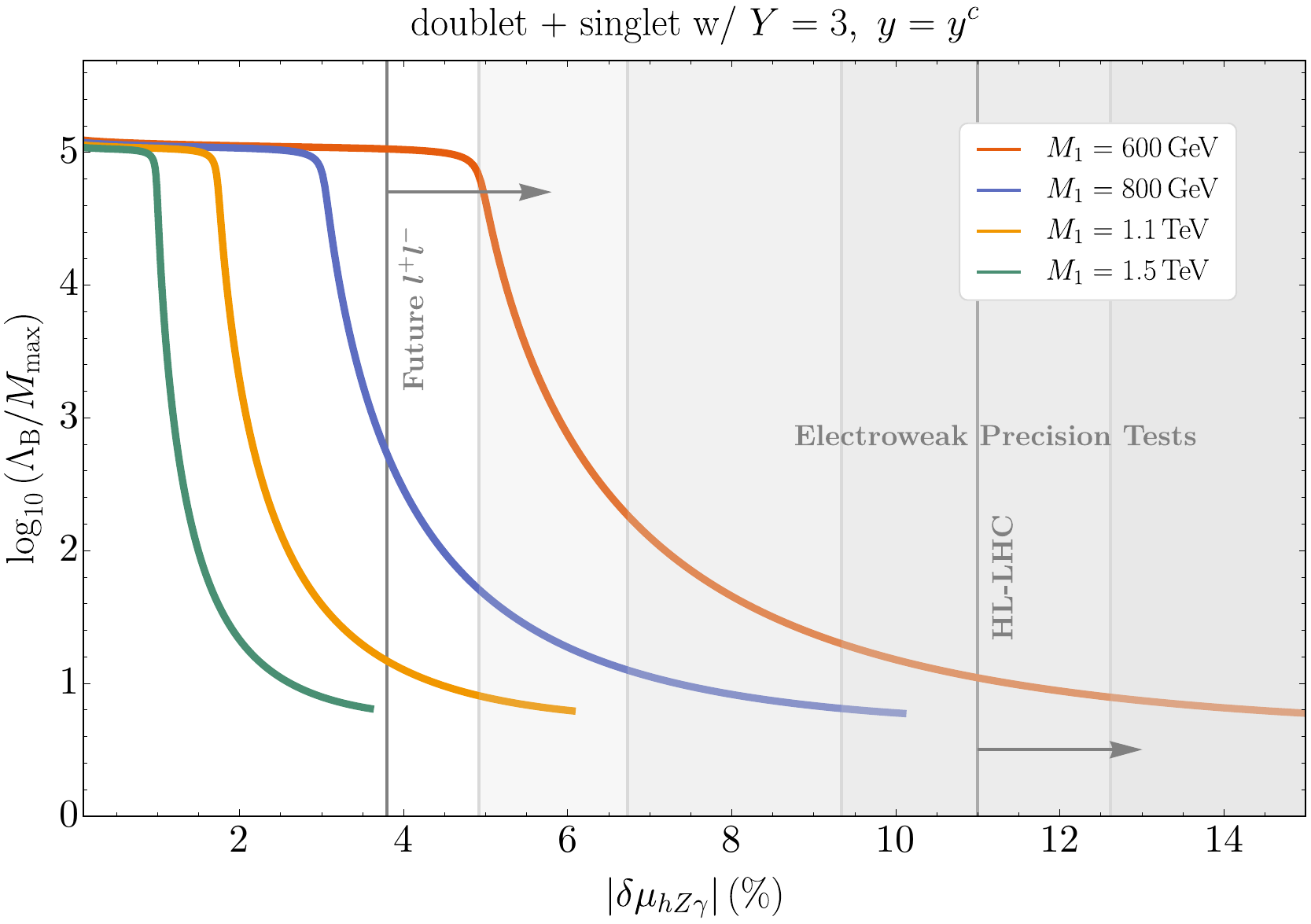}\hfill
\end{center}
\caption{Upper bound on the scale of new bosons $\Lambda_\tn{B}$ as a function of the relative coupling deviation $\delta \mu_{hZ\gamma}$ in the models $(r=1, n=3, N_\mathrm{F}=1)$ with $Y = 2, 3$ (left and right panels, respectively). The choice $y=(-1)^n y^c$ indicated in the title of each plot maximizes $\Lambda_\tn{B}$. $M_\mathrm{max}$ is the largest of the VLF masses, while $M_1$ is the smallest one. The gray shaded areas represent the constraint from EWPTs which, at lower $\delta \mu_{hZ\gamma}$, is on the line with lowest $M_1$.}
\label{hZgam_Y}
\end{figure}

\begin{figure}[t]
\begin{center}
\includegraphics[width=7.5cm]{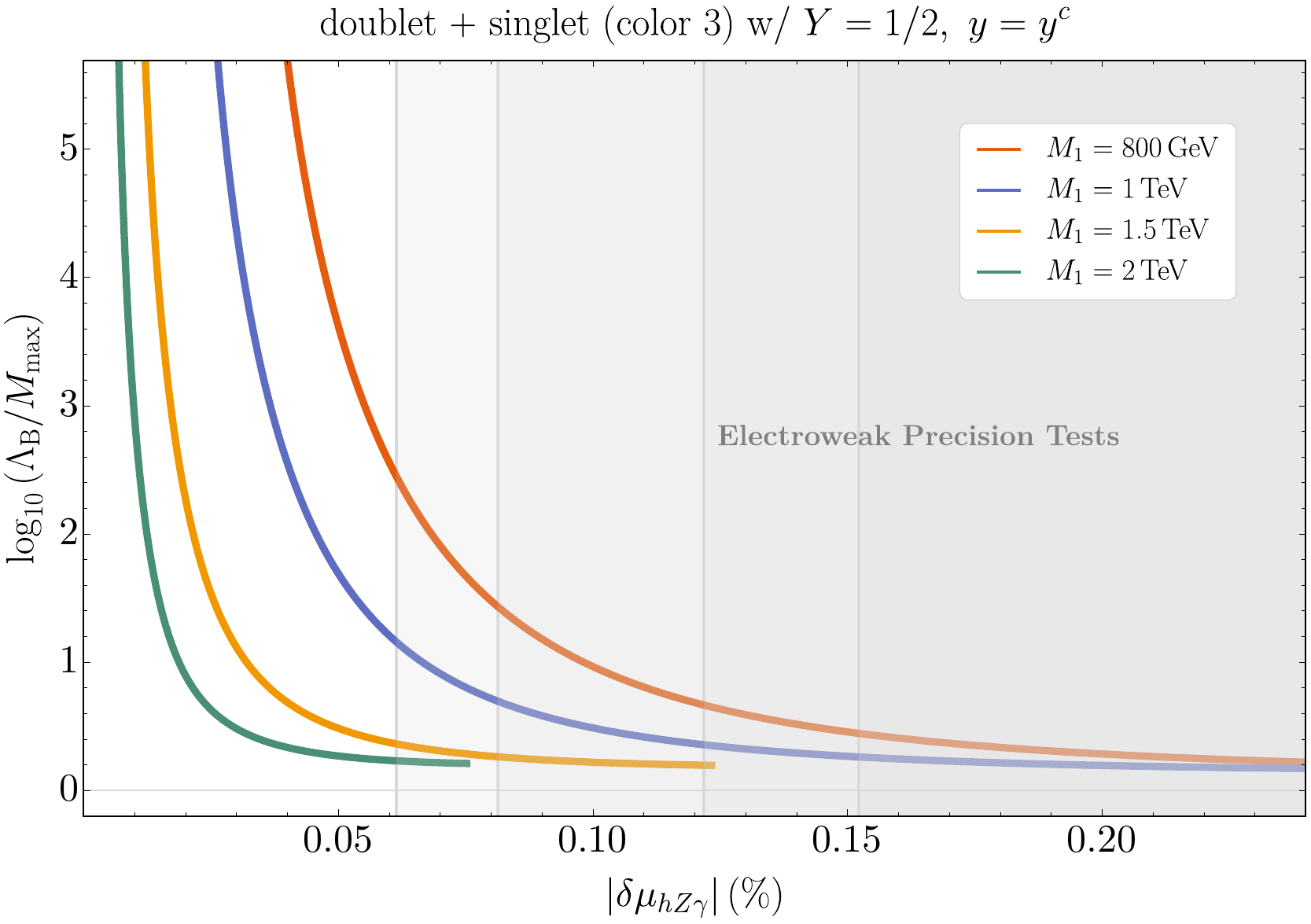}\hfill
\includegraphics[width=7.5cm]{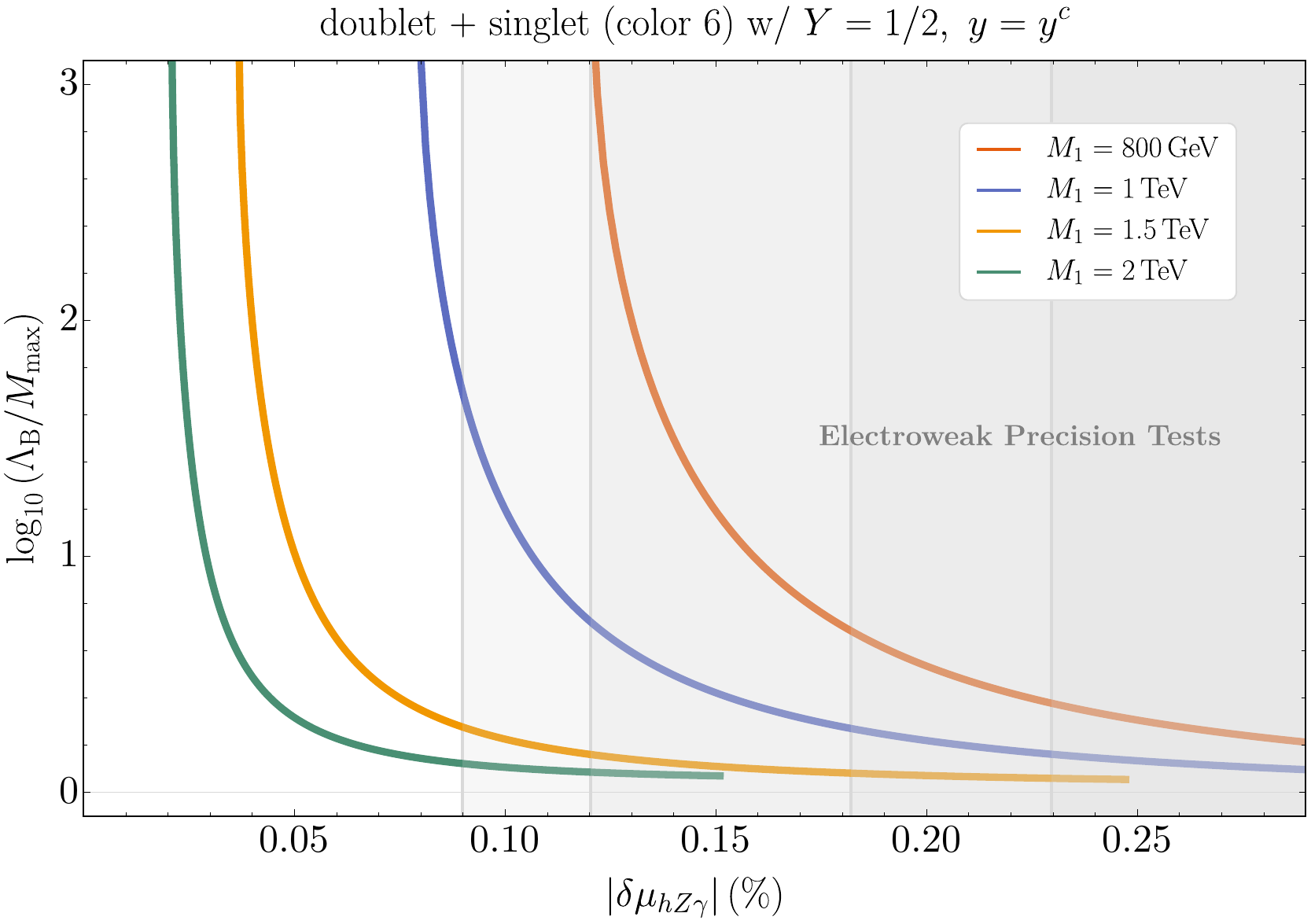}\hfill
\includegraphics[width=7.5cm]{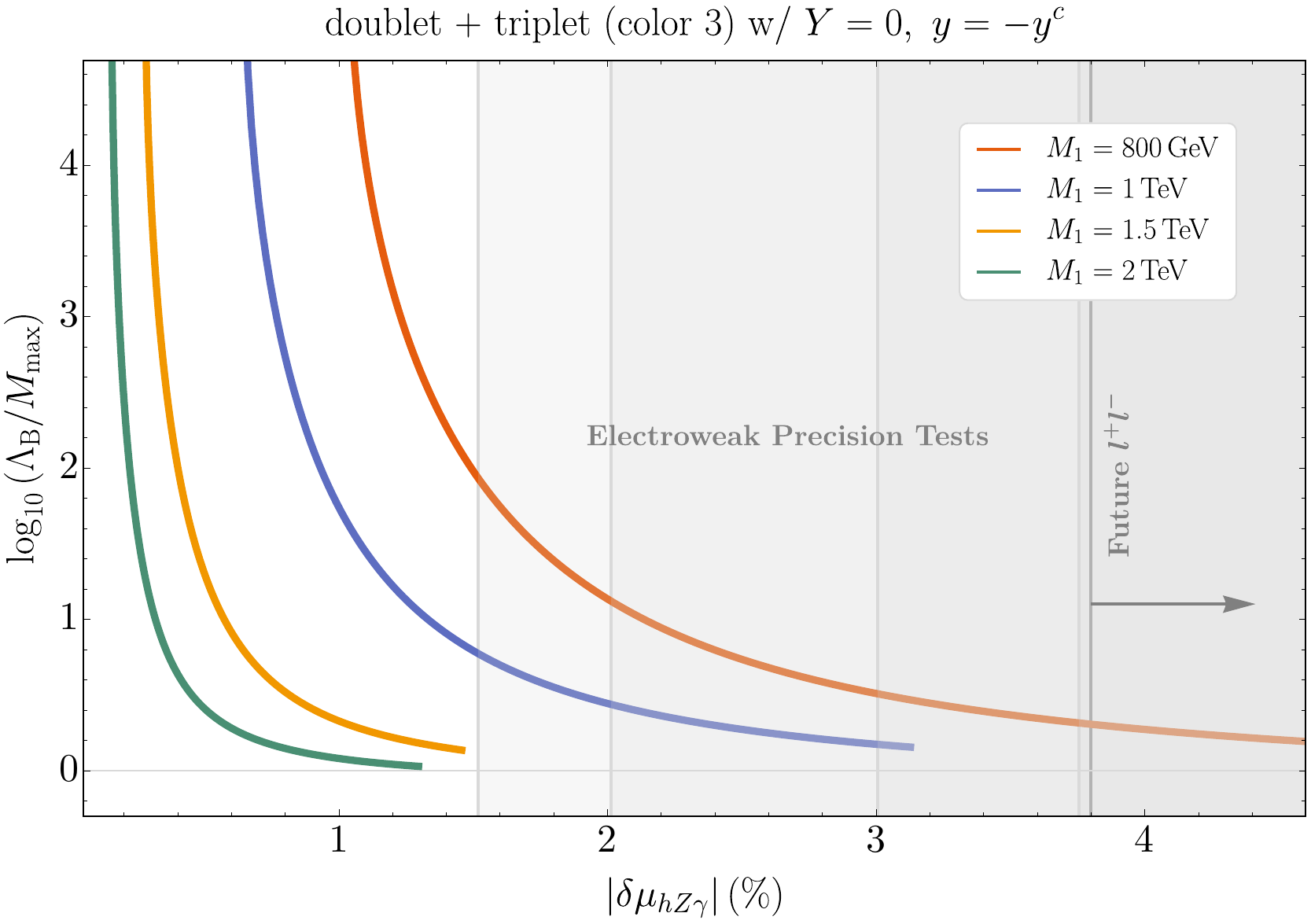}\hfill
\end{center}
\caption{Upper bound on the scale of new bosons $\Lambda_\tn{B}$ as a function of the relative coupling deviation $\delta \mu_{hZ\gamma}$.
The choice $y= (-1)^n y^c$ indicated in the title of each plot maximizes $\Lambda_\tn{B}$. $M_\mathrm{max}$ is the largest of the VLF masses, while $M_1$ is the smallest one. The gray shaded areas represent the constraint from EWPTs which, at lower $\delta \mu_{hZ\gamma}$, is on the line with lowest $M_1$.
Top-left: Model $(r=3, n=2, Y=1/2, N_\mathrm{F} = 1)$. Top-right: Model $(r=6, n=2, Y=1/2, N_\mathrm{F}=1)$. Bottom: Model $(r=3, n=3, Y=0, N_\mathrm{F} = 1)$.}
\label{hZgamCol}
\end{figure}

\begin{figure}[t]
\begin{center}
\includegraphics[width=7.5cm]{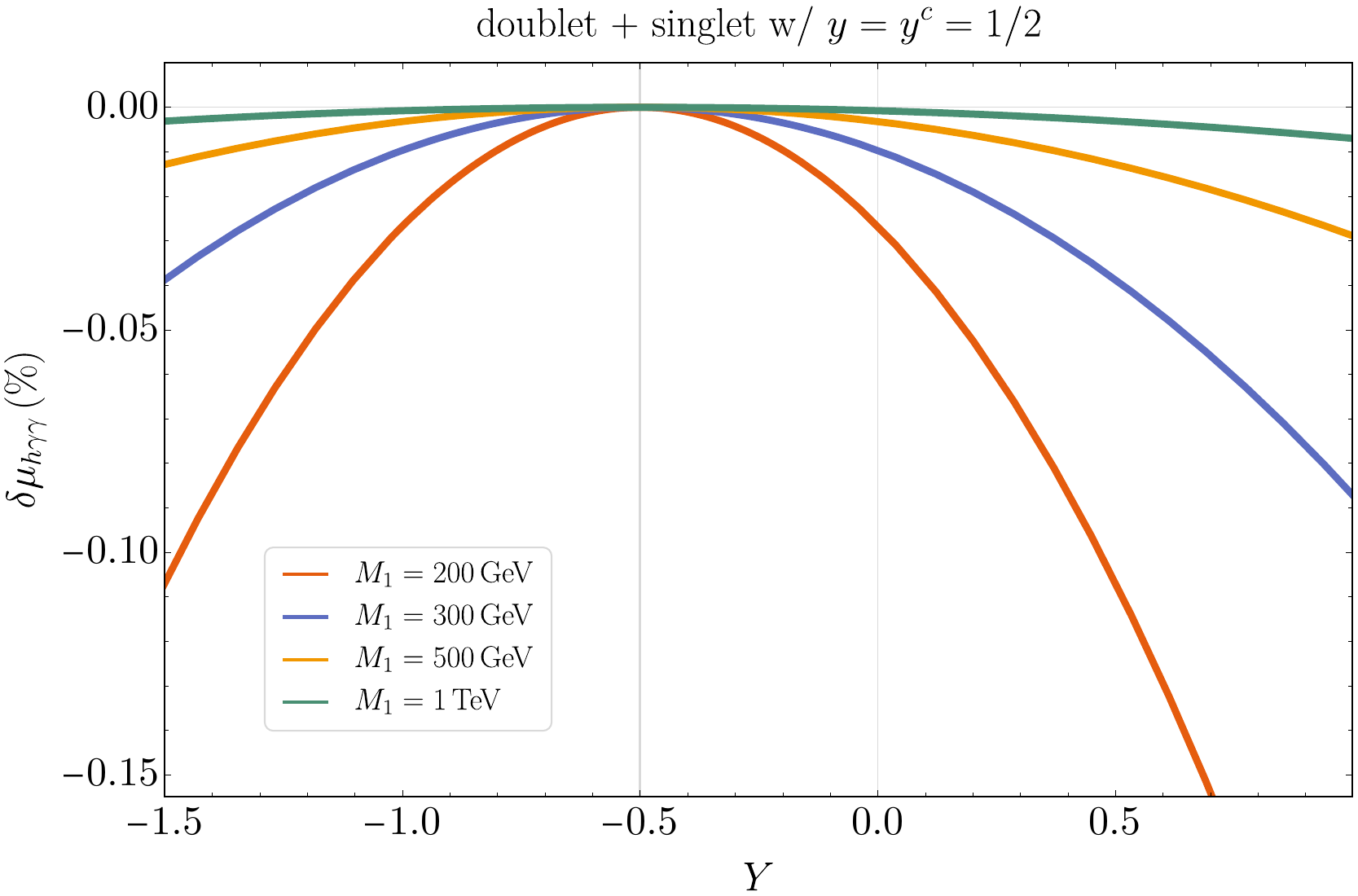}\hfill
\includegraphics[width=7.5cm]{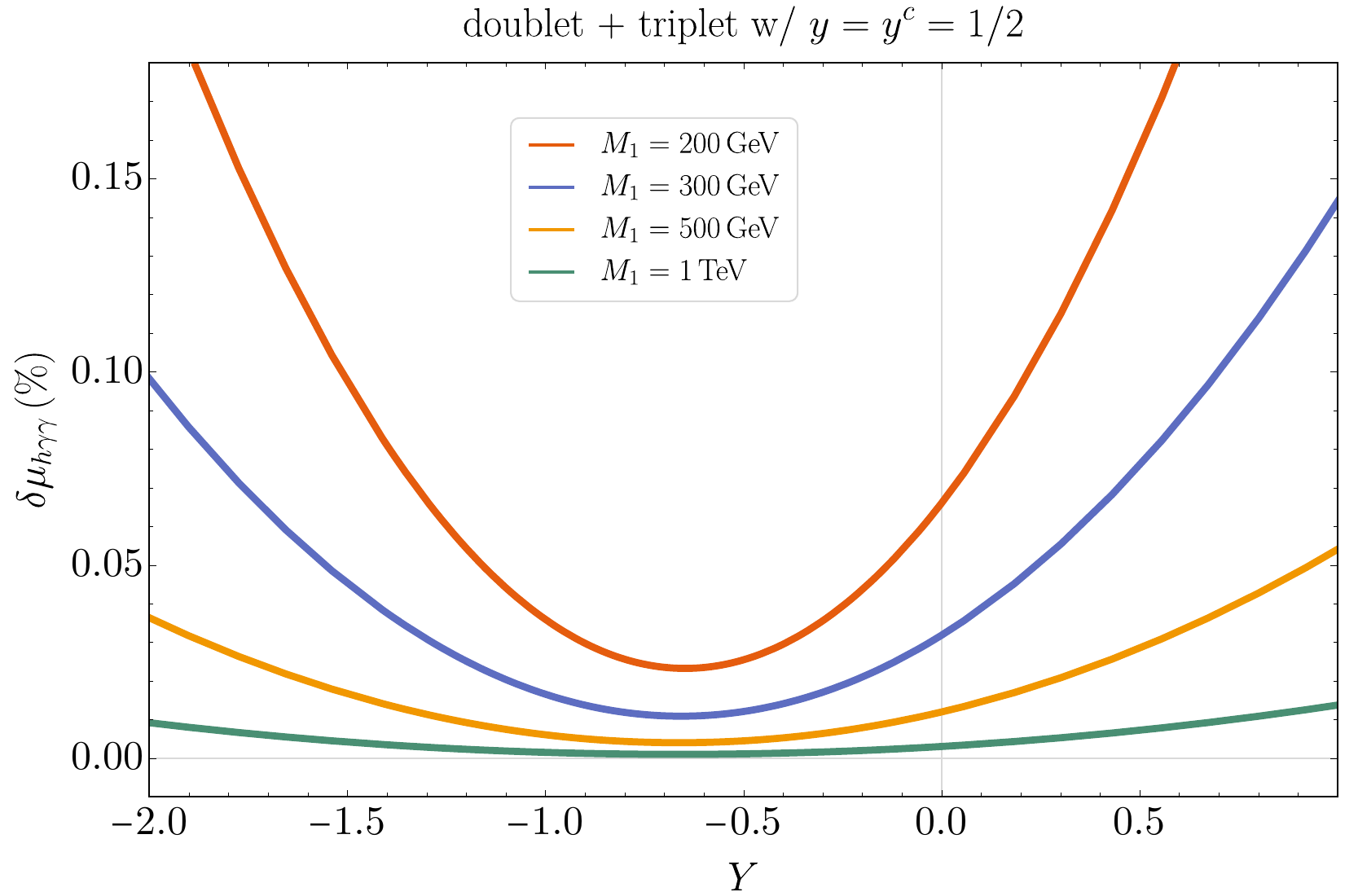}\hfill
\includegraphics[width=7.5cm]{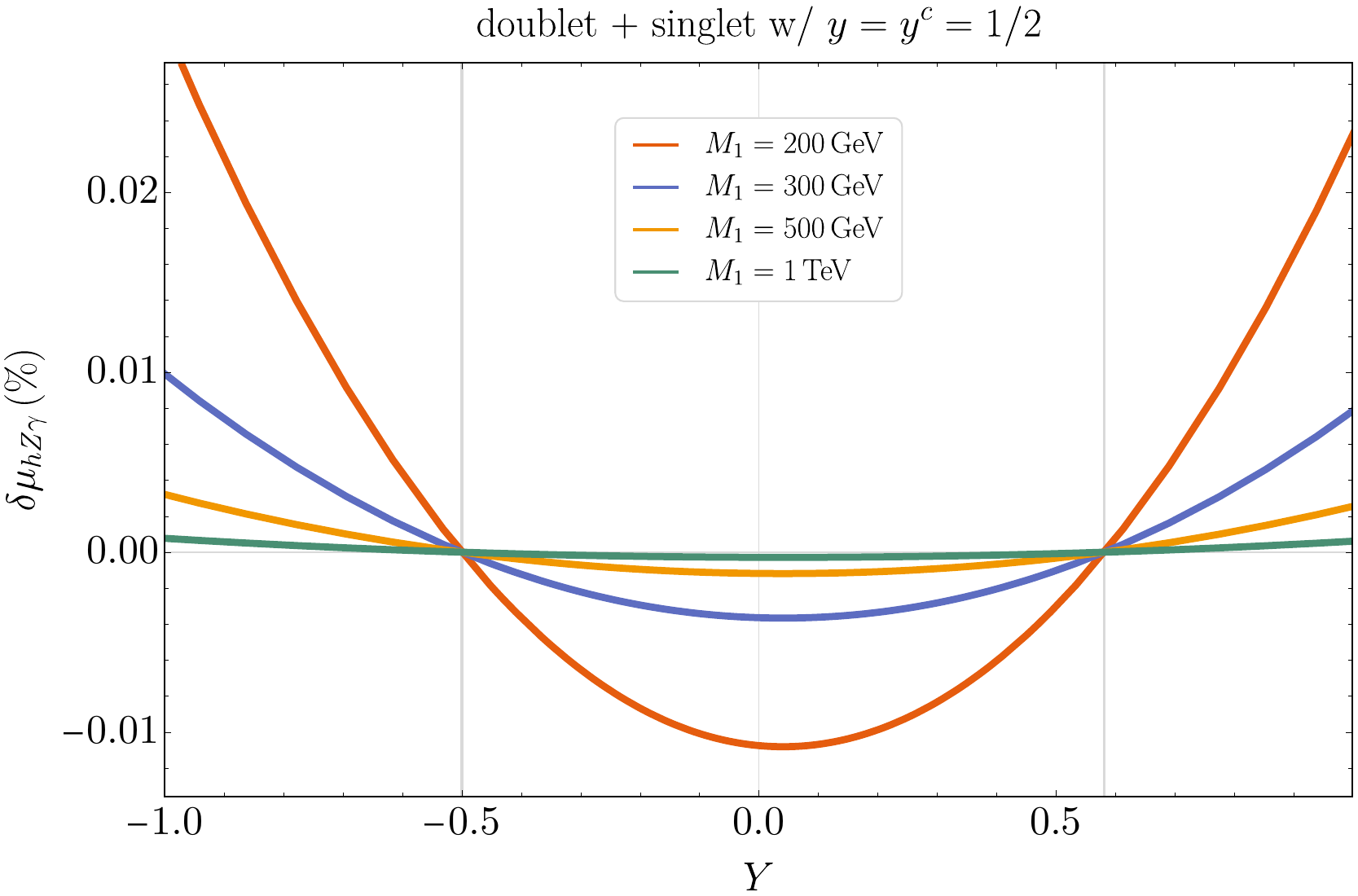}\hfill
\includegraphics[width=7.5cm]{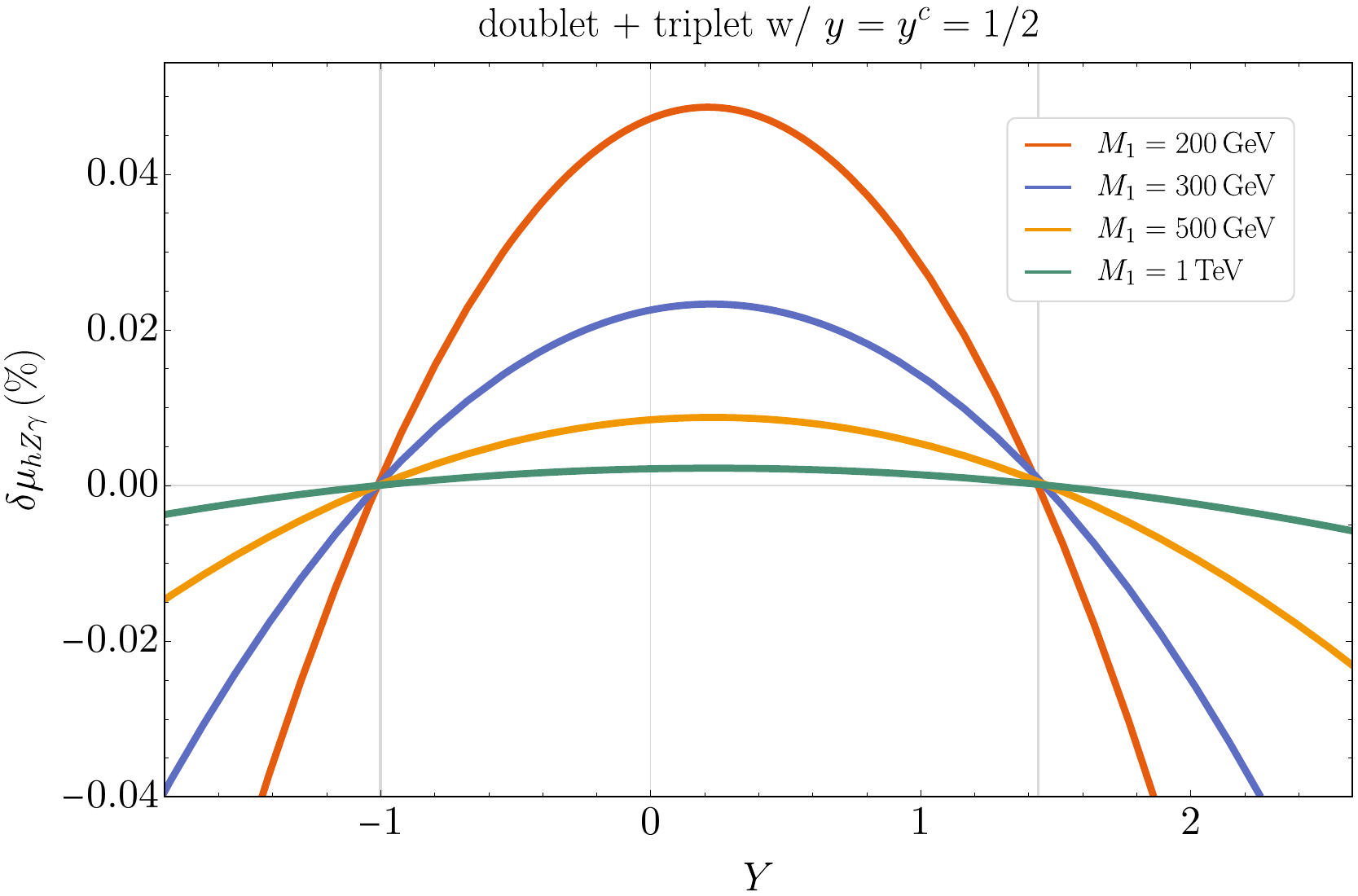}\hfill
\end{center}
\caption{Relative coupling deviation $\delta \mu_{h\gamma\gamma}$ (top panels) and $\delta \mu_{hZ\gamma}$ (bottom panels) as function of $Y$, for different values of $M_1$ (the mass of the lightest VLF in the spectrum), for $n=2$ (left panels) and $n=3$ (right panels). The choice $y=(-1)^n y^c$ indicated in the title of each plot maximizes the coupling deviation.}
\label{hVV_Y}
\end{figure}

\subsection{Preliminary Discussion}
In Section~\ref{results_sec}, we analyze the relative deviations $\delta \mu_{hVV^\prime}$ of the three loop-induced Higgs couplings at 1-loop (see Fig.~\ref{H_decays}) to derive an upper bound on the mass scale of new bosons, $\Lambda_\tn{B}$. In Appendix~\ref{scaling_arg}, we argue that no clear scaling relation exists to determine $\Lambda_\tn{B}$ for large values of $r$, $n$, $Y$, or $N_\mathrm{F}$. The pure SM decay widths $\Gamma^\mathrm{SM}(h \rightarrow gg)$, $\Gamma^\mathrm{SM}(h \rightarrow \gamma \gamma)$, and $\Gamma^\mathrm{SM}(h \rightarrow Z \gamma)$ are available in the review~\cite{Djouadi:2005gi}. For the SM loops, we include only the dominant contributions from the $W$-boson and the $t$-quark.

In the computation of the amplitudes, we neglect the running of the couplings between the weak scale and the new fermion scale $\Lambda_\tn{F}$. As discussed in Ref.~\cite{Blum:2015rpa}, a leading-order (LO) analysis for VLFs with $r=3$ is justified, as the dominant next-to-leading-order (NLO) effect is a multiplicative factor that cancels out in the ratio of Eq.~\eqref{delta_HVV}. However, for $r \geq 6$, this approximation no longer holds, and we anticipate some modifications to our results, potentially on the order of $10\%$. A full NLO analysis, however, is beyond the scope of this study. To determine $\Lambda_\tn{B}$ from the criteria of vacuum stability and the absence of Landau poles, we compute the running of the couplings using the 2-loop renormalization group equations (RGEs), which we derive with the \texttt{Wolfram Mathematica} extension \texttt{SARAH}~\cite{Staub:2008uz, Staub:2009bi, Staub:2010jh, Staub:2012pb, Staub:2013tta, Staub:2015kfa, Staub:2015iza, Staub:2016sms, Goodsell:2018tti}.

The relative coupling deviations $\delta \mu_{hVV^\prime}$ are compared with the latest projections for the $1\sigma$ sensitivity of FLCs\footnote{We use the most optimistic projections, assuming the realization of both the FCC-ee and a muon collider (the latter with a run at the Higgs pole and $\sqrt{s} = 10$ TeV)~\cite{deBlas:2022ofj}.}. These projections are derived from a global SMEFT fit~\cite{deBlas:2022ofj}. Additionally, $2\sigma$ constraints from LHC experiments~\cite{CMS:2022dwd, ATLAS:2022vkf, ParticleDataGroup:2024cfk} are indicated by vertical lines in the plots\footnote{Lines not displayed in the plots correspond to constraints outside the range shown.}. In Figs.~\ref{hgg_plots}–\ref{hZgamCol}, we present plots of $\Lambda_\tn{B}$ as a function of the relative coupling deviations for different VLF representations. We find that $\Lambda_\tn{B}$ is typically determined by the vacuum stability scale, except for cases involving large hypercharges $Y$ (we discuss this exception later). In the same plots, we compare different values of the lightest particle mass $M_1$ in the VLF spectrum (depicted as red, blue, yellow, and green lines). The relation $y = (-1)^n y^c$ is chosen to maximize $\Lambda_\tn{B}$ for a given $\delta \mu_{hVV^\prime}$. The colored lines in the plots are truncated where perturbative control of the model is lost, specifically where the Higgs quartic coupling becomes rapidly large and negative beyond the vacuum stability threshold.

\subsection{Discussion of the Results}
\label{results_sec}
Below, we provide the leading $1/M_L^2$ behavior of the 1-loop amplitudes for $v y^{(c)}/M_L\ll 1$. However, for all numerical applications, we use the full 1-loop analytic expressions for the amplitudes, computed with \texttt{Package-X}~\cite{Patel:2015tea, Patel:2016fam}. An important observation is that $\Lambda_\tn{B}$ is typically determined by the vacuum stability scale, except in cases involving higher exotic hypercharges (we will address this when relevant).

\subsubsection{Higgs Coupling to Digluons}
\label{ggH}
The $h \to gg$ decay is induced at 1-loop by all SM quarks, as well as by colored VLFs that couple to the $h$-boson. In addition to the SM diagrams, the Feynman diagram shown in Fig.~\ref{H_decays} (left) must also be considered. The 2-body decay width is given by
\begin{equation}
\Gamma(h\rightarrow gg)=\frac{m_h^3}{8\pi} \left| C_{hgg}^{\text{SM}}+C_{hgg}^{\text{VLF}} \right|^2 \, ,
\end{equation}
where we have displayed separately the SM and VLF contributions, $C_{h gg}^{\text{SM}}$ and $C_{h gg}^{\text{VLF}}$, respectively. The VLF contribution, in the limit $M_L \gg y^{(c)} v$, is given by
\begin{equation}
C_{hgg}^{\text{VLF}}= (-1)^{n-1} N_\mathrm{F} \, T(r) \, n \, \frac{g_s^2 y y^c v}{12 \pi^2 M_L^2} \, , \ \ \ T(r)\delta^{ab} = \text{Tr}[T_r^{a}T_r^{b}] \, ,  
\label{C_hgg}
\end{equation}
where the $T_r^{a}$ are the generators of $SU(3)_C$ in the representation $r$ of the VLFs in the loop. The $1\sigma$ sensitivities of the HL-LHC and FLCs are $1.8\%$ and $0.37\%$, respectively~\cite{deBlas:2022ofj}, while the $2\sigma$ constraint from the LHC is $17\%$~\cite{CMS:2022dwd, ATLAS:2022vkf, ParticleDataGroup:2024cfk}.

The results for different models are presented in the plots in Fig.~\ref{hgg_plots}. From these, the following points can be deduced:

\begin{itemize}
\item The model $(r=3, n=2, Y=1/2, N_\mathrm{F}=1)$ can produce a deviation $\delta \mu_{hgg}$ that falls within the sensitivity of the HL-LHC projections without conflicting with the EWPT constraints. For the lightest VLF with a mass $M_1 \simeq 1$ TeV, a significant hierarchy between the new fermions and new bosons can still be achieved, i.e., $\Lambda_\tn{B} \gg M_\mathrm{max}$, where $M_\mathrm{max}$ is the mass of the heaviest VLF. This mass region is beginning to be constrained by the LHC, while $M_1 < 1$ TeV is excluded by direct searches. However, as $M_1$ increases above 1 TeV, $\Lambda_\tn{B}$ quickly approaches $M_\mathrm{max}$: it is then meaningless to consider a model with only new fermions and no bosons to explain a potential observed deviation $\delta \mu_{hgg}$.

\item For the higher color representation $(r=6, n=2, Y=1/2, N_\mathrm{F}=1)$, achieving a sizable $\delta \mu_{hgg}$ at the TeV scale becomes easier. This deviation can fall within the sensitivity of the HL-LHC, while maintaining a significant hierarchy $\Lambda_\tn{B} \gg M_\mathrm{max}$. With $M_1 \simeq 1$ TeV, it is even possible to accommodate an anomaly in $hgg$ at the LHC, should one appear during Run 3, though the hierarchy with new bosons would be somewhat limited. For TeV-scale VLFs, there is no tension with EWPTs.

\item If $n$ is increased instead of $r$, for example in the model $(r=3, n=3, Y=0, N_\mathrm{F}=1)$, the situation improves slightly compared to the $n=2$ case. This results in a larger hierarchy for fixed $M_1$ and $\delta \mu_{hgg}$.
\end{itemize}
Therefore, the current collider bounds and EWPTs still allow for colored VLFs at the TeV scale that can produce a deviation $\delta \mu_{hgg}$ within the projected sensitivity of the HL-LHC. Such deviations could even be indirectly observed at the LHC Run 3 for $r \geq 6$. Consequently, there is (a priori) no need to introduce multiple flavors of VLFs to accommodate a potential future anomaly without adding new bosons.

\subsubsection{Higgs Coupling to Diphotons}
\label{Hgamgam}
The $h \rightarrow \gamma \gamma$ decay is mediated at 1-loop by all electrically charged particles in the SM, as well as by electrically charged VLFs that couple to the $h$-boson. In addition to the SM diagrams, the Feynman diagram shown in Fig.~\ref{H_decays} (middle) must also be included. The 2-body decay width is given by
\begin{equation}
\Gamma (h \rightarrow \gamma \gamma) = \dfrac{m_h^3}{64 \pi} \left| C_{h \gamma \gamma}^{\text{SM}} + C_{h \gamma \gamma}^{\text{VLF}} \right|^2 \, ,
\end{equation}
where we have distinguished the SM and VLF contributions, $C_{h \gamma \gamma}^{\text{SM}}$ and  $C_{h \gamma \gamma}^{\text{VLF}}$, respectively.
The VLF contribution, in the limit $M_L \gg y^{(c)} v$, is given by
\begin{equation}
C_{h \gamma \gamma}^{\text{VLF}}=(-1)^{n-1} \, N_\mathrm{F} \, r \,  \frac{e^2 y y^c v}{144 \pi^2 M_L^2}\left[n^2-1+4(n+1)Y+12Y^2\right] \, .
\label{C_hgamgam}
\end{equation}
The $1\sigma$ sensitivities of the HL-LHC and FLCs are $2.4\%$ and $0.69\%$, respectively~\cite{deBlas:2022ofj}, while the $2\sigma$ constraint from the LHC is $33\%$~\cite{CMS:2022dwd, ATLAS:2022vkf, ParticleDataGroup:2024cfk}. In Figs.~\ref{hgamgam_12_NF}–\ref{hgamgam_Col}, we present plots of $\Lambda_\tn{B}$ as a function of $|\delta \mu_{h\gamma\gamma}|$ for different VLF models.

In Fig.~\ref{hgamgam_12_NF}, we consider the model $(r=1, n=2, Y=1/2)$ for $N_\mathrm{F} = 1, 3, 5$. The results show that any future deviation observed at the LHC is unlikely to be explained by a small number of VLFs, as they would need to be near the EW scale and would be in tension with EWPTs. However, within the sensitivity of the HL-LHC, an anomaly can be accommodated with new VLFs at the EW scale while maintaining a large hierarchy $M_\mathrm{max} \gg \Lambda_\tn{B}$. For $M_1 \gtrsim 1$ TeV, generating an anomaly observable at the HL-LHC would require new bosons at the same scale as the new fermions $(M_\mathrm{max} \sim \Lambda_\tn{B})$. Additionally, increasing $N_\mathrm{F} \sim 1$ does not significantly alter this discussion. In fact, larger values of $N_\tn{F}$ allow for larger deviations at a fixed hierarchy, consistent with the scaling of RGEs and coupling deviations. This behavior is a general feature of our analysis across all the couplings considered. If an anomalous coupling were to be detected experimentally, these models would warrant further exploration despite their nonminimal content.

In Fig.~\ref{hgamgam_23_NF}, we consider the model $(r=1, n=3, Y=0)$ for small numbers of VLF flavors, specifically $N_\mathrm{F} = 1, 3, 5$. This model cannot explain any deviation observed at the LHC without being in significant tension with EWPTs. However, it can achieve the HL-LHC sensitivity, performing slightly better than the $n=2$ model, while allowing $\Lambda_\tn{B} \gg M_\mathrm{max}$ for $M_1$ on the order of a few hundred GeV. These features improve when multiple flavors are considered. For $M_1 \gtrsim 1$ TeV, however, $\Lambda_\tn{B} \sim M_\mathrm{max}$, and the model becomes irrelevant.

Fig.~\ref{hgamgam_Y} illustrates the impact of selecting a higher hypercharge in the model $(r=1, n=2, N_\mathrm{F}=1)$. For $Y=2$, it is still possible to achieve a small hierarchy $\Lambda_\tn{B} \gg M_\mathrm{max}$ for $M_1 \sim 1$ TeV and an anomaly observable at the HL-LHC, which was not feasible for $Y=1/2$. For $Y=3$, there is a plateau near the HL-LHC sensitivity, with a hierarchy $\Lambda_\tn{B} / M_\mathrm{max} \sim 10^5$. This corresponds to a Landau pole instability rather than a vacuum instability. Thus, exotic VLFs with higher hypercharges remain viable minimal models to explain an anomaly at the HL-LHC. However, the trade-off is stronger collider constraints from long-lived charged particles around 1 TeV. Since such particles cannot simply mix with SM fermions to decay, achieving this requires additional model-building.

Regarding the impact of colored representations, Fig.~\ref{hgamgam_Col} highlights the following points:

\begin{itemize}
\item The model $(r=3, n=2, Y=1/2, N_\tn{F}=1)$ does not allow for a hierarchy between new bosons and VLFs for a TeV-scale $M_1$ (consistent with current collider bounds), although such a hierarchy becomes possible for FLCs.

\item Increasing the color representation to $(r=6, n=2, Y=1/2, N_\tn{F}=1)$ enables a hierarchy $\Lambda_\tn{B} \gg M_\mathrm{max}$ for $M_1 \simeq 1$ TeV within the HL-LHC sensitivity. However, this hierarchy collapses rapidly with a slight increase in $M_1$.

\item Alternatively, increasing $n$, as in the model $(r=3, n=3, Y=1/2, N_\tn{F}=1)$, results in a worse situation compared to $n=2$ for maintaining a hierarchy.

\end{itemize}

\subsubsection{Higgs Coupling to a \texorpdfstring{$\bm{Z}$}{Z}-Boson and a Photon}
\label{HZgam}

The $h \rightarrow Z \gamma$ decay occurs at 1-loop through virtual SM bosons and fermions, with additional contributions from VLFs as represented by the Feynman diagram in Fig.~\ref{H_decays} (right). The 2-body decay width is given by
\begin{equation}
\Gamma (h \rightarrow Z \gamma) = \dfrac{m_h^3}{32 \pi} \left( 1- \dfrac{m_Z^2}{m_h^2} \right)^3 \left| C_{h Z \gamma}^{\text{SM}} + C_{h Z \gamma}^{\text{VLF}} \right|^2 \, ,
\end{equation}
where we have again separated the SM contribution $C_{h Z \gamma}^{SM}$ and the VLF contribution $C_{h Z \gamma}^{VLF}$. In the limit $M_L \gg y^{(c)} v$, one gets
\begin{equation}
\begin{aligned}
C_{h Z \gamma}^{\text{VLF}}
&= (-1)^{n-1} \, N_\mathrm{F} \, r \, n \, \dfrac{g^2 y y^c v}{288\pi^2 M_L^2} \, \tan \theta_W \left\{n(n-1)-2(1+3Y+6Y^2) \right. \\
&\phantom{{}={}}\left. + \cos 2 \theta_W \left[n^2-1+4(n+1)Y+12Y^2 \right] \right\} \, .
\label{C_hZgam}
\end{aligned}
\end{equation}
The $1\sigma$ sensitivities of the HL-LHC and FLCs are $11\%$ and $3.8\%$, respectively~\cite{deBlas:2022ofj}, while no precise measurement is available from the LHC yet~\cite{ATLAS:2023yqk}. It is interesting to note that, unlike $\delta \mu_{h\gamma\gamma}$, we found that $\delta \mu_{hZ\gamma}$ can change sign as a function of $|Y|$, as shown in Fig.~\ref{hVV_Y}. For sufficiently large $|Y|$, however, $\delta \mu_{hZ\gamma}$ becomes an increasing function of $|Y|$.

In Figs.~\ref{hzgam_12_NF}–\ref{hzgam_23_NF}, we illustrate the effect of the number of flavors $N_\tn{F} = 1, 3, 5$ for the models $(r=1, n=2, Y=1/2)$ and $(r=1, n=3, Y=0)$. In the case $n=2$, future collider experiments are not expected to be sensitive to the loop-induced deviations caused by these VLFs, even with an increase in $N_\tn{F} \sim 1$. For $n=3$, the values of $M_1$ that allow a deviation to be observable at FLCs while maintaining a (mild) hierarchy $\Lambda_\tn{B} > M_\mathrm{max}$ are in tension with EWPTs. However, an optimistic window of $M_1 \sim 300$–$500$ GeV for $N_\tn{F} = 3, 5$ remains allowed. In contrast, the values of $M_1$ required to produce a deviation observable at the HL-LHC are in significant tension with EWPTs. Furthermore, increasing the dimension of the color representation, as shown in Fig.~\ref{hZgamCol}, significantly lowers $\Lambda_\tn{B}$, making the situation worse than for colorless representations.

The most interesting case for the $hZ\gamma$ coupling arises when the hypercharge is increased, as shown in Fig.~\ref{hZgam_Y} for the model $(r=1, n=2, N_\tn{F}=1)$. While the value $Y=2$ does not produce a deviation observable at future colliders without conflicting with EWPTs, the value $Y=3$ is sufficient to generate a deviation detectable at FLCs. This scenario allows for a modest boson/fermion hierarchy with $M_1$ at the TeV scale.

To conclude this section, let us note that the ATLAS and CMS experiments have recently reported evidence of the $hZ\gamma$ coupling with an observed signal yield of $2.2 \pm 0.7$ times the SM value~\cite{ATLAS:2023yqk}. Since the result is a $1.9 \sigma$ deviation from the SM prediction, it does not constitute solid evidence for BSM physics. If this deviation is confirmed, it is clear from our results that a purely fermionic extension of the SM cannot fit such a large deviation without being in great tension with EWPTs, and without the requirement of new bosons at almost the same scale.

\section{Conclusion \& Outlook}
\label{conclusion}
In this article, we have examined pure fermionic extensions of the SM and their potential to produce deviations in the three loop-induced Higgs boson couplings—$hgg$, $h\gamma\gamma$, and $hZ\gamma$—at future colliders. If these fermions have sufficiently strong Yukawa couplings to the Higgs boson, they can induce an instability in the Higgs potential or lead to a Landau pole. These instabilities must be resolved by introducing new bosons. We have investigated whether it is possible to generate observable deviations in these couplings while maintaining a sufficiently large hierarchy with the new bosons, ensuring that they do not need to be explicitly included in the EFT:
\begin{itemize}
\item For the $hgg$ coupling, it is relatively straightforward to produce a deviation within the sensitivity of the HL-LHC, using relatively small representations of the gauge groups, and without violating EWPT constraints or collider bounds. In several scenarios, the upper bound on the mass scale of new bosons can be sufficiently high to neglect their contributions in the EFT.

\item For the $h\gamma\gamma$ coupling, due to the experimental constraints, increasing the hypercharge and/or the number of flavors is the most effective strategy for achieving a deviation at the HL-LHC while maintaining a fermion/boson hierarchy. In other cases, the hierarchy quickly diminishes as $M_1$ approaches 1 TeV. However, this is not a concern for achieving a deviation observable at FLCs.

\item For the $hZ\gamma$ coupling, future colliders have a lower sensitivity compared to the other two loop-induced couplings. There is a narrow mass window around 1 TeV where fermions with higher hypercharges can produce a deviation observable at FLCs while maintaining a hierarchy, consistent with experimental constraints. The key takeaway is that any deviation observed at the HL-LHC is highly unlikely to be caused solely by new fermions, as it would be in significant tension with EWPTs.
\end{itemize}

As a disclaimer, it is important to emphasize that our results were derived within a simplified framework designed to enable a general analysis of the dimensions of gauge representations. We employed approximate (but conservative) estimates for the experimental bounds, and several assumptions—such as the absence of mixing between different flavors and vanishing $CP$ phases—would need to be relaxed in a more detailed analysis. If a deviation is observed at a future collider, a dedicated analysis focusing on a specific fermion content would be required to determine precise collider, EW, and flavor constraints. Thus, our study should be regarded as a guide for model-builders, offering insights into the best strategies for constructing minimal models to explain a potential anomaly.

\acknowledgments
We are especially grateful to Raffaele Tito D'Agnolo for valuable advice, as well as for his work on our companion article~\cite{DAgnolo:2023rnh}. Thanks also to Luc Darmé for useful discussions. G.~R. acknowledges funding from the European Union’s Horizon 2020 research and innovation program under the Marie Skłodowska-Curie actions Grant Agreement no 945298-ParisRegionFP.

\appendix

\section{\texorpdfstring{$\bm{SU(2)}$}{SU(2)} Irreducible Representations as Symmetric Tensors}
\label{su2irreps}

In this appendix, we review the treatment of irreducible representations (irreps) of $SU(2)$ as totally symmetric tensors, establishing the notation used in Appendix~\ref{app_Lag}.

The standard approach to $SU(2)$ irreps involves a $(2j+1)$-dimensional vector space with a basis given by the set of states ${\left|j,m\right\rangle}$. These states are eigenvectors of the mutually diagonalizable operators $\vec{J}^2$ and $J_3$, satisfying $\vec{J}^2\left|j,m\right\rangle = j(j+1)\left|j,m\right\rangle$ and $J_3\left|j,m\right\rangle = m\left|j,m\right\rangle$, where $m$ takes values in $\llbracket -j, j \rrbracket$, making the irrep $(2j+1)$-dimensional.

Alternatively, it is sometimes more convenient to describe the $SU(2)$ irrep of spin-$j$ as a completely symmetric rank-$2j$ tensor, denoted by $\Psi_{(j,m)}$, with components
\begin{equation}
\Psi_{(j,m)}^{i_1\cdots i_{2j}},\qquad i_1,\cdots,i_{2j}=1,\,2\, .
\end{equation}
These tensors can be constructed from spin-$1/2$ states as follows: for any pair of quantum numbers $(j,m)$, a direct product of spin-$1/2$ states can be formed to represent the spin-$j$ irrep. Let us denote the orthonormal spin-$1/2$ states as $\zeta_{\pm} = \ket{1/2,\pm 1/2}$, defined abstractly through the actions of $J_3$ and $J_{\pm}$. Specifically, we have
\begin{equation}
J_3 \zeta_{\pm}=\pm \frac{1}{2}\zeta_{\pm},\quad J_{\pm}\zeta_{\pm}=0\quad\textnormal{and}\quad J_{\pm}\zeta_{\mp}=\zeta_{\pm}\, .
\end{equation}
Now, consider a collection of $2j$ spin-$1/2$ states, each described by $\zeta^{(i)}{\pm}$, where the superscript $i \in \llbracket 1, 2j \rrbracket$ labels the individual spins. The entire collection can be expressed as a tensor product of these individual states:
\begin{equation}
\Phi_{(j,q)} = \zeta_+^{(1)} \otimes \cdots \otimes \zeta_+^{(q)} \otimes \zeta_-^{(q+1)} \otimes \cdots \otimes \zeta_-^{(2j)}\, .
\label{first definition of SU(2) tensor}
\end{equation}

This tensor is invariant under permutations of the $\zeta_+$ vectors among themselves and the $\zeta_-$ vectors among themselves. However, replacing a $\zeta_+$ with a $\zeta_-$, or vice versa, generates a new tensor, as this operation changes the value of $m$. The $2j$ labels of the vectors $\zeta_{\pm}$ in $\Phi_{(j,q)}$ can be permuted in $(2j)!$ different ways. To avoid overcounting, we only consider permutations that result in distinct tensors. The number of such distinct tensors is given by
\begin{equation}
\binom{2j}{q}=\frac{(2j)!}{q!(2j-q)!}\, ,
\end{equation}
where each class of tensors $\Phi_{(j,q)}$ contains $q!(2j-q)!$ elements. Moreover, tensors belonging to different classes are orthogonal to each other.

Let us now examine how the spin-$j$ angular momentum operators $J_3$ and $J_{\pm}$ act on these tensors. Starting with $J_3$, it is defined as $J_3=J_3^{(1)}+\cdots+ J_3^{(2j)}$, where $J_3^{(k)}$ acts exclusively on the $\zeta_{\pm}^{(k)}$ vector while leaving all other $\zeta$ vectors in $\Phi_{(j,q)}$ unaffected. Applying $J_3$ to $\Phi_{(j,q)}$ yields
\begin{equation}
J_3 \Phi_{(j,q)}=\sum_{k=1}^{2j}J_3^{(k)}\Phi_{(j,q)}=(q-j)\Phi_{(j,q)}\, ,
\end{equation}
which motivates introducing $m=q-j$, the eigenvalue of $J_3$ corresponding to $\Phi_{(j,q)}$. From now on, we update the notation of the tensor to $\Phi_{(j,m)}$.

Now, the operator $J_{\pm}^{(k)}$ acts exclusively on the vector labeled by $k$, annihilating $\zeta_{\pm}^{(k)}$. Specifically, for $\Phi_{j,q}$, we have
\begin{equation}
J_{\pm}^{(k)}\Phi_{j,q}=\Phi_{(j,q\pm 1)}\quad \text{if }(k)\text{ corresponds to a }\mp\text{ index}\, ,
\end{equation}
i.e., the resulting state is either zero or belongs to a class of states with eigenvalue $m\pm 1$ of $J_3$. It follows that the collective operators $J_{\pm}=J_{\pm}^{(1)}+\cdots+J_{\pm}^{(2j)}$, acting on $\Phi_{(j,m)}$, yield a sum of $j\mp m$ different states, each with an eigenvalue of $J_3$ equal to $m\pm 1$. Consequently, we extend our description from a single tensor product $\Phi_{(j,m)}$, as defined in Eq. $\eqref{first definition of SU(2) tensor}$, to a symmetrized sum over all $(2j)!$ permutations of the vectors. We define
\begin{equation}
\varphi_{(j,m)}=\sum_{\sigma\in \mathfrak{S}_{2j}}\zeta_+^{\sigma(1)}\otimes\cdots\otimes\zeta_+^{\sigma(j+m)}\otimes \zeta_-^{\sigma(j+m+1)}\otimes\cdots \otimes\zeta_-^{\sigma(2j)}\, ,
\end{equation}
where $\mathfrak{S}_{2j}$ is the symmetric group of rank $2j$. In terms of tensor components, this is expressed as
\begin{equation}
\varphi^{i_1\cdots i_{2j}}_{(j,m)}=\sum_{\sigma\in\mathfrak{S}_{2j}}\zeta_+^{\sigma(i_1)}\cdots \zeta_+^{\sigma(i_{j+m})}\zeta_-^{\sigma(i_{j+m+1})}\cdots\zeta_-^{\sigma(i_{2j})}\, ,
\end{equation}
for a particular value of $m$. These tensor components, however, are not normalized. For a given value of $m$, the norm evaluates to
\begin{equation}
\left\langle \varphi_{(j,m)}\left| \right.\varphi_{(j,m)}\right\rangle\equiv\varphi_{(j,m)}^{ i_1\cdots i_{2j}}\varphi_{(j,m)\,\, i_1\cdots i_{2j}}^*=\binom{2j}{j+m}\, .
\end{equation}
Thus, the normalized tensors are
\begin{equation}
\Psi_{(j,m)}=\binom{2j}{j+m}^{-1/2}\varphi_{(j,m)}\, .
\end{equation}
Finally, we verify that these tensors satisfy the expected action of the angular momentum operators:
\begin{equation}
J_{\pm}\Psi_{(j,m)}=\sqrt{j(j+1)-m(m\pm 1)}\Psi_{(j,m\pm 1)},\quad J_3\Psi_{(j,m)}=m\Psi_{(j,m)}\, .
\end{equation}
This completes the construction of totally symmetric rank-$2j$ tensors as spin-$j$ representations of $SU(2)$.

\section{Fermion Mass Spectrum \& Couplings to Bosons}
\label{app_Lag}
In this appendix, we provide the computation of the fermion mass spectrum and couplings to the SM bosons.

\subsection{Flavor Basis Lagrangian}
We consider the BSM model with the following Lagrangian:
\begin{align}
    \mathcal{L}_{\mathrm{BSM}}
    &=\mathcal{L}_{\text{gauge}+\text{kin.}}-\left(M_L L L^c + M_E E E^c + y L H E^c + y^c L^c H^{\dagger} E+\text{H.c.}\right) \, \nonumber \\
    &=\mathcal{L}_{\text{gauge}+\text{kin.}}-V \, .
\end{align}
Here, $L$ and $L^c$ belong to the representation $n$ of $SU(2)_W$, while $E$ and $E^c$ belong to its representation $n-1$. The Lagrangian can be formulated in terms of the $SU(2)$ tensor components, denoted as, for instance, $L_{(t,m)}$, where $n=2t+1$ and $m$ represents the eigenvalue of $J_3$ (see Appendix~\ref{su2irreps}). For convenience, we simplify the notations by writing $L_{(t,m)}$ as $L_m$, leaving the isospin label $t$ implicit. Recall that if $L$ and $L^c$ have isospin-$t$, then $E$ and $E^c$ have isospin-$(t-1/2)$.

In terms of these components, the fermion potential of the BSM model after EWSB can be written as
\begin{equation}
V=M_L L_{-t} L^c_{t}+\sum_{m=-(t-1)}^{t}\mathcal{V}_{m}^T
\begin{pmatrix}
M_L &  \alpha(t,m) v y /\sqrt{2}\\
(-1)^{2t+1}\alpha(t,m) v y^c /\sqrt{2} & M_E\\
\end{pmatrix}
\mathcal{U}_{m}+\textnormal{H.c.} \, ,
\end{equation}
where for $m\in\llbracket -(t-1),t\rrbracket$ we have defined
\begin{equation}
\mathcal{V}_m=
\begin{pmatrix}
L_{m}\\
E_{m-1/2}
\end{pmatrix},\quad \mathcal{U}_m=
\begin{pmatrix}
L^c_{-m}\\
E^c_{-m+1/2}
\end{pmatrix},\quad \alpha(t,m)=\sqrt{1-\frac{t-m}{2t}}\, .
\end{equation}
We can split this potential into a $h$-boson part and a VEV part, which will correspond to the mass Lagrangian. The former reads
\begin{equation}
-\mathcal{L}_{h\ell\ell}=\sum_{m=-(t-1)}^{t}\frac{\alpha(t,m)}{\sqrt{2}}h\left[y L_{m}E^c_{-m+1/2}+(-1)^{2t+1}y^c L^c_{-m} E_{m-1/2} +\textnormal{H.c.}\right] \, ,
\end{equation}
while the latter reads
\begin{equation}
-\mathcal{L}_{\rm mass}=M_L L_{-t}L^c_{t}+\sum_{m=-(t-1)}^{t} \mathcal{V}_m^T \mathcal{M}_m \mathcal{U}_{m} + \textnormal{H.c.} \, ,
\end{equation}
where we defined the $m$-th mass matrix as
\begin{equation}
\mathcal{M}_m=
\begin{pmatrix}
M_L &  \alpha(t,m)v y /\sqrt{2}\\
(-1)^{2t+1} \alpha(t,m)v y^c/\sqrt{2} & M_E\\
\end{pmatrix}.
\end{equation}
For the $SU(2)_W\times U(1)_Y$ sector, the gauge-kinetic term for $L$ and $L^c$ can be written as
\begin{align}
iL^{\dagger}\bar{\sigma}^{\mu}D_{\mu}L&= \sum_{m=-t}^{t} \left[ i L_{-m}^{\dagger}\bar{\sigma}^{\mu}\partial_{\mu}L_{m}+g' Y B_{\mu} L_{-m}^{\dagger}\bar{\sigma}^{\mu}L_{m}+g \mathcal{T}^3(t,m) W_{\mu}^3 L_{-m}^{\dagger}\bar{\sigma}^{\mu}L_{m} \right]\nonumber \\
&+\frac{g}{\sqrt{2}}\sum_{m=-(t-1)}^{t}\left[\mathcal{T}^{+}(t,-m)W_{\mu}^+L_{m}^{\dagger}\bar{\sigma}^{\mu}L_{-m+1} +\mathcal{T}^{-}(t,m)W_{\mu}^-L_{-m}^{\dagger}\bar{\sigma}^{\mu}L_{m-1}\right]\, ,
\end{align}
and
\begin{align}
iL^{c}\sigma^{\mu}D_{\mu}L^{c\dagger}&= \sum_{m=-t}^{t} \left[ i L^c_{-m}\sigma^{\mu}\partial_{\mu}L_{m}^{c\dagger}+g' Y B_{\mu} L^c_{-m}\sigma^{\mu}L_{m}^{c\dagger}+g \mathcal{T}^3(t,m) W_{\mu}^3L^c_{-m}\sigma^{\mu}L_{m}^{c\dagger}\right]\nonumber \\
&+\frac{g}{\sqrt{2}}\sum_{m=-(t-1)}^{t}\left[\mathcal{T}^{+}(t,-m)W_{\mu}^+L^c_{m}\sigma^{\mu}L_{-m+1}^{c\dagger} +\mathcal{T}^{-}(t,m)W_{\mu}^-L_{-m}^{c}\sigma^{\mu}L_{m-1}^{c\dagger}\right]\, ,
\end{align}
where we introduced
\begin{equation}
\mathcal{T}^{3}(t,m)=m ,\qquad \mathcal{T}^{\pm}(t,m)=\sqrt{t(t+1)-m(m\pm 1)} \, .
\end{equation}
The expressions for the gauge-kinetic terms of $E$ and $E^c$ can be easily obtained from the ones above by adapting to the right quantum numbers.

We can thus obtain $J_3^{\mu}$ and $J_{\rm Y}^{\mu}$ in terms of these tensor components, which will be useful to build $J_{\rm EM}^{\mu}$ and $J_Z^{\mu}$ from
\begin{equation}
J_{\rm EM}^{\mu}=J_{\rm Y}^{\mu}+J_3^{\mu},\qquad J_Z^{\mu}=J_3^{\mu}-\sin^2\theta_W J_{\rm EM}^{\mu} \, ,
\end{equation} 
defined such that
\begin{equation}
\mathcal{L}_{A,Z}=e A_{\mu}J^{\mu}_{\rm EM}+g_Z Z_{\mu} J^{\mu}_Z,\qquad g_Z=e/\cos\theta_W \, .
\end{equation}
We have
\begin{align}
J_{\rm Y}^{\mu}
&=Y \sum_{m=-t}^{t}\left[L_{-m}^{\dagger}\bar{\sigma}^{\mu}L_{m}+L_{-m}^c\sigma^{\mu}L_{m}^{c\dagger} \right]+Y^\prime \sum_{m=-(t-1/2)}^{t-1/2}\left[E_{-m}^{\dagger}\bar{\sigma}^{\mu}E_{m}+E^c_{-m}\sigma^{\mu}E_{m}^{c\dagger} \right] \, ,
\end{align}
and
\begin{align}
J_3^{\mu}
&= \sum_{m=-t}^{t}\mathcal{T}^3(t,m)\left[L_{-m}^{\dagger}\bar{\sigma}^{\mu}L_{m}+L_{-m}^c\sigma^{\mu}L_{m}^{c\dagger} \right] \nonumber\\ &+\sum_{m=-(t-1/2)}^{t-1/2}\mathcal{T}^3(t-1/2,m)\left[E_{-m}^{\dagger}\bar{\sigma}^{\mu}E_{m}+E^c_{-m}\sigma^{\mu}E_{m}^{c\dagger} \right] \, .
\end{align}
Thus, we obtain
\begin{align}
J^{\mu}_{\rm EM}&=\left(Y-t\right)\left[L_{t}^{\dagger}\bar{\sigma}^{\mu}L_{-t}+L_{t}^c\sigma^{\mu}L_{-t}^{c\dagger} \right] \nonumber \\
&+\sum_{m=-(t-1)}^{t}Q(t,m)\left[L_{-m}^{\dagger}\bar{\sigma}^{\mu}L_{m}+E_{-m+1/2}^{\dagger}\bar{\sigma}^{\mu}E_{m-1/2}\right. \nonumber\\
&\left. \qquad \qquad \qquad \qquad \ \  +L^c_{-m}\sigma^{\mu}L_{m}^{c\dagger}+E^c_{-m+1/2}\sigma^{\mu}E_{m-1/2}^{c\dagger} \right] \, ,
\end{align}
where
\begin{equation}
Q(t,m)=Y+\mathcal{T}^3(t,m)=Y+m \, .
\end{equation}
Now, we also have
\begin{align}
J_Z^{\mu}&=\Xi(t,-t)\left[L^{\dagger}_{t}\bar{\sigma}^{\mu}L_{-t}+L^c_{t}\sigma^{\mu}L^{c\dagger}_{-t} \right] \nonumber \\
&+\sum_{m=-(t-1)}^{t}\left[\Xi(t,m)\left(L_{-m}^{\dagger}\bar{\sigma}^{\mu}L_m+L^c_{-m}\sigma^{\mu}L_m^{c\dagger}\right) \right. \nonumber\\
&\left. \qquad \qquad \qquad \qquad \ \ \ \ 
+\tilde{\Xi}(t,m)\left(E_{-m}^{\dagger}\bar{\sigma}^{\mu}E_m+E^{c}_{-m+1/2}\sigma^{\mu}E_{m-1/2}^{c\dagger} \right) \right] \, ,
\end{align}
where we introduced
\begin{equation}
\Xi(t,m)=\mathcal{T}^3(t,m)-Q(t,m) \sin^2 \theta_W \, , \quad \widetilde{\Xi}(t,m)=\Xi(t,m)-\frac{1}{2} \, .
\end{equation}

\subsection{Mass Basis Lagrangian}
In the previous section, we wrote the Lagrangian in the flavor basis, where the fermion mass matrices are not diagonal. To perform computations of observables, it is simpler to go to the mass basis.

\subsubsection{Mass Spectrum}
Let us focus on the $m$-th mass matrix $\mathcal{M}_m$ for $m\in\llbracket -(t-1),t\rrbracket$ defined as
\begin{equation}
\mathcal{M}_m=
\begin{pmatrix}
M_L & \alpha(t,m)v y/\sqrt{2} \\
(-1)^{2t+1} \alpha(t,m) v y^c/\sqrt{2} & M_E\\
\end{pmatrix}.
\end{equation}
We have to perform a singular value decomposition to diagonalize these matrices in order to always have positive eigenmasses~\cite{Dreiner:2008tw}. We can rewrite the corresponding mass term in the Lagrangian as
\begin{equation}
\mathcal{V}_m^T \mathcal{M}_m \mathcal{U}_m=\mathcal{V}_m^T \mathcal{O}_L^{(m)}\left[\mathcal{O}_L^{(m)}\right]^T\mathcal{M}_m \mathcal{O}_R^{(m)}\left[\mathcal{O}_R^{(m)}\right]^T \mathcal{U}_m\equiv \mathscr{L}_m^T M_D^{(m)}\mathscr{R}_m \, ,
\end{equation}
where $\mathcal{O}_L^{(m)}$ and $\mathcal{O}_R^{(m)}$ are $O(2)$ matrices that diagonalize $\mathcal{M}_m$ as
\begin{equation}
M_D^{(m)}\equiv \left[\mathcal{O}_L^{(m)}\right]^T\mathcal{M}_m \mathcal{O}_R^{(m)}=
\begin{pmatrix}
M_1^{(m)} & 0\\
0 & M_2^{(m)}\\
\end{pmatrix},\quad M_1^{(m)}\leq M_2^{(m)},\quad m\in \llbracket -(t-1) , t \rrbracket \, , 
\end{equation}
and
\begin{align}
\begin{pmatrix}
L_{m}\\
E_{m-1/2}\\
\end{pmatrix}=&~\mathcal{O}_L^{(m)}
\begin{pmatrix}
\xi_1^{(m)}\\
\xi_2^{(m)}\\
\end{pmatrix}=
\begin{pmatrix}
\mathcal{O}_{L,11}^{(m)}\xi_1^{(m)}+\mathcal{O}_{L,12}^{(m)}\xi_2^{(m)}\\
\mathcal{O}_{L,21}^{(m)}\xi_1^{(m)}+\mathcal{O}_{L,22}^{(m)}\xi_2^{(m)}
\end{pmatrix}\\
\begin{pmatrix}
L^c_{-m}\\
E^c_{-m+1/2}\\
\end{pmatrix}=&~\mathcal{O}_R^{(m)}
\begin{pmatrix}
\rho_1^{(m)}\\
\rho_2^{(m)}\\
\end{pmatrix}=
\begin{pmatrix}
\mathcal{O}_{R,11}^{(m)}\rho_1^{(m)}+\mathcal{O}_{R,12}^{(m)}\rho_2^{(m)}\\
\mathcal{O}_{R,21}^{(m)}\rho_1^{(m)}+\mathcal{O}_{R,22}^{(m)}\rho_2^{(m)}
\end{pmatrix}.
\end{align}
The components of $\mathcal{O}_{L/R}$ are constrained by $\mathcal{O}^T_{L/R}\mathcal{O}_{L/R}=\mathds{1}_2$, which translates in terms of components as
\begin{align}
1=&~\mathcal{O}_{L/R,11}^2+\mathcal{O}_{L/R,21}^2 \, ,\\
1=&~\mathcal{O}_{L/R,12}^2+\mathcal{O}_{L/R,22}^2 \, ,\\
0=&~\mathcal{O}_{L/R,11}\mathcal{O}_{L/R,12}+\mathcal{O}_{L/R,21}\mathcal{O}_{L/R,22} \, .
\end{align}
With these definitions, the mass Lagrangian becomes
\begin{equation}
-\mathcal{L}_{\rm mass}=M_L L_{-t}L^c_{t}+\sum_{m=-(t-1)}^{t}\sum_{i=1,2}\xi_i^{(m)}M_i^{(m)}\rho_i^{(m)}+\textnormal{H.c.} \, .
\end{equation}
We can introduce Dirac spinors as follows:
\begin{equation}
\Phi=
\begin{pmatrix}
L_{-t}\\
L^{c\dagger}_{-t}
\end{pmatrix},\quad \Psi_i^{(m)}=
\begin{pmatrix}
\xi_i^{(m)}\\
\rho_i^{(m)\dagger}
\end{pmatrix},\quad m\in \llbracket -(t-1),t\rrbracket,\quad i=1,2 \, ,
\end{equation}
such that
\begin{equation}
-\mathcal{L}_{\rm mass}=M_L \bar{\Phi}\Phi+\sum_{m=-(t-1)}^{t}\sum_{i=1,2}M_i^{(m)}\bar{\Psi}_i^{(m)}\Psi_i^{(m)} \, .
\end{equation}

\subsubsection{Couplings to Gluons}

We begin by presenting the coupling of the VLFs to gluons, described by the Lagrangian
\begin{equation}
\mathcal{L}_{g\ell\ell}=g_S\bar{\Phi}A^{a}_{\mu}\gamma^{\mu}T^{a}_r\Phi+\sum_{m=-(t-1)}^{t}\sum_{i=1,2}g_S\bar{\Psi}_i^{(m)}A_{\mu}^{a}\gamma^{\mu}T^{a}_r\Psi_i^{(m)} \, ,
\end{equation}
where $T^{a}_r$ are the generators of $SU(3)_C$ in the representation $r$ of the VLFs. 

To compute triangle diagrams for the process $h\rightarrow gg$, it is necessary to have an expression for $\text{Tr}[T^{a}_r,T^{b}_r]=T(r)\delta^{ab}$, where $T(r)$ is the Dynkin index of the representation $r$. The expression of $T(r)$ can be determined in terms of the quadratic Casimir $C_2(r)$, the dimension of the Lie algebra, and the dimension of the representation $r$, as follows:
\begin{equation}
    T(r)=\frac{\text{dim}(r)C_2(r)}{\text{dim}~\mathfrak{su}(3)}\, ,
    \label{formula Dynkin index}
\end{equation}
where, for $SU(3)$, $\text{dim}~\mathfrak{su}(3)=8$. Representations of $SU(3)$ are classified by two integers, $r=(p,q)$, and the dimension of the representation and the quadratic Casimir are given by
\begin{equation}
    \text{dim}(r)=\frac{(p+1)(q+1)(p+q+2)}{2},\qquad C_2(r)=\frac{p^2+q^2+3p+3q+pq}{3}\, .
\end{equation}
From these expressions, we can determine the value of the Dynkin index $T(r)$ of the representation $r=(p,q)$.

\subsubsection{Couplings to the Photon}
Now, it is straightforward to write the electromagnetic Lagrangian for our VLFs. It is given by
\begin{equation}
\mathcal{L}_{A\ell\ell}=eQ(t,-t)\bar{\Phi}A_{\mu}\gamma^{\mu}\Phi+\sum_{m=-(t-1)}^{t}\sum_{i=1,2}eQ(t,m)\bar{\Psi}_i^{(m)}A_{\mu}\gamma^{\mu}\Psi_i^{(m)} \, ,
\end{equation}
with
\begin{equation}
Q(t,m)=Y+\mathcal{T}^3(t,m)=Y+m \, .
\end{equation}

\subsubsection{\boldmath Couplings to the \texorpdfstring{$\bm{Z}$}{Z}-Boson}

We introduce the combinations $\Xi$ and $\Tilde{\Xi}$, defined by
\begin{equation}
    \Xi(t,m)=\mathcal{T}^3(t,m)-\sin^2\theta_W \, Q(t,m),\quad\text{and}\quad \Tilde{\Xi}(t,m)=\Xi(t,m)-\frac{1}{2} \, .
\end{equation}
The interaction Lagrangian between VLFs and the $Z$-boson is given by
\begin{equation}
\mathcal{L}_{Z\ell\ell}=g_{\phi,Z}\Bar{\Phi}Z_{\mu}\gamma^{\mu}\Phi+\sum_{a=L,R}\sum_{i,j\in\{1,2\}}\sum_{m=-(t-1)}^{t}g_{ija,Z}^{(m)}\Bar{\Psi}^{(m)}_iZ_{\mu}\gamma^{\mu}P_a\Psi_j^{(m)} \, .
\end{equation}
The expressions of the couplings are given in the following table:
\begin{center}
\begin{tabular}{ |p{1.5cm}||p{8cm}|  }
 \hline
 \multicolumn{2}{|c|}{Couplings between the $Z$-boson and the fermions} \\
 \hline
 Coupling & \qquad\qquad Expression\\
 \hline
 $g_{11a,Z}^{(m)}$ & $g_Z\left[\Xi(t,m)\left[\mathcal{O}_{a11}^{(m)}\right]^2+\Tilde{\Xi}(t,m)\left[\mathcal{O}_{a21}^{(m)}\right]^2\right]$  \\
 \hline
 $g_{22a,Z}^{(m)}$ & $g_Z\left[\Xi(t,m)\left[\mathcal{O}_{a12}^{(m)}\right]^2+\Tilde{\Xi}(t,m)\left[\mathcal{O}_{a22}^{(m)}\right]^2\right]$  \\
 \hline
 $g_{12a,Z}^{(m)}$ & $ g_Z \mathcal{O}_{a11}^{(m)}\mathcal{O}_{a12}^{(m)}\left[\Xi(t,m)-\Tilde{\Xi}(t,m) \right]$  \\
 \hline
 $g_{21a,Z}^{(m)}$ & $ g_Z \mathcal{O}_{a11}^{(m)}\mathcal{O}_{a12}^{(m)}\left[\Xi(t,m)-\Tilde{\Xi}(t,m) \right]$  \\
 \hline
 $g_{\phi,Z}$ & $-g_Z\left(t\cos^2\theta_W+Y\sin^2\theta_W\right)$   \\
 \hline
\end{tabular}
\end{center}
with $g_Z=e/\cos\theta_W$.

\subsubsection{\boldmath Couplings to the \texorpdfstring{$\bm{h}$}{h}-Boson}
The interaction Lagrangian between VLFs and the $h$-boson is
\begin{equation}
    \mathcal{L}_{h\ell\ell}=\sum_{m=-(t-1)}^{t}\sum_{i=1,2}g_{i,h}^{(m)}h\Bar{\Psi}_i^{(m)}\Psi_i^{(m)}+\sum_{i\ne j\in\{1,2\}}\sum_{m=-(t-1)}^{t}\left[g_{ij,h}^{(m)}h\Bar{\Psi}^{(m)}_iP_L\Psi_j^{(m)}+\text{H.c.}\right] \, ,
\end{equation}
where the expressions of the couplings are given in the following table:
\begin{center}
\begin{tabular}{ |p{2cm}||p{7.5cm}|  }
 \hline
 \multicolumn{2}{|c|}{Couplings between the Higgs boson and the leptons} \\
 \hline
 Coupling & \qquad\qquad\qquad Expression\\
 \hline
 $g_{1,h}^{(m)}$ & $-\alpha_m\left[y \mathcal{O}_{L,11}^{(m)}\mathcal{O}_{R,21}^{(m)}+y^c(-1)^{2t+1}\mathcal{O}_{L,21}^{(m)}\mathcal{O}_{R,11}^{(m)} \right]$ \\
 \hline
 $g_{2,h}^{(m)}$ & $-\alpha_m\left[y \mathcal{O}_{L,12}^{(m)}\mathcal{O}_{R,22}^{(m)}+y^c(-1)^{2t+1}\mathcal{O}_{L,22}^{(m)}\mathcal{O}_{R,12}^{(m)} \right]$   \\
 \hline
 $g_{12,h}^{(m)}$ & $-\alpha_m\left[y \mathcal{O}_{L,12}^{(m)}\mathcal{O}_{R,21}^{(m)}+y^c(-1)^{2t+1}\mathcal{O}_{L,22}^{(m)}\mathcal{O}_{R,11}^{(m)}\right]$  \\
 \hline
 $g_{21,h}^{(m)}$ & $-\alpha_m\left[y \mathcal{O}_{L,11}^{(m)}\mathcal{O}_{R,22}^{(m)}+y^c(-1)^{2t+1}\mathcal{O}_{L,21}^{(m)}\mathcal{O}_{R,12}^{(m)} \right]$  \\
 \hline
\end{tabular}
\end{center}
where we defined $\alpha_m\equiv \alpha(t,m)/\sqrt{2}$ for simplicity.

\section{No Conservative Scaling Argument}
\label{scaling_arg}
In our companion article~\cite{DAgnolo:2023rnh}, we used scaling arguments in certain cases to derive bounds on the dimensions of the representations that need to be considered. However, for the three loop-induced couplings studied in this article, we find that such straightforward arguments are insufficient to establish strong bounds. The discussion in this appendix leads us to systematically consider several representations in our analysis.

We begin by examining the $h\gamma\gamma$ and $hZ\gamma$ couplings, focusing on the dimension $r$ of the color representation of the VLFs (while treating other representations as subdominant). Using Eqs.~\eqref{C_hgamgam} and~\eqref{C_hZgam}, we observe that both couplings exhibit the same scaling behavior,
\begin{equation}
C_{h\gamma\gamma},C_{hZ\gamma} \sim r \, \dfrac{y^2}{M_L^2} \, ,
\end{equation}
where $y = (-1)^n y^c$ to maximize the coupling deviations. Furthermore, the RGEs yield the same $\Lambda_\tn{B}$ for large $r$, regardless of the value of $r$, as long as the rescaled ’t Hooft-like Yukawa couplings satisfy $y_c = r^{\gamma_c} y$. Here, $\gamma_c = 1/2$ or $1/4$ depending on whether $\Lambda_\tn{B}$ is determined by a Landau pole or a vacuum instability, respectively. Thus,
\begin{equation}
C_{h\gamma\gamma},C_{hZ
\gamma} \sim r^{1-2\gamma_c} \, \dfrac{y_c^2}{M_L^2} \, ,
\end{equation}
where $1 - 2\gamma_c = 0$ for a Landau pole and $1/2$ for a vacuum instability. Since we have found that $\Lambda_\tn{B}$ is always determined by a vacuum instability within the range of parameters of interest, it follows that larger values of $r$ lead to greater coupling deviations. However, $M_\mathrm{exp}$ is larger than for colorless VLFs. Determining the parametric dependence of $M_\mathrm{exp}$ on $r$ is challenging because the bounds from direct searches depend on the specific details of the model. Nevertheless, we know that $M_\mathrm{exp}$ will increase with $r$, as larger $r$ results in a greater production cross-section at the LHC. To remain conservative, we include colored representations in our analysis. Regarding the $hgg$ coupling for large $r$, we have from Eq.~\eqref{C_hgg}
\begin{equation}
C_{hgg} = T(r) \, \frac{y^2}{M_L} \, ,
\end{equation}
where $T(r) \sim r^{5/3}$ or $r^2$, depending on the representation, as we can see from Eq.~$\eqref{formula Dynkin index}$. Using the previously rescaled $y_c$, it is not possible to draw a definitive conclusion without precise knowledge of the $r$-dependence of $M_\mathrm{exp}$, which is typically model-dependent.

The same argument applies to a large number of flavors, $N_\mathrm{F}$, since
\begin{equation}
C_{hgg} \sim N_\mathrm{F} \, \dfrac{y^2}{M_L^2} \, , \quad
C_{h\gamma\gamma} \sim N_\mathrm{F} \, \dfrac{y^2}{M_L^2} \, \quad
C_{hZ\gamma} \sim N_\mathrm{F} \, \dfrac{y^2}{M_L^2} \, ,
\end{equation}
as derived from Eqs.~\eqref{C_hgg},~\eqref{C_hgamgam},~and~\eqref{C_hZgam}. The rescaled Yukawa couplings are given by $y_N = N_\mathrm{F}^{\gamma_N} y$, where $\gamma_N = 1/2$ for a Landau pole and $1/4$ for a vacuum instability.

For the dimension of the weak isospin representations $n$, Eqs.~\eqref{C_hgg},~\eqref{C_hgamgam} and~\eqref{C_hZgam} give (for large $n)$
\begin{equation}
C_{hgg} \sim n \, \dfrac{y^2}{M_L^2} \, , \quad
C_{h\gamma\gamma} \sim n^2 \, \dfrac{y^2}{M_L^2} \, \quad
C_{hZ\gamma} \sim n^3 \, \dfrac{y^2}{M_L^2} \, .
\end{equation}
The Yukawa couplings are rescaled as $y_n = n^{\gamma_n} y$, where $\gamma_n = 1/2$ for a Landau pole and $1/4$ for a vacuum instability. The same considerations as for $r$ apply here, and we have verified that, in our analysis, $\Lambda_\tn{B}$ is always determined by a vacuum instability.

As for the hypercharge $Y$, $\delta \mu_{hgg}$ at 1-loop does not explicitly depend on it in Eq.~\eqref{C_hgg}. Therefore, it is clear that $Y$ only affects the RGEs by reducing $\Lambda_\tn{B}$ for a given $\delta \mu_{hgg}$. A larger $|Y|$ is also associated with a larger $M_\mathrm{exp}$, implying that it does not allow for a larger $\Lambda_\tn{B}$. This is one of the few cases where a clear argument can be made. For $h\gamma\gamma$ and $hZ\gamma$, there is no obvious scaling argument from the couplings in the RGEs, except in the case where $\Lambda_\tn{B}$ arises from a Landau pole in the hypercharge coupling, as previously noted in our companion article~\cite{DAgnolo:2023rnh}. However, this is not sufficient to systematically exclude higher hypercharges from our analysis.

\bibliographystyle{JHEP}
\bibliography{biblio}

\providecommand{\href}[2]{#2}\begingroup\raggedright\begin{thebibliography}{100}

\bibitem{ATLAS:2012yve}
{\scshape ATLAS} collaboration, \emph{{Observation of a new particle in the
  search for the Standard Model Higgs boson with the ATLAS detector at the
  LHC}}, \href{https://doi.org/10.1016/j.physletb.2012.08.020}{\emph{Phys.
  Lett. B} {\bfseries 716} (2012) 1}
  [\href{https://arxiv.org/abs/1207.7214}{{\ttfamily 1207.7214}}].

\bibitem{CMS:2012qbp}
{\scshape CMS} collaboration, \emph{{Observation of a new boson at a mass of
  125 GeV with the CMS experiment at the LHC}},
  \href{https://doi.org/10.1016/j.physletb.2012.08.021}{\emph{Phys. Lett. B}
  {\bfseries 716} (2012) 30} [\href{https://arxiv.org/abs/1207.7235}{{\ttfamily
  1207.7235}}].

\bibitem{Glashow:1961tr}
S.L.~Glashow, \emph{{Partial-symmetries of weak interactions}},
  \href{https://doi.org/10.1016/0029-5582(61)90469-2}{\emph{Nucl. Phys.}
  {\bfseries 22} (1961) 579}.

\bibitem{Weinberg:1967tq}
S.~Weinberg, \emph{{A Model of Leptons}},
  \href{https://doi.org/10.1103/PhysRevLett.19.1264}{\emph{Phys. Rev. Lett.}
  {\bfseries 19} (1967) 1264}.

\bibitem{Salam:1968rm}
A.~Salam, \emph{{Weak and Electromagnetic Interactions}},
  \href{https://doi.org/10.1142/9789812795915_0034}{\emph{Conf. Proc. C}
  {\bfseries 680519} (1968) 367}.

\bibitem{Quevedo:2024kmy}
F.~Quevedo and A.~Schachner, \emph{{Cambridge Lectures on The Standard Model}},
   \href{https://arxiv.org/abs/2409.09211}{{\ttfamily 2409.09211}}.

\bibitem{Nambu:1960tm}
Y.~Nambu, \emph{{Quasi-Particles and Gauge Invariance in the Theory of
  Superconductivity}},
  \href{https://doi.org/10.1103/PhysRev.117.648}{\emph{Phys. Rev.} {\bfseries
  117} (1960) 648}.

\bibitem{Schwinger:1962tn}
J.S.~Schwinger, \emph{{Gauge Invariance and Mass}},
  \href{https://doi.org/10.1103/PhysRev.125.397}{\emph{Phys. Rev.} {\bfseries
  125} (1962) 397}.

\bibitem{Anderson:1963pc}
P.W.~Anderson, \emph{{Plasmons, Gauge Invariance, and Mass}},
  \href{https://doi.org/10.1103/PhysRev.130.439}{\emph{Phys. Rev.} {\bfseries
  130} (1963) 439}.

\bibitem{Higgs:1964ia}
P.W.~Higgs, \emph{{Broken symmetries, massless particles and gauge fields}},
  \href{https://doi.org/10.1016/0031-9163(64)91136-9}{\emph{Phys. Lett.}
  {\bfseries 12} (1964) 132}.

\bibitem{Englert:1964et}
F.~Englert and R.~Brout, \emph{{Broken Symmetry and the Mass of Gauge Vector
  Mesons}}, \href{https://doi.org/10.1103/PhysRevLett.13.321}{\emph{Phys. Rev.
  Lett.} {\bfseries 13} (1964) 321}.

\bibitem{Higgs:1964pj}
P.W.~Higgs, \emph{{Broken Symmetries and the Masses of Gauge Bosons}},
  \href{https://doi.org/10.1103/PhysRevLett.13.508}{\emph{Phys. Rev. Lett.}
  {\bfseries 13} (1964) 508}.

\bibitem{Guralnik:1964eu}
G.S.~Guralnik, C.R.~Hagen and T.W.B.~Kibble, \emph{{Global Conservation Laws
  and Massless Particles}},
  \href{https://doi.org/10.1103/PhysRevLett.13.585}{\emph{Phys. Rev. Lett.}
  {\bfseries 13} (1964) 585}.

\bibitem{Higgs:1966ev}
P.W.~Higgs, \emph{{Spontaneous Symmetry Breakdown without Massless Bosons}},
  \href{https://doi.org/10.1103/PhysRev.145.1156}{\emph{Phys. Rev.} {\bfseries
  145} (1966) 1156}.

\bibitem{Migdal:1966tq}
A.A.~Migdal and A.M.~Polyakov, \emph{{Spontaneous breakdown of strong
  interaction symmetry and absence of massless particles}},
  {\emph{Sov.Phys.JETP} {\bfseries 24} (1967) 91}.

\bibitem{Kibble:1967sv}
T.W.B.~Kibble, \emph{{Symmetry Breaking in Non-Abelian Gauge Theories}},
  \href{https://doi.org/10.1103/PhysRev.155.1554}{\emph{Phys. Rev.} {\bfseries
  155} (1967) 1554}.

\bibitem{Guralnik:2011zz}
G.S.~Guralnik, \emph{{Gauge Invariance and the Goldstone Theorem}},
  \href{https://doi.org/10.1142/S0217732311036188}{\emph{Mod. Phys. Lett. A}
  {\bfseries 26} (2011) 1381}
  [\href{https://arxiv.org/abs/1107.4592}{{\ttfamily 1107.4592}}].

\bibitem{CMS:2022dwd}
{\scshape CMS} collaboration, \emph{{A portrait of the Higgs boson by the CMS
  experiment ten years after the discovery}},
  \href{https://doi.org/10.1038/s41586-022-04892-x}{\emph{Nature} {\bfseries
  607} (2022) 60} [\href{https://arxiv.org/abs/2207.00043}{{\ttfamily
  2207.00043}}].

\bibitem{ATLAS:2022vkf}
{\scshape ATLAS} collaboration, \emph{{A detailed map of Higgs boson
  interactions by the ATLAS experiment ten years after the discovery}},
  \href{https://doi.org/10.1038/s41586-022-04893-w}{\emph{Nature} {\bfseries
  607} (2022) 52} [\href{https://arxiv.org/abs/2207.00092}{{\ttfamily
  2207.00092}}].

\bibitem{ParticleDataGroup:2024cfk}
{\scshape Particle Data Group} collaboration, \emph{{Review of particle
  physics}}, \href{https://doi.org/10.1103/PhysRevD.110.030001}{\emph{Phys.
  Rev. D} {\bfseries 110} (2024) 030001}.

\bibitem{Gunion:1989we}
J.F.~Gunion, H.E.~Haber, G.L.~Kane and S.~Dawson, \emph{{The Higgs Hunter's
  Guide}}, \href{https://doi.org/10.1201/9780429496448}{\emph{Front.Phys.}
  {\bfseries 80} (2000) 1}.

\bibitem{Djouadi:2005gi}
A.~Djouadi, \emph{{The anatomy of electroweak symmetry breaking: Tome I: The
  Higgs boson in the Standard Model}},
  \href{https://doi.org/10.1016/j.physrep.2007.10.004}{\emph{Phys. Rept.}
  {\bfseries 457} (2008) 1}
  [\href{https://arxiv.org/abs/hep-ph/0503172}{{\ttfamily hep-ph/0503172}}].

\bibitem{Dawson:2018dcd}
S.~Dawson, C.~Englert and T.~Plehn, \emph{{Higgs physics: It ain’t over till
  it is over}},
  \href{https://doi.org/10.1016/j.physrep.2019.05.001}{\emph{Phys. Rept.}
  {\bfseries 816} (2019) 1} [\href{https://arxiv.org/abs/1808.01324}{{\ttfamily
  1808.01324}}].

\bibitem{Dawson:2022zbb}
S.~Dawson et~al., \emph{{Report of the Topical Group on Higgs Physics for
  Snowmass 2021: The Case for Precision Higgs Physics}},  in \emph{{2022
  Snowmass Summer Study}}, 9, 2022
  [\href{https://arxiv.org/abs/2209.07510}{{\ttfamily 2209.07510}}].

\bibitem{Cho:2007cb}
A.~Cho, \emph{{Physicists' Nightmare Scenario: The Higgs and Nothing Else}},
  \href{https://doi.org/10.1126/science.315.5819.1657}{\emph{Science}
  {\bfseries 315} (2007) 1657}.

\bibitem{Wilson:1970ag}
K.G.~Wilson, \emph{{Renormalization Group and Strong Interactions}},
  \href{https://doi.org/10.1103/PhysRevD.3.1818}{\emph{Phys. Rev. D} {\bfseries
  3} (1971) 1818}.

\bibitem{Weinberg:1975gm}
S.~Weinberg, \emph{{Implications of dynamical symmetry breaking}},
  \href{https://doi.org/10.1103/PhysRevD.13.974}{\emph{Phys. Rev. D} {\bfseries
  13} (1976) 974}.

\bibitem{Gildener:1976ai}
E.~Gildener, \emph{{Gauge-symmetry hierarchies}},
  \href{https://doi.org/10.1103/PhysRevD.14.1667}{\emph{Phys. Rev. D}
  {\bfseries 14} (1976) 1667}.

\bibitem{Susskind:1978ms}
L.~Susskind, \emph{{Dynamics of spontaneous symmetry breaking in the
  Weinberg-Salam theory}},
  \href{https://doi.org/10.1103/PhysRevD.20.2619}{\emph{Phys. Rev. D}
  {\bfseries 20} (1979) 2619}.

\bibitem{tHooft:1979rat}
G.~'t~Hooft, \emph{{Naturalness, chiral symmetry, and spontaneous chiral
  symmetry breaking}},
  \href{https://doi.org/10.1007/978-1-4684-7571-5_9}{\emph{NATO Sci. Ser. B}
  {\bfseries 59} (1980) 135}.

\bibitem{Veltman:1980mj}
M.J.G.~Veltman, \emph{{The Infrared-Ultraviolet Connection}}, {\emph{Acta Phys.
  Polon. B} {\bfseries 12} (1981) 437}.

\bibitem{Kolda:2000wi}
C.F.~Kolda and H.~Murayama, \emph{{The Higgs mass and new physics scales in the
  minimal standard model}},
  \href{https://doi.org/10.1088/1126-6708/2000/07/035}{\emph{JHEP} {\bfseries
  07} (2000) 035} [\href{https://arxiv.org/abs/hep-ph/0003170}{{\ttfamily
  hep-ph/0003170}}].

\bibitem{Giudice:2008bi}
G.F.~Giudice, \emph{{Naturally Speaking: The Naturalness Criterion and Physics
  at the LHC}},  \href{https://arxiv.org/abs/0801.2562}{{\ttfamily 0801.2562}}.

\bibitem{Giudice:2013yca}
G.F.~Giudice, \emph{{Naturalness after LHC8}},
  \href{https://doi.org/10.22323/1.180.0163}{\emph{PoS} {\bfseries EPS-HEP2013}
  (2013) 163} [\href{https://arxiv.org/abs/1307.7879}{{\ttfamily 1307.7879}}].

\bibitem{Giudice:2017pzm}
G.F.~Giudice, \emph{{The Dawn of the Post-Naturalness Era}},
  \href{https://arxiv.org/abs/1710.07663}{{\ttfamily 1710.07663}}.

\bibitem{Craig:2022uua}
N.~Craig, \emph{{Naturalness: A Snowmass White Paper}},  in \emph{{Snowmass
  2021}}, 5, 2022 [\href{https://arxiv.org/abs/2205.05708}{{\ttfamily
  2205.05708}}].

\bibitem{Falkowski:2023hsg}
A.~Falkowski, \emph{{Lectures on SMEFT}},
  \href{https://doi.org/10.1140/epjc/s10052-023-11821-3}{\emph{Eur. Phys. J. C}
  {\bfseries 83} (2023) 656}.

\bibitem{Bizot:2015zaa}
N.~Bizot and M.~Frigerio, \emph{{Fermionic extensions of the Standard Model in
  light of the Higgs couplings}},
  \href{https://doi.org/10.1007/JHEP01(2016)036}{\emph{JHEP} {\bfseries 01}
  (2016) 036} [\href{https://arxiv.org/abs/1508.01645}{{\ttfamily
  1508.01645}}].

\bibitem{Branchina:2013jra}
V.~Branchina and E.~Messina, \emph{{Stability, Higgs Boson Mass, and New
  Physics}}, \href{https://doi.org/10.1103/PhysRevLett.111.241801}{\emph{Phys.
  Rev. Lett.} {\bfseries 111} (2013) 241801}
  [\href{https://arxiv.org/abs/1307.5193}{{\ttfamily 1307.5193}}].

\bibitem{Branchina:2014usa}
V.~Branchina, E.~Messina and A.~Platania, \emph{{Top mass determination, Higgs
  inflation, and vacuum stability}},
  \href{https://doi.org/10.1007/JHEP09(2014)182}{\emph{JHEP} {\bfseries 09}
  (2014) 182} [\href{https://arxiv.org/abs/1407.4112}{{\ttfamily 1407.4112}}].

\bibitem{Branchina:2015nda}
V.~Branchina and E.~Messina, \emph{{Stability and UV completion of the Standard
  Model}}, \href{https://doi.org/10.1209/0295-5075/117/61002}{\emph{EPL}
  {\bfseries 117} (2017) 61002}
  [\href{https://arxiv.org/abs/1507.08812}{{\ttfamily 1507.08812}}].

\bibitem{Branchina:2016bws}
V.~Branchina, E.~Messina and D.~Zappala, \emph{{Impact of gravity on vacuum
  stability}}, \href{https://doi.org/10.1209/0295-5075/116/21001}{\emph{EPL}
  {\bfseries 116} (2016) 21001}
  [\href{https://arxiv.org/abs/1601.06963}{{\ttfamily 1601.06963}}].

\bibitem{Bentivegna:2017qry}
E.~Bentivegna, V.~Branchina, F.~Contino and D.~Zappal\`a, \emph{{Impact of New
  Physics on the EW vacuum stability in a curved spacetime background}},
  \href{https://doi.org/10.1007/JHEP12(2017)100}{\emph{JHEP} {\bfseries 12}
  (2017) 100} [\href{https://arxiv.org/abs/1708.01138}{{\ttfamily
  1708.01138}}].

\bibitem{Branchina:2018xdh}
V.~Branchina, F.~Contino and A.~Pilaftsis, \emph{{Protecting the stability of
  the electroweak vacuum from Planck-scale gravitational effects}},
  \href{https://doi.org/10.1103/PhysRevD.98.075001}{\emph{Phys. Rev. D}
  {\bfseries 98} (2018) 075001}
  [\href{https://arxiv.org/abs/1806.11059}{{\ttfamily 1806.11059}}].

\bibitem{Branchina:2018qlf}
V.~Branchina, F.~Contino and P.M.~Ferreira, \emph{{Electroweak vacuum lifetime
  in two Higgs doublet models}},
  \href{https://doi.org/10.1007/JHEP11(2018)107}{\emph{JHEP} {\bfseries 11}
  (2018) 107} [\href{https://arxiv.org/abs/1807.10802}{{\ttfamily
  1807.10802}}].

\bibitem{Branchina:2019tyy}
V.~Branchina, E.~Bentivegna, F.~Contino and D.~Zappal\`a, \emph{{Direct
  Higgs-gravity interaction and stability of our Universe}},
  \href{https://doi.org/10.1103/PhysRevD.99.096029}{\emph{Phys. Rev. D}
  {\bfseries 99} (2019) 096029}
  [\href{https://arxiv.org/abs/1905.02975}{{\ttfamily 1905.02975}}].

\bibitem{Gogoladze:2008ak}
I.~Gogoladze, N.~Okada and Q.~Shafi, \emph{{Higgs boson mass bounds in the
  Standard Model with type III and type I seesaw}},
  \href{https://doi.org/10.1016/j.physletb.2008.08.023}{\emph{Phys. Lett. B}
  {\bfseries 668} (2008) 121}
  [\href{https://arxiv.org/abs/0805.2129}{{\ttfamily 0805.2129}}].

\bibitem{Chen:2012faa}
C.-S.~Chen and Y.~Tang, \emph{{Vacuum stability, neutrinos, and dark matter}},
  \href{https://doi.org/10.1007/JHEP04(2012)019}{\emph{JHEP} {\bfseries 04}
  (2012) 019} [\href{https://arxiv.org/abs/1202.5717}{{\ttfamily 1202.5717}}].

\bibitem{Joglekar:2012vc}
A.~Joglekar, P.~Schwaller and C.E.M.~Wagner, \emph{{Dark Matter and enhanced $h
  \rightarrow \gamma \gamma$ rate from vector-like Leptons}},
  \href{https://doi.org/10.1007/JHEP12(2012)064}{\emph{JHEP} {\bfseries 12}
  (2012) 064} [\href{https://arxiv.org/abs/1207.4235}{{\ttfamily 1207.4235}}].

\bibitem{Kearney:2012zi}
J.~Kearney, A.~Pierce and N.~Weiner, \emph{{Vectorlike fermions and Higgs
  couplings}}, \href{https://doi.org/10.1103/PhysRevD.86.113005}{\emph{Phys.
  Rev. D} {\bfseries 86} (2012) 113005}
  [\href{https://arxiv.org/abs/1207.7062}{{\ttfamily 1207.7062}}].

\bibitem{Reece:2012gi}
M.~Reece, \emph{{Vacuum instabilities with a wrong-sign Higgs–gluon–gluon
  amplitude}}, \href{https://doi.org/10.1088/1367-2630/15/4/043003}{\emph{New
  J. Phys.} {\bfseries 15} (2013) 043003}
  [\href{https://arxiv.org/abs/1208.1765}{{\ttfamily 1208.1765}}].

\bibitem{Batell:2012ca}
B.~Batell, S.~Gori and L.-T.~Wang, \emph{{Higgs couplings and precision
  electroweak data}},
  \href{https://doi.org/10.1007/JHEP01(2013)139}{\emph{JHEP} {\bfseries 01}
  (2013) 139} [\href{https://arxiv.org/abs/1209.6382}{{\ttfamily 1209.6382}}].

\bibitem{Fairbairn:2013xaa}
M.~Fairbairn and P.~Grothaus, \emph{{Baryogenesis and dark matter with
  vector-like fermions}},
  \href{https://doi.org/10.1007/JHEP10(2013)176}{\emph{JHEP} {\bfseries 10}
  (2013) 176} [\href{https://arxiv.org/abs/1307.8011}{{\ttfamily 1307.8011}}].

\bibitem{Altmannshofer:2013zba}
W.~Altmannshofer, M.~Bauer and M.~Carena, \emph{{Exotic leptons: Higgs, flavor
  and collider phenomenology}},
  \href{https://doi.org/10.1007/JHEP01(2014)060}{\emph{JHEP} {\bfseries 01}
  (2014) 060} [\href{https://arxiv.org/abs/1308.1987}{{\ttfamily 1308.1987}}].

\bibitem{Xiao:2014kba}
M.-L.~Xiao and J.-H.~Yu, \emph{{Stabilizing electroweak vacuum in a vectorlike
  fermion model}},
  \href{https://doi.org/10.1103/PhysRevD.90.014007}{\emph{Phys. Rev. D}
  {\bfseries 90} (2014) 014007}
  [\href{https://arxiv.org/abs/1404.0681}{{\ttfamily 1404.0681}}].

\bibitem{Ellis:2014dza}
S.A.R.~Ellis, R.M.~Godbole, S.~Gopalakrishna and J.D.~Wells, \emph{{Survey of
  vector-like fermion extensions of the Standard Model and their
  phenomenological implications}},
  \href{https://doi.org/10.1007/JHEP09(2014)130}{\emph{JHEP} {\bfseries 09}
  (2014) 130} [\href{https://arxiv.org/abs/1404.4398}{{\ttfamily 1404.4398}}].

\bibitem{Angelescu:2016mhl}
A.~Angelescu and G.~Arcadi, \emph{{Dark matter phenomenology of SM and enlarged
  Higgs sectors extended with vector-like leptons}},
  \href{https://doi.org/10.1140/epjc/s10052-017-5015-2}{\emph{Eur. Phys. J. C}
  {\bfseries 77} (2017) 456}
  [\href{https://arxiv.org/abs/1611.06186}{{\ttfamily 1611.06186}}].

\bibitem{Goswami:2018jar}
S.~Goswami, K.N.~Vishnudath and N.~Khan, \emph{{Constraining the minimal
  type-III seesaw model with naturalness, lepton flavor violation, and
  electroweak vacuum stability}},
  \href{https://doi.org/10.1103/PhysRevD.99.075012}{\emph{Phys. Rev. D}
  {\bfseries 99} (2019) 075012}
  [\href{https://arxiv.org/abs/1810.11687}{{\ttfamily 1810.11687}}].

\bibitem{Gopalakrishna:2018uxn}
S.~Gopalakrishna and A.~Velusamy, \emph{{Higgs vacuum stability with vectorlike
  fermions}}, \href{https://doi.org/10.1103/PhysRevD.99.115020}{\emph{Phys.
  Rev. D} {\bfseries 99} (2019) 115020}
  [\href{https://arxiv.org/abs/1812.11303}{{\ttfamily 1812.11303}}].

\bibitem{Borah:2020nsz}
D.~Borah, R.~Roshan and A.~Sil, \emph{{Sub-TeV singlet scalar dark matter and
  electroweak vacuum stability with vectorlike fermions}},
  \href{https://doi.org/10.1103/PhysRevD.102.075034}{\emph{Phys. Rev. D}
  {\bfseries 102} (2020) 075034}
  [\href{https://arxiv.org/abs/2007.14904}{{\ttfamily 2007.14904}}].

\bibitem{Bandyopadhyay:2020djh}
P.~Bandyopadhyay, S.~Jangid and M.~Mitra, \emph{{Scrutinizing vacuum stability
  in IDM with Type-III inverse seesaw}},
  \href{https://doi.org/10.1007/JHEP02(2021)075}{\emph{JHEP} {\bfseries 02}
  (2021) 075} [\href{https://arxiv.org/abs/2008.11956}{{\ttfamily
  2008.11956}}].

\bibitem{Hiller:2022rla}
G.~Hiller, T.~H\"ohne, D.F.~Litim and T.~Steudtner, \emph{{Portals into Higgs
  vacuum stability}},
  \href{https://doi.org/10.1103/PhysRevD.106.115004}{\emph{Phys. Rev. D}
  {\bfseries 106} (2022) 115004}
  [\href{https://arxiv.org/abs/2207.07737}{{\ttfamily 2207.07737}}].

\bibitem{Arsenault:2022xty}
A.~Arsenault, K.Y.~Cingiloglu and M.~Frank, \emph{{Vacuum stability in the
  Standard Model with vectorlike fermions}},
  \href{https://doi.org/10.1103/PhysRevD.107.036018}{\emph{Phys. Rev. D}
  {\bfseries 107} (2023) 036018}
  [\href{https://arxiv.org/abs/2207.10332}{{\ttfamily 2207.10332}}].

\bibitem{Cingiloglu:2023ylm}
K.Y.~Cingiloglu and M.~Frank, \emph{{Vacuum stability and electroweak precision
  in the two-Higgs-doublet model with vectorlike quarks}},
  \href{https://doi.org/10.1103/PhysRevD.109.036016}{\emph{Phys. Rev. D}
  {\bfseries 109} (2024) 036016}
  [\href{https://arxiv.org/abs/2309.03700}{{\ttfamily 2309.03700}}].

\bibitem{Adhikary:2024esf}
A.~Adhikary, M.~Olechowski, J.~Rosiek and M.~Ryczkowski, \emph{{Theoretical
  constraints on models with vectorlike fermions}},
  \href{https://doi.org/10.1103/PhysRevD.110.075029}{\emph{Phys. Rev. D}
  {\bfseries 110} (2024) 075029}
  [\href{https://arxiv.org/abs/2406.16050}{{\ttfamily 2406.16050}}].

\bibitem{Cingiloglu:2024vdh}
K.Y.~Cingiloglu and M.~Frank, \emph{{Stability of the standard model vacuum
  with vectorlike leptons: A critical examination}},
  \href{https://doi.org/10.1103/PhysRevD.111.016025}{\emph{Phys. Rev. D}
  {\bfseries 111} (2025) 016025}
  [\href{https://arxiv.org/abs/2408.10898}{{\ttfamily 2408.10898}}].

\bibitem{Arkani-Hamed:2012dcq}
N.~Arkani-Hamed, K.~Blum, R.T.~D'Agnolo and J.~Fan, \emph{{2:1 for naturalness
  at the LHC?}}, \href{https://doi.org/10.1007/JHEP01(2013)149}{\emph{JHEP}
  {\bfseries 01} (2013) 149} [\href{https://arxiv.org/abs/1207.4482}{{\ttfamily
  1207.4482}}].

\bibitem{Blum:2015rpa}
K.~Blum, R.T.~D'Agnolo and J.~Fan, \emph{{Vacuum stability bounds on Higgs
  coupling deviations in the absence of new bosons}},
  \href{https://doi.org/10.1007/JHEP03(2015)166}{\emph{JHEP} {\bfseries 03}
  (2015) 166} [\href{https://arxiv.org/abs/1502.01045}{{\ttfamily
  1502.01045}}].

\bibitem{DAgnolo:2023rnh}
R.T.~D'Agnolo, F.~Nortier, G.~Rigo and P.~Sesma, \emph{{The two scales of new
  physics in Higgs couplings}},
  \href{https://doi.org/10.1007/JHEP08(2023)019}{\emph{JHEP} {\bfseries 08}
  (2023) 019} [\href{https://arxiv.org/abs/2305.19325}{{\ttfamily
  2305.19325}}].

\bibitem{deBlas:2022ofj}
J.~de~Blas, Y.~Du, C.~Grojean, J.~Gu, V.~Miralles, M.E.~Peskin et~al.,
  \emph{{Global SMEFT Fits at Future Colliders}},  in \emph{{Snowmass 2021}},
  6, 2022 [\href{https://arxiv.org/abs/2206.08326}{{\ttfamily 2206.08326}}].

\bibitem{ATLAS:2023yqk}
{\scshape ATLAS, CMS} collaboration, \emph{{Evidence for the Higgs Boson Decay
  to a Z Boson and a Photon at the LHC}},
  \href{https://doi.org/10.1103/PhysRevLett.132.021803}{\emph{Phys. Rev. Lett.}
  {\bfseries 132} (2024) 021803}
  [\href{https://arxiv.org/abs/2309.03501}{{\ttfamily 2309.03501}}].

\bibitem{Djouadi:2007fm}
A.~Djouadi and G.~Moreau, \emph{{Higgs production at the LHC in warped
  extra-dimensional models}},
  \href{https://doi.org/10.1016/j.physletb.2007.11.034}{\emph{Phys. Lett. B}
  {\bfseries 660} (2008) 67} [\href{https://arxiv.org/abs/0707.3800}{{\ttfamily
  0707.3800}}].

\bibitem{Krauss:2007bz}
F.~Krauss, T.E.J.~Underwood and R.~Zwicky, \emph{{Process $gg \rightarrow h_0
  \rightarrow \gamma\gamma$ in the Lee-Wick standard model}},
  \href{https://doi.org/10.1103/PhysRevD.83.019902}{\emph{Phys. Rev. D}
  {\bfseries 77} (2008) 015012}
  [\href{https://arxiv.org/abs/0709.4054}{{\ttfamily 0709.4054}}].

\bibitem{Cacciapaglia:2009ky}
G.~Cacciapaglia, A.~Deandrea and J.~Llodra-Perez, \emph{{$H \rightarrow
  \gamma\gamma$ beyond the Standard Model}},
  \href{https://doi.org/10.1088/1126-6708/2009/06/054}{\emph{JHEP} {\bfseries
  06} (2009) 054} [\href{https://arxiv.org/abs/0901.0927}{{\ttfamily
  0901.0927}}].

\bibitem{Bouchart:2009vq}
C.~Bouchart and G.~Moreau, \emph{{Higgs boson phenomenology and vacuum
  expectation value shift in the Randall-Sundrum scenario}},
  \href{https://doi.org/10.1103/PhysRevD.80.095022}{\emph{Phys. Rev. D}
  {\bfseries 80} (2009) 095022}
  [\href{https://arxiv.org/abs/0909.4812}{{\ttfamily 0909.4812}}].

\bibitem{Gopalakrishna:2009yz}
S.~Gopalakrishna, S.J.~Lee and J.D.~Wells, \emph{{Dark matter and Higgs boson
  collider implications of fermions in an Abelian-gauged hidden sector}},
  \href{https://doi.org/10.1016/j.physletb.2009.08.010}{\emph{Phys. Lett. B}
  {\bfseries 680} (2009) 88} [\href{https://arxiv.org/abs/0904.2007}{{\ttfamily
  0904.2007}}].

\bibitem{Bhattacharyya:2009nb}
G.~Bhattacharyya and T.S.~Ray, \emph{{Probing warped extra dimension via $gg
  \rightarrow h$ and $h \rightarrow \gamma\gamma$ at LHC}},
  \href{https://doi.org/10.1016/j.physletb.2009.03.069}{\emph{Phys. Lett. B}
  {\bfseries 675} (2009) 222}
  [\href{https://arxiv.org/abs/0902.1893}{{\ttfamily 0902.1893}}].

\bibitem{Casagrande:2010si}
S.~Casagrande, F.~Goertz, U.~Haisch, M.~Neubert and T.~Pfoh, \emph{{The
  custodial Randall-Sundrum model: from precision tests to Higgs physics}},
  \href{https://doi.org/10.1007/JHEP09(2010)014}{\emph{JHEP} {\bfseries 09}
  (2010) 014} [\href{https://arxiv.org/abs/1005.4315}{{\ttfamily 1005.4315}}].

\bibitem{Azatov:2010pf}
A.~Azatov, M.~Toharia and L.~Zhu, \emph{{Higgs boson production from gluon
  fusion in warped extra dimensions}},
  \href{https://doi.org/10.1103/PhysRevD.82.056004}{\emph{Phys. Rev. D}
  {\bfseries 82} (2010) 056004}
  [\href{https://arxiv.org/abs/1006.5939}{{\ttfamily 1006.5939}}].

\bibitem{Alves:2011kc}
A.~Alves, E.~Ramirez~Barreto, A.G.~Dias, C.A.~de~S.~Pires, F.S.~Queiroz and
  P.S.~Rodrigues~da Silva, \emph{{Probing 3-3-1 models in diphoton Higgs boson
  decay}}, \href{https://doi.org/10.1103/PhysRevD.84.115004}{\emph{Phys. Rev.
  D} {\bfseries 84} (2011) 115004}
  [\href{https://arxiv.org/abs/1109.0238}{{\ttfamily 1109.0238}}].

\bibitem{Azatov:2011qy}
A.~Azatov and J.~Galloway, \emph{{Light Custodians and Higgs Physics in
  Composite Models}},
  \href{https://doi.org/10.1103/PhysRevD.85.055013}{\emph{Phys. Rev. D}
  {\bfseries 85} (2012) 055013}
  [\href{https://arxiv.org/abs/1110.5646}{{\ttfamily 1110.5646}}].

\bibitem{Goertz:2011hj}
F.~Goertz, U.~Haisch and M.~Neubert, \emph{{Bounds on Warped Extra Dimensions
  from a Standard Model-like Higgs Boson}},
  \href{https://doi.org/10.1016/j.physletb.2012.05.024}{\emph{Phys. Lett. B}
  {\bfseries 713} (2012) 23} [\href{https://arxiv.org/abs/1112.5099}{{\ttfamily
  1112.5099}}].

\bibitem{Ishiwata:2011hr}
K.~Ishiwata and M.B.~Wise, \emph{{Higgs Properties and Fourth Generation
  Leptons}}, \href{https://doi.org/10.1103/PhysRevD.84.055025}{\emph{Phys. Rev.
  D} {\bfseries 84} (2011) 055025}
  [\href{https://arxiv.org/abs/1107.1490}{{\ttfamily 1107.1490}}].

\bibitem{Azatov:2012rj}
A.~Azatov, O.~Bondu, A.~Falkowski, M.~Felcini, S.~Gascon-Shotkin, D.K.~Ghosh
  et~al., \emph{{Higgs boson production via vectorlike top-partner decays:
  Diphoton or multilepton plus multijets channels at the LHC}},
  \href{https://doi.org/10.1103/PhysRevD.85.115022}{\emph{Phys. Rev. D}
  {\bfseries 85} (2012) 115022}
  [\href{https://arxiv.org/abs/1204.0455}{{\ttfamily 1204.0455}}].

\bibitem{Bonne:2012im}
N.~Bonne and G.~Moreau, \emph{{Reproducing the Higgs boson data with
  vector-like quarks}},
  \href{https://doi.org/10.1016/j.physletb.2012.09.063}{\emph{Phys. Lett. B}
  {\bfseries 717} (2012) 409}
  [\href{https://arxiv.org/abs/1206.3360}{{\ttfamily 1206.3360}}].

\bibitem{Moreau:2012da}
G.~Moreau, \emph{{Constraining extra fermion(s) from the Higgs boson data}},
  \href{https://doi.org/10.1103/PhysRevD.87.015027}{\emph{Phys. Rev. D}
  {\bfseries 87} (2013) 015027}
  [\href{https://arxiv.org/abs/1210.3977}{{\ttfamily 1210.3977}}].

\bibitem{Carena:2012xa}
M.~Carena, I.~Low and C.E.M.~Wagner, \emph{{Implications of a modified Higgs to
  diphoton decay width}},
  \href{https://doi.org/10.1007/JHEP08(2012)060}{\emph{JHEP} {\bfseries 08}
  (2012) 060} [\href{https://arxiv.org/abs/1206.1082}{{\ttfamily 1206.1082}}].

\bibitem{Wang:2012gm}
L.~Wang and X.-F.~Han, \emph{{The recent Higgs boson data and Higgs triplet
  model with vector-like quark}},
  \href{https://doi.org/10.1103/PhysRevD.86.095007}{\emph{Phys. Rev. D}
  {\bfseries 86} (2012) 095007}
  [\href{https://arxiv.org/abs/1206.1673}{{\ttfamily 1206.1673}}].

\bibitem{Ajaib:2012eb}
M.A.~Ajaib, I.~Gogoladze and Q.~Shafi, \emph{{Higgs Boson Production and Decay:
  Effects from Light Third Generation and Vectorlike Matter}},
  \href{https://doi.org/10.1103/PhysRevD.86.095028}{\emph{Phys. Rev. D}
  {\bfseries 86} (2012) 095028}
  [\href{https://arxiv.org/abs/1207.7068}{{\ttfamily 1207.7068}}].

\bibitem{Voloshin:2012tv}
M.B.~Voloshin, \emph{{$CP$ violation in Higgs boson diphoton decay in models
  with vectorlike heavy fermions}},
  \href{https://doi.org/10.1103/PhysRevD.86.093016}{\emph{Phys. Rev. D}
  {\bfseries 86} (2012) 093016}
  [\href{https://arxiv.org/abs/1208.4303}{{\ttfamily 1208.4303}}].

\bibitem{Frank:2012nb}
M.~Frank, B.~Korutlu and M.~Toharia, \emph{{Saving the fourth generation Higgs
  with radion mixing}},
  \href{https://doi.org/10.1103/PhysRevD.85.115025}{\emph{Phys. Rev. D}
  {\bfseries 85} (2012) 115025}
  [\href{https://arxiv.org/abs/1204.5944}{{\ttfamily 1204.5944}}].

\bibitem{Carmi:2012yp}
D.~Carmi, A.~Falkowski, E.~Kuflik and T.~Volansky, \emph{{Interpreting LHC
  Higgs Results from Natural New Physics Perspective}},
  \href{https://doi.org/10.1007/JHEP07(2012)136}{\emph{JHEP} {\bfseries 07}
  (2012) 136} [\href{https://arxiv.org/abs/1202.3144}{{\ttfamily 1202.3144}}].

\bibitem{Basso:2012nh}
L.~Basso, O.~Fischer and J.J.~van~der Bij, \emph{{A singlet-triplet extension
  for the Higgs search at LEP and LHC}},
  \href{https://doi.org/10.1209/0295-5075/101/51004}{\emph{EPL} {\bfseries 101}
  (2013) 51004} [\href{https://arxiv.org/abs/1212.5560}{{\ttfamily
  1212.5560}}].

\bibitem{Kumar:2012ww}
K.~Kumar, R.~Vega-Morales and F.~Yu, \emph{{Effects from New Colored States and
  the Higgs Portal on Gluon Fusion and Higgs Decays}},
  \href{https://doi.org/10.1103/PhysRevD.86.113002}{\emph{Phys. Rev. D}
  {\bfseries 86} (2012) 113002}
  [\href{https://arxiv.org/abs/1205.4244}{{\ttfamily 1205.4244}}].

\bibitem{Feng:2013mea}
W.-Z.~Feng and P.~Nath, \emph{{Higgs diphoton rate and mass enhancement with
  vectorlike leptons and the scale of supersymmetry}},
  \href{https://doi.org/10.1103/PhysRevD.87.075018}{\emph{Phys. Rev. D}
  {\bfseries 87} (2013) 075018}
  [\href{https://arxiv.org/abs/1303.0289}{{\ttfamily 1303.0289}}].

\bibitem{Chen:2013dh}
C.-S.~Chen, C.-Q.~Geng, D.~Huang and L.-H.~Tsai, \emph{{Correlation of
  $h\rightarrow \gamma\gamma$ and $Z\gamma$ in Type-II seesaw neutrino model}},
  \href{https://doi.org/10.1016/j.physletb.2013.05.007}{\emph{Phys. Lett. B}
  {\bfseries 723} (2013) 156}
  [\href{https://arxiv.org/abs/1302.0502}{{\ttfamily 1302.0502}}].

\bibitem{Frank:2013un}
M.~Frank, N.~Pourtolami and M.~Toharia, \emph{{Higgs Bosons in Warped Space,
  from the Bulk to the Brane}},
  \href{https://doi.org/10.1103/PhysRevD.87.096003}{\emph{Phys. Rev. D}
  {\bfseries 87} (2013) 096003}
  [\href{https://arxiv.org/abs/1301.7692}{{\ttfamily 1301.7692}}].

\bibitem{Englert:2013tya}
C.~Englert and M.~McCullough, \emph{{Modified Higgs Sectors and NLO Associated
  Production}}, \href{https://doi.org/10.1007/JHEP07(2013)168}{\emph{JHEP}
  {\bfseries 07} (2013) 168} [\href{https://arxiv.org/abs/1303.1526}{{\ttfamily
  1303.1526}}].

\bibitem{Malm:2013jia}
R.~Malm, M.~Neubert, K.~Novotny and C.~Schmell, \emph{{5D perspective on Higgs
  production at the boundary of a warped extra dimension}},
  \href{https://doi.org/10.1007/JHEP01(2014)173}{\emph{JHEP} {\bfseries 01}
  (2014) 173} [\href{https://arxiv.org/abs/1303.5702}{{\ttfamily 1303.5702}}].

\bibitem{Dey:2013cqa}
U.K.~Dey and T.S.~Ray, \emph{{Constraining minimal and nonminimal universal
  extra dimension models with Higgs couplings}},
  \href{https://doi.org/10.1103/PhysRevD.88.056016}{\emph{Phys. Rev. D}
  {\bfseries 88} (2013) 056016}
  [\href{https://arxiv.org/abs/1305.1016}{{\ttfamily 1305.1016}}].

\bibitem{Hahn:2013nza}
J.~Hahn, C.~H\"orner, R.~Malm, M.~Neubert, K.~Novotny and C.~Schmell,
  \emph{{Higgs decay into two photons in a warped extra dimension}},
  \href{https://doi.org/10.1140/epjc/s10052-014-2857-8}{\emph{Eur. Phys. J. C}
  {\bfseries 74} (2014) 2857}
  [\href{https://arxiv.org/abs/1312.5731}{{\ttfamily 1312.5731}}].

\bibitem{Dermisek:2013gta}
R.~Dermisek and A.~Raval, \emph{{Explanation of the Muon g-2 Anomaly with
  Vectorlike Leptons and its Implications for Higgs Decays}},
  \href{https://doi.org/10.1103/PhysRevD.88.013017}{\emph{Phys. Rev. D}
  {\bfseries 88} (2013) 013017}
  [\href{https://arxiv.org/abs/1305.3522}{{\ttfamily 1305.3522}}].

\bibitem{Aguilar-Saavedra:2013qpa}
J.A.~Aguilar-Saavedra, R.~Benbrik, S.~Heinemeyer and M.~P\'erez-Victoria,
  \emph{{Handbook of vectorlike quarks: Mixing and single production}},
  \href{https://doi.org/10.1103/PhysRevD.88.094010}{\emph{Phys. Rev. D}
  {\bfseries 88} (2013) 094010}
  [\href{https://arxiv.org/abs/1306.0572}{{\ttfamily 1306.0572}}].

\bibitem{Carmona:2013cq}
A.~Carmona and F.~Goertz, \emph{{Custodial Leptons and Higgs Decays}},
  \href{https://doi.org/10.1007/JHEP04(2013)163}{\emph{JHEP} {\bfseries 04}
  (2013) 163} [\href{https://arxiv.org/abs/1301.5856}{{\ttfamily 1301.5856}}].

\bibitem{Delaunay:2013iia}
C.~Delaunay, C.~Grojean and G.~Perez, \emph{{Modified Higgs Physics from
  Composite Light Flavors}},
  \href{https://doi.org/10.1007/JHEP09(2013)090}{\emph{JHEP} {\bfseries 09}
  (2013) 090} [\href{https://arxiv.org/abs/1303.5701}{{\ttfamily 1303.5701}}].

\bibitem{Malm:2014gha}
R.~Malm, M.~Neubert and C.~Schmell, \emph{{Higgs couplings and phenomenology in
  a warped extra dimension}},
  \href{https://doi.org/10.1007/JHEP02(2015)008}{\emph{JHEP} {\bfseries 02}
  (2015) 008} [\href{https://arxiv.org/abs/1408.4456}{{\ttfamily 1408.4456}}].

\bibitem{Dey:2015pba}
U.K.~Dey and T.S.~Ray, \emph{{Higgs-Gluon Coupling in Warped Extra Dimensional
  Models with Brane Kinetic Terms}},
  \href{https://doi.org/10.1103/PhysRevD.93.011901}{\emph{Phys. Rev. D}
  {\bfseries 93} (2016) 011901}
  [\href{https://arxiv.org/abs/1507.04357}{{\ttfamily 1507.04357}}].

\bibitem{Angelescu:2015kga}
A.~Angelescu, A.~Djouadi and G.~Moreau, \emph{{Vector-like top/bottom-quark
  partners and Higgs physics at the LHC}},
  \href{https://doi.org/10.1140/epjc/s10052-016-3950-y}{\emph{Eur. Phys. J. C}
  {\bfseries 76} (2016) 99} [\href{https://arxiv.org/abs/1510.07527}{{\ttfamily
  1510.07527}}].

\bibitem{Lalak:2015xea}
Z.~Lalak, M.~Lewicki and J.D.~Wells, \emph{{Higgs boson mass and
  high-luminosity LHC probes of supersymmetry with vectorlike top quark}},
  \href{https://doi.org/10.1103/PhysRevD.91.095022}{\emph{Phys. Rev. D}
  {\bfseries 91} (2015) 095022}
  [\href{https://arxiv.org/abs/1502.05702}{{\ttfamily 1502.05702}}].

\bibitem{Dermisek:2015hue}
R.~Dermisek, E.~Lunghi and S.~Shin, \emph{{New decay modes of heavy Higgs
  bosons in a two Higgs doublet model with vectorlike leptons}},
  \href{https://doi.org/10.1007/JHEP05(2016)148}{\emph{JHEP} {\bfseries 05}
  (2016) 148} [\href{https://arxiv.org/abs/1512.07837}{{\ttfamily
  1512.07837}}].

\bibitem{Arhrib:2016rlj}
A.~Arhrib, R.~Benbrik, S.J.D.~King, B.~Manaut, S.~Moretti and C.S.~Un,
  \emph{{Phenomenology of 2HDM with vectorlike quarks}},
  \href{https://doi.org/10.1103/PhysRevD.97.095015}{\emph{Phys. Rev. D}
  {\bfseries 97} (2018) 095015}
  [\href{https://arxiv.org/abs/1607.08517}{{\ttfamily 1607.08517}}].

\bibitem{Hashimoto:2017jvc}
M.~Hashimoto, \emph{{Revisiting vectorlike quark models with enhanced top
  Yukawa coupling}},
  \href{https://doi.org/10.1103/PhysRevD.96.035020}{\emph{Phys. Rev. D}
  {\bfseries 96} (2017) 035020}
  [\href{https://arxiv.org/abs/1704.02615}{{\ttfamily 1704.02615}}].

\bibitem{Chen:2017hak}
C.-Y.~Chen, S.~Dawson and E.~Furlan, \emph{{Vectorlike fermions and Higgs
  effective field theory revisited}},
  \href{https://doi.org/10.1103/PhysRevD.96.015006}{\emph{Phys. Rev. D}
  {\bfseries 96} (2017) 015006}
  [\href{https://arxiv.org/abs/1703.06134}{{\ttfamily 1703.06134}}].

\bibitem{Poh:2017tfo}
Z.~Poh and S.~Raby, \emph{{Vectorlike leptons: Muon $g-2$ anomaly, lepton
  flavor violation, Higgs boson decays, and lepton nonuniversality}},
  \href{https://doi.org/10.1103/PhysRevD.96.015032}{\emph{Phys. Rev. D}
  {\bfseries 96} (2017) 015032}
  [\href{https://arxiv.org/abs/1705.07007}{{\ttfamily 1705.07007}}].

\bibitem{Aboubrahim:2018hll}
A.~Aboubrahim, T.~Ibrahim, A.~Itani and P.~Nath, \emph{{Observables of
  low-lying supersymmetric vectorlike leptonic generations via loop
  corrections}}, \href{https://doi.org/10.1103/PhysRevD.98.075009}{\emph{Phys.
  Rev. D} {\bfseries 98} (2018) 075009}
  [\href{https://arxiv.org/abs/1808.00071}{{\ttfamily 1808.00071}}].

\bibitem{Barducci:2023zml}
D.~Barducci, L.~Di~Luzio, M.~Nardecchia and C.~Toni, \emph{{Closing in on new
  chiral leptons at the LHC}},
  \href{https://doi.org/10.1007/JHEP12(2023)154}{\emph{JHEP} {\bfseries 12}
  (2023) 154} [\href{https://arxiv.org/abs/2311.10130}{{\ttfamily
  2311.10130}}].

\bibitem{Isidori:2001bm}
G.~Isidori, G.~Ridolfi and A.~Strumia, \emph{{On the metastability of the
  Standard Model vacuum}},
  \href{https://doi.org/10.1016/S0550-3213(01)00302-9}{\emph{Nucl. Phys. B}
  {\bfseries 609} (2001) 387}
  [\href{https://arxiv.org/abs/hep-ph/0104016}{{\ttfamily hep-ph/0104016}}].

\bibitem{Elias-Miro:2011sqh}
J.~Elias-Miro, J.R.~Espinosa, G.F.~Giudice, G.~Isidori, A.~Riotto and
  A.~Strumia, \emph{{Higgs mass implications on the stability of the
  electroweak vacuum}},
  \href{https://doi.org/10.1016/j.physletb.2012.02.013}{\emph{Phys. Lett. B}
  {\bfseries 709} (2012) 222}
  [\href{https://arxiv.org/abs/1112.3022}{{\ttfamily 1112.3022}}].

\bibitem{Degrassi:2012ry}
G.~Degrassi, S.~Di~Vita, J.~Elias-Miro, J.R.~Espinosa, G.F.~Giudice, G.~Isidori
  et~al., \emph{{Higgs mass and vacuum stability in the Standard Model at
  NNLO}}, \href{https://doi.org/10.1007/JHEP08(2012)098}{\emph{JHEP} {\bfseries
  08} (2012) 098} [\href{https://arxiv.org/abs/1205.6497}{{\ttfamily
  1205.6497}}].

\bibitem{Buttazzo:2013uya}
D.~Buttazzo, G.~Degrassi, P.P.~Giardino, G.F.~Giudice, F.~Sala, A.~Salvio
  et~al., \emph{{Investigating the near-criticality of the Higgs boson}},
  \href{https://doi.org/10.1007/JHEP12(2013)089}{\emph{JHEP} {\bfseries 12}
  (2013) 089} [\href{https://arxiv.org/abs/1307.3536}{{\ttfamily 1307.3536}}].

\bibitem{Devoto:2022qen}
F.~Devoto, S.~Devoto, L.~Di~Luzio and G.~Ridolfi, \emph{{False vacuum decay: an
  introductory review}},
  \href{https://doi.org/10.1088/1361-6471/ac7f24}{\emph{J. Phys. G} {\bfseries
  49} (2022) 103001} [\href{https://arxiv.org/abs/2205.03140}{{\ttfamily
  2205.03140}}].

\bibitem{Dreiner:2008tw}
H.K.~Dreiner, H.E.~Haber and S.P.~Martin, \emph{{Two-component spinor
  techniques and Feynman rules for quantum field theory and supersymmetry}},
  \href{https://doi.org/10.1016/j.physrep.2010.05.002}{\emph{Phys. Rept.}
  {\bfseries 494} (2010) 1} [\href{https://arxiv.org/abs/0812.1594}{{\ttfamily
  0812.1594}}].

\bibitem{Milagre:2024wcg}
A.~Milagre and L.~Lavoura, \emph{{Unitarity constraints on large multiplets of
  arbitrary gauge groups}},
  \href{https://doi.org/10.1016/j.nuclphysb.2024.116542}{\emph{Nucl. Phys. B}
  {\bfseries 1004} (2024) 116542}
  [\href{https://arxiv.org/abs/2403.12914}{{\ttfamily 2403.12914}}].

\bibitem{Wells:2005vk}
J.D.~Wells, \emph{{Introduction to Precision Electroweak Analysis}},  in
  \emph{{TASI 2004}: {Physics in D $\geqq$ 4}}, World Scientific Publishing Co
  Pte Ltd, 7, 2006, \href{https://doi.org/10.1142/9789812773579_0002}{DOI}
  [\href{https://arxiv.org/abs/hep-ph/0512342}{{\ttfamily hep-ph/0512342}}].

\bibitem{Albergaria:2023nby}
F.~Albergaria, D.~Jur\v{c}iukonis and L.~Lavoura, \emph{{The oblique parameters
  from arbitrary new fermions}},
  \href{https://doi.org/10.1007/JHEP05(2024)190}{\emph{JHEP} {\bfseries 05}
  (2024) 190} [\href{https://arxiv.org/abs/2312.09099}{{\ttfamily
  2312.09099}}].

\bibitem{Haller:2018nnx}
J.~Haller, A.~Hoecker, R.~Kogler, K.~M\"onig, T.~Peiffer and J.~Stelzer,
  \emph{{Update of the global electroweak fit and constraints on
  two-Higgs-doublet models}},
  \href{https://doi.org/10.1140/epjc/s10052-018-6131-3}{\emph{Eur. Phys. J. C}
  {\bfseries 78} (2018) 675}
  [\href{https://arxiv.org/abs/1803.01853}{{\ttfamily 1803.01853}}].

\bibitem{Patel:2015tea}
H.H.~Patel, \emph{{Package-X: A Mathematica package for the analytic
  calculation of one-loop integrals}},
  \href{https://doi.org/10.1016/j.cpc.2015.08.017}{\emph{Comput. Phys. Commun.}
  {\bfseries 197} (2015) 276}
  [\href{https://arxiv.org/abs/1503.01469}{{\ttfamily 1503.01469}}].

\bibitem{Patel:2016fam}
H.H.~Patel, \emph{{Package-X 2.0: A Mathematica package for the analytic
  calculation of one-loop integrals}},
  \href{https://doi.org/10.1016/j.cpc.2017.04.015}{\emph{Comput. Phys. Commun.}
  {\bfseries 218} (2017) 66}
  [\href{https://arxiv.org/abs/1612.00009}{{\ttfamily 1612.00009}}].

\bibitem{Sikivie:1980hm}
P.~Sikivie, L.~Susskind, M.B.~Voloshin and V.I.~Zakharov, \emph{{Isospin
  breaking in technicolor models}},
  \href{https://doi.org/10.1016/0550-3213(80)90214-X}{\emph{Nucl. Phys. B}
  {\bfseries 173} (1980) 189}.

\bibitem{Appelquist:1974tg}
T.~Appelquist and J.~Carazzone, \emph{{Infrared singularities and massive
  fields}}, \href{https://doi.org/10.1103/PhysRevD.11.2856}{\emph{Phys. Rev. D}
  {\bfseries 11} (1975) 2856}.

\bibitem{Wilson:1983xri}
K.G.~Wilson, \emph{{The renormalization group and critical phenomena}},
  \href{https://doi.org/10.1103/RevModPhys.55.583}{\emph{Rev. Mod. Phys.}
  {\bfseries 55} (1983) 583}.

\bibitem{ATLAS:2024fdw}
{\scshape ATLAS} collaboration, \emph{{Exploration at the high-energy frontier:
  ATLAS Run 2 searches investigating the exotic jungle beyond the Standard
  Model}},  \href{https://arxiv.org/abs/2403.09292}{{\ttfamily 2403.09292}}.

\bibitem{CMS:2024bni}
{\scshape CMS} collaboration, \emph{{Review of searches for vector-like quarks,
  vector-like leptons, and heavy neutral leptons in proton-proton collisions at
  $\sqrt{s}$ = 13 TeV at the CMS experiment}},
  \href{https://arxiv.org/abs/2405.17605}{{\ttfamily 2405.17605}}.

\bibitem{CMS:2024nhn}
{\scshape CMS} collaboration, \emph{{Search for heavy long-lived charged
  particles with large ionization energy loss in proton-proton collisions at
  $\sqrt{s}$ = 13 TeV}},  \href{https://arxiv.org/abs/2410.09164}{{\ttfamily
  2410.09164}}.

\bibitem{ATLAS:2023sbu}
{\scshape ATLAS} collaboration, \emph{{Search for third-generation vector-like
  leptons in $pp$ collisions at $\sqrt{s} = 13\,\text{TeV}$ with the ATLAS
  detector}}, \href{https://doi.org/10.1007/JHEP07(2023)118}{\emph{JHEP}
  {\bfseries 07} (2023) 118}
  [\href{https://arxiv.org/abs/2303.05441}{{\ttfamily 2303.05441}}].

\bibitem{CMS:2018szt}
{\scshape CMS} collaboration, \emph{{Combined search for electroweak production
  of charginos and neutralinos in proton-proton collisions at $\sqrt{s} =$ 13
  TeV}}, \href{https://doi.org/10.1007/JHEP03(2018)160}{\emph{JHEP} {\bfseries
  03} (2018) 160} [\href{https://arxiv.org/abs/1801.03957}{{\ttfamily
  1801.03957}}].

\bibitem{CMS:2020bfa}
{\scshape CMS} collaboration, \emph{{Search for supersymmetry in final states
  with two oppositely charged same-flavor leptons and missing transverse
  momentum in proton-proton collisions at $\sqrt{s} =$ 13 TeV}},
  \href{https://doi.org/10.1007/JHEP04(2021)123}{\emph{JHEP} {\bfseries 04}
  (2021) 123} [\href{https://arxiv.org/abs/2012.08600}{{\ttfamily
  2012.08600}}].

\bibitem{CMS:2021edw}
{\scshape CMS} collaboration, \emph{{Search for supersymmetry in final states
  with two or three soft leptons and missing transverse momentum in
  proton-proton collisions at $ \sqrt{s} $ = 13 TeV}},
  \href{https://doi.org/10.1007/JHEP04(2022)091}{\emph{JHEP} {\bfseries 04}
  (2022) 091} [\href{https://arxiv.org/abs/2111.06296}{{\ttfamily
  2111.06296}}].

\bibitem{CMS:2021cox}
{\scshape CMS} collaboration, \emph{{Search for electroweak production of
  charginos and neutralinos in proton-proton collisions at $ \sqrt{s} $ = 13
  TeV}}, \href{https://doi.org/10.1007/JHEP04(2022)147}{\emph{JHEP} {\bfseries
  04} (2022) 147} [\href{https://arxiv.org/abs/2106.14246}{{\ttfamily
  2106.14246}}].

\bibitem{CMS:2022sfi}
{\scshape CMS} collaboration, \emph{{Search for electroweak production of
  charginos and neutralinos at s=13TeV in final states containing hadronic
  decays of WW, WZ, or WH and missing transverse momentum}},
  \href{https://doi.org/10.1016/j.physletb.2022.137460}{\emph{Phys. Lett. B}
  {\bfseries 842} (2023) 137460}
  [\href{https://arxiv.org/abs/2205.09597}{{\ttfamily 2205.09597}}].

\bibitem{ATLAS:2018eui}
{\scshape ATLAS} collaboration, \emph{{Search for chargino-neutralino
  production using recursive jigsaw reconstruction in final states with two or
  three charged leptons in proton-proton collisions at $\sqrt{s}=13$ TeV with
  the ATLAS detector}},
  \href{https://doi.org/10.1103/PhysRevD.98.092012}{\emph{Phys. Rev. D}
  {\bfseries 98} (2018) 092012}
  [\href{https://arxiv.org/abs/1806.02293}{{\ttfamily 1806.02293}}].

\bibitem{ATLAS:2019lff}
{\scshape ATLAS} collaboration, \emph{{Search for electroweak production of
  charginos and sleptons decaying into final states with two leptons and
  missing transverse momentum in $\sqrt{s}=13$ TeV $pp$ collisions using the
  ATLAS detector}},
  \href{https://doi.org/10.1140/epjc/s10052-019-7594-6}{\emph{Eur. Phys. J. C}
  {\bfseries 80} (2020) 123}
  [\href{https://arxiv.org/abs/1908.08215}{{\ttfamily 1908.08215}}].

\bibitem{ATLAS:2020pgy}
{\scshape ATLAS} collaboration, \emph{{Search for direct production of
  electroweakinos in final states with one lepton, missing transverse momentum
  and a Higgs boson decaying into two $b$-jets in $pp$ collisions at
  $\sqrt{s}=13$ TeV with the ATLAS detector}},
  \href{https://doi.org/10.1140/epjc/s10052-020-8050-3}{\emph{Eur. Phys. J. C}
  {\bfseries 80} (2020) 691}
  [\href{https://arxiv.org/abs/1909.09226}{{\ttfamily 1909.09226}}].

\bibitem{ATLAS:2019lng}
{\scshape ATLAS} collaboration, \emph{{Searches for electroweak production of
  supersymmetric particles with compressed mass spectra in $\sqrt{s}=$ 13 TeV
  $pp$ collisions with the ATLAS detector}},
  \href{https://doi.org/10.1103/PhysRevD.101.052005}{\emph{Phys. Rev. D}
  {\bfseries 101} (2020) 052005}
  [\href{https://arxiv.org/abs/1911.12606}{{\ttfamily 1911.12606}}].

\bibitem{ATLAS:2021moa}
{\scshape ATLAS} collaboration, \emph{{Search for
  chargino\textendash{}neutralino pair production in final states with three
  leptons and missing transverse momentum in $\sqrt{s} = 13$~TeV pp collisions
  with the ATLAS detector}},
  \href{https://doi.org/10.1140/epjc/s10052-021-09749-7}{\emph{Eur. Phys. J. C}
  {\bfseries 81} (2021) 1118}
  [\href{https://arxiv.org/abs/2106.01676}{{\ttfamily 2106.01676}}].

\bibitem{ATLAS:2021yqv}
{\scshape ATLAS} collaboration, \emph{{Search for charginos and neutralinos in
  final states with two boosted hadronically decaying bosons and missing
  transverse momentum in $pp$ collisions at $\sqrt {s}$ = 13\,\,TeV with the
  ATLAS detector}},
  \href{https://doi.org/10.1103/PhysRevD.104.112010}{\emph{Phys. Rev. D}
  {\bfseries 104} (2021) 112010}
  [\href{https://arxiv.org/abs/2108.07586}{{\ttfamily 2108.07586}}].

\bibitem{ATLAS:2022zwa}
{\scshape ATLAS} collaboration, \emph{{Searches for new phenomena in events
  with two leptons, jets, and missing transverse momentum in 139~fb$^{-1}$ of
  $\sqrt{s}=13$~TeV $pp$ collisions with the ATLAS detector}},
  \href{https://doi.org/10.1140/epjc/s10052-023-11434-w}{\emph{Eur. Phys. J. C}
  {\bfseries 83} (2023) 515}
  [\href{https://arxiv.org/abs/2204.13072}{{\ttfamily 2204.13072}}].

\bibitem{ATLAS:2022hbt}
{\scshape ATLAS} collaboration, \emph{{Search for direct pair production of
  sleptons and charginos decaying to two leptons and neutralinos with mass
  splittings near the W-boson mass in $ \sqrt{s} $ = 13 TeV pp collisions with
  the ATLAS detector}},
  \href{https://doi.org/10.1007/JHEP06(2023)031}{\emph{JHEP} {\bfseries 06}
  (2023) 031} [\href{https://arxiv.org/abs/2209.13935}{{\ttfamily
  2209.13935}}].

\bibitem{ATLAS:2023zxo}
{\scshape ATLAS} collaboration, \emph{{Search for heavy long-lived
  multi-charged particles in the full LHC Run 2 pp collision data at s=13 TeV
  using the ATLAS detector}},
  \href{https://doi.org/10.1016/j.physletb.2023.138316}{\emph{Phys. Lett. B}
  {\bfseries 847} (2023) 138316}
  [\href{https://arxiv.org/abs/2303.13613}{{\ttfamily 2303.13613}}].

\bibitem{CMS:2016ybj}
{\scshape CMS} collaboration, \emph{{Search for heavy stable charged particles
  with $12.9~\mathrm{fb}^{-1}$ of 2016 data}},  2016,
  \href{https://inspirehep.net/literature/1479657}{https://inspirehep.net/literature/1479657}.

\bibitem{ATLAS:2018yey}
{\scshape ATLAS} collaboration, \emph{{Reinterpretation of searches for
  supersymmetry in models with variable $R$-parity-violating coupling strength
  and long-lived $R$-hadrons}},  2018,
  \href{https://inspirehep.net/literature/1662545}{https://inspirehep.net/literature/1662545}.

\bibitem{Staub:2008uz}
F.~Staub, \emph{{SARAH}},  \href{https://arxiv.org/abs/0806.0538}{{\ttfamily
  0806.0538}}.

\bibitem{Staub:2009bi}
F.~Staub, \emph{{From Superpotential to Model Files for FeynArts and
  CalcHep/CompHep}},
  \href{https://doi.org/10.1016/j.cpc.2010.01.011}{\emph{Comput. Phys. Commun.}
  {\bfseries 181} (2010) 1077}
  [\href{https://arxiv.org/abs/0909.2863}{{\ttfamily 0909.2863}}].

\bibitem{Staub:2010jh}
F.~Staub, \emph{{Automatic Calculation of supersymmetric Renormalization Group
  Equations and Self Energies}},
  \href{https://doi.org/10.1016/j.cpc.2010.11.030}{\emph{Comput. Phys. Commun.}
  {\bfseries 182} (2011) 808}
  [\href{https://arxiv.org/abs/1002.0840}{{\ttfamily 1002.0840}}].

\bibitem{Staub:2012pb}
F.~Staub, \emph{{SARAH 3.2: Dirac Gauginos, UFO output, and more}},
  \href{https://doi.org/10.1016/j.cpc.2013.02.019}{\emph{Comput. Phys. Commun.}
  {\bfseries 184} (2013) 1792}
  [\href{https://arxiv.org/abs/1207.0906}{{\ttfamily 1207.0906}}].

\bibitem{Staub:2013tta}
F.~Staub, \emph{{SARAH 4: A tool for (not only SUSY) model builders}},
  \href{https://doi.org/10.1016/j.cpc.2014.02.018}{\emph{Comput. Phys. Commun.}
  {\bfseries 185} (2014) 1773}
  [\href{https://arxiv.org/abs/1309.7223}{{\ttfamily 1309.7223}}].

\bibitem{Staub:2015kfa}
F.~Staub, \emph{{Exploring new models in all detail with SARAH}},
  \href{https://doi.org/10.1155/2015/840780}{\emph{Adv. High Energy Phys.}
  {\bfseries 2015} (2015) 840780}
  [\href{https://arxiv.org/abs/1503.04200}{{\ttfamily 1503.04200}}].

\bibitem{Staub:2015iza}
F.~Staub, \emph{{Introduction to SARAH and related tools}},
  \href{https://doi.org/10.22323/1.263.0027}{\emph{PoS} {\bfseries CORFU2015}
  (2016) 027} [\href{https://arxiv.org/abs/1509.07061}{{\ttfamily
  1509.07061}}].

\bibitem{Staub:2016sms}
F.~Staub, \emph{{Tutorial to SARAH}},
  \href{https://doi.org/10.22323/1.263.0058}{\emph{PoS} {\bfseries CORFU2015}
  (2016) 058} [\href{https://arxiv.org/abs/1603.05958}{{\ttfamily
  1603.05958}}].

\bibitem{Goodsell:2018tti}
M.D.~Goodsell and F.~Staub, \emph{{Unitarity constraints on general scalar
  couplings with SARAH}},
  \href{https://doi.org/10.1140/epjc/s10052-018-6127-z}{\emph{Eur. Phys. J. C}
  {\bfseries 78} (2018) 649}
  [\href{https://arxiv.org/abs/1805.07306}{{\ttfamily 1805.07306}}].

\end{thebibliography}\endgroup

\end{document}